\documentclass[a4paper,oldversion]{aa}
\usepackage{psfig}
\usepackage{pifont}
\usepackage{lscape}
\usepackage{amssymb}

\newcommand{\lya}{\ifmmode {\rm Ly}\alpha \else Ly$\alpha$\fi}
\def\micron{$\mu$m}

\def\erg{ergs s$^{-1}$ cm$^{-2}$ \AA$^{-1}$}
\def\ergs{ergs s$^{-1}$}

\def\msun{\ifmmode M_{\odot} \else M$_{\odot}$\fi}
\def\zsun{\ifmmode Z_{\odot} \else Z$_{\odot}$\fi}
\def\lsun{\ifmmode L_{\odot} \else L$_{\odot}$\fi}

\def\aap{A\&A}

\def\aj{AJ}
\def\apj{ApJ}
\def\apjl{ApJ}
\def\apjs{ApJS}
\def\mnras{MNRAS}

\begin{document}

   \title{
Constraining the population of $6 \lesssim z \lesssim 10$ star-forming
galaxies with deep near-IR images of lensing clusters 
   \thanks{Based on observations collected at the Very Large Telescope (Antu/UT1), European Southern Observatory, Paranal, Chile (ESO Programs 69.A-0508, 70.A-0355, 271.A-5013, 272.A-5049, 73.A-0471) and the NASA/ESA \textit{Hubble Space Telescope} obtained at the Space Telescope Science Institute, which is operated by AURA under NASA contract NAS5-26555}}
   \titlerunning{Abundance of $z\sim6-10$ galaxies from lensing clusters}
   \author{J. Richard\inst{1,5}\and R. Pell\'o\inst{1}\and D. Schaerer\inst{2,1}
\and J.-F. Le Borgne\inst{1}\and J.-P. Kneib\inst{3,4}}
   \authorrunning{J. Richard et al.\ }
   \offprints{Johan Richard, johan@astro.caltech.edu}

   \institute{ Observatoire Midi-Pyr\'en\'ees, Laboratoire
d'Astrophysique, UMR 5572, 14 Avenue E. Belin, F-31400 Toulouse, France
\and Geneva Observatory, 51 Ch. des Maillettes, CH--1290 Sauverny, Switzerland
\and OAMP, Laboratoire d'Astrophysique de Marseille, UMR 6110 traverse du Siphon, 13012 Marseille, France
\and Caltech Astronomy, MC105-24, Pasadena, CA 91125, USA 
\and Present address : Caltech Astronomy, MC105-24, Pasadena, CA 91125, USA 
}
   \date{Received \today }

 \abstract{
We present the first results of our deep survey of lensing
clusters aimed at constraining the abundance of star-forming galaxies at
$z\sim6-10$, using lensing magnification to improve the search
efficiency and subsequent spectroscopic studies.  
Deep near-IR photometry of two lensing clusters (A1835 and AC114) was obtained
with ISAAC/VLT. These images, combined with existing data in the optical
bands including HST images, were used to select very high redshift candidates
at $z\gtrsim 6$ among the optical-dropouts. Photometric selection criteria
have been defined based on the well-proven dropout technique, specifically
tuned to target star-forming galaxies in this redshift domain. 

We have identified 18(8) first and second-category optical dropouts in A1835 (AC114),
detected in more than one filter up to $H$ (Vega) $\sim 23.8$ (AB $\sim 25.2$,
uncorrected for lensing).
Among them, 8(5) exhibit homogeneous SEDs compatible with star-forming galaxies at $z\gtrsim
6$, and 5(1) are more likely intermediate-redshift EROs
based on luminosity considerations. We have also
identified a number of fainter sources in these fields fulfilling our
photometric selection and located around the critical lines. 
We use all these data to make a first attempt at constraining the density
of star-forming galaxies present at $6\lesssim z \lesssim10$ using lensing clusters.
Magnification effects and sample incompleteness are addressed
through a careful modeling of the lensing clusters. A correction was also
introduced to account for the expected fraction of false-positive detections
among this photometric sample. 

It appears that 
the number of candidates found in these lensing fields, corrected for
magnification, incompleteness and false-positive detections, is 
higher than the one achieved in blank fields with similar photometric
depth in the near-IR. 
The luminosity function derived for $z\gtrsim 6$ candidates appears 
compatible with that of LBGs at $z\simeq 3$, without any renormalization. 
The turnover observed by Bouwens et al. (2005) towards the bright end relative
to the $z\sim 3$ LF is not observed in this sample.
Also the upper limit for the
UV SFR density at $z\sim6-10$, integrated down to
$L_{1500}=0.3\ L^{*}_{z=3}$, of $\rho_\star=7.4\ 10^{-2}$ M$_{\odot}$
yr$^{-1}$ Mpc$^{-3}$ is compatible with the usual values derived at $z \simeq
5-6$, but higher than the estimates obtained in the NICMOS Ultra Deep
Field (UDF). The same holds for the upper limit of the SFR density in the $z
\simeq 8-10$ interval ($\rho_\star=1.1\ 10^{-1}$). 
This systematic trend towards the bright end of the LF with respect to blank
fields could be due to field-to-field variance, a positive
magnification bias from intermediate-redshift EROs, 
and/or residual contamination. Given the low S/N ratio of the high-$z$
candidates, and the large correction factors applied to this sample,
increasing the number of blank and lensing fields with ultra-deep
near-IR photometry is essential to obtain more accurate constraints on the abundance
of $z \gtrsim 6$ galaxies.

      }

\maketitle
\section{\label{intro}Introduction}
During the last decade considerable advances have been made in the
exploration of the early Universe, from the discovery and detailed
studies of redshift $z \sim 3$ galaxies (the so-called Lyman break galaxies,
LBGs, e.g. Steidel et al.\ 2003),
to $z \sim$ 4--5 galaxies found from numerous deep multi-wavelength
surveys, to galaxies at $z \sim$ 6--7, close to what is believed to be the end of
reionization epoch of the Universe (e.g. Hu et al.\ 2002, Kodaira et al.\ 2003, 
Cuby et al.\ 2003, Kneib et al. 2004, Stanway et al.\ 2004, Bouwens et al.\ 2004b).
Extending the searches beyond $z\simeq$ 6.5 and back to ages where the
Universe re-ionized (cf.\ Fan et al.\ 2002) requires extremely deep
observations in the near-IR bands. Astounding depths can be reached in
ultra-deep fields, as demonstrated e.g.\ with $J$ and $H$ imaging of
the NICMOS Ultra-Deep Field (UDF; Thompson et al.\ 2005; Bouwens et al. 2004a)
from which 5 faint ($H_{AB} \sim$ 27) candidates at
$z \sim$ 7--8 have been identified (Bouwens et al.\ 2004b).

On the other hand, recent {\tt WMAP} results seem to place the first building
blocks of the Universe at redshifts up to $z\sim 10-15$ (Spergel et al.\ 2006) and very distant
star-forming systems could have been responsible for a significant part of the
cosmic reionization. The end of this epoch is suggested to be at $z\sim
6.0-6.5$ from the spectrum of high redshift quasars (Fan et al.\ 2002). Therefore,
constraining the abundance of $z>7$ sources is an important challenge  
of modern cosmology.

Photometric selection of high-redshift galaxies, such as the
Lyman-break technique (e.g. Steidel et al.\ 1995), has been
successful in identifying star-forming objects at $z \sim$ 2-4 (cf.\
Steidel et al.\ 2003, Shapley et al.\ 2003)
and up to $z\sim 6$ (Bunker et al.\ 2003, Mobasher et al.\ 2005).
At redshifts $z\gtrsim 5$, only $\approx$ 30
galaxies are currently known with confirmed redshifts
(cf.\ review by Spinrad 2003).
For now, more than 5 galaxies with $z\sim
6.5$ (Hu et al.\ 2002, Kodaira et al.\ 2003, Cuby et al.\ 2003, Kneib et al.\ 2004)
are known with secure redshifts. These objects are generally found through
their Lyman-$\alpha$ emission which is redshifted into the $\sim$ 9200 \AA\
window, the ``reddest'' window relatively free of skylines in the optical.
The abundance of $z\sim10$ galaxies was recently discussed
by Bouwens et al.\ (2005a) using NICMOS-UDF $J_{110}$ and $H_{160}$
data. Their conclusion is that strong evolution exists between $z \sim 7-8$
and $z \sim 3-4$, the SFR density being much lower at very high $z$ 
down to the
limits of their survey ($L_{1500}=0.3L^{*}_{z=3}$). However, it is crucial to
increase both the size and the depth of the surveyed field to set strong
constraints on the star-formation at $z\gtrsim 7$, as field-to-field variance can be important.

In this paper we present the first results of our deep survey of lensing
clusters aimed at constraining the abundance of star-forming galaxies at
$z\sim6-10$, using lensing magnification to improve the search
efficiency and subsequent spectroscopic studies. The motivations are the
following. 
On the one hand, our understanding of the first generation of stars and
galaxies, the so-called Population III objects (cf. review of Loeb
\& Barkana 2001), has improved with the
development of new models for these low-metallicity starbursts 
(Tumlinson \& Shull 2000, Bromm et al.\ 2001, Schaerer 2002, Schaerer 2003). 
The observational properties inferred 
from this modeling show us that it is now possible to detect some
of these objects at $z\sim 8-10$, thanks to the intensity of
their emission lines, using very deep near-IR surveys on 8 m-class
telescopes (e.g. Schaerer \& Pell\'o 2001, Barton et al.\ 2004).

Our project is based on the photometric pre-selection of candidates making use
of the natural magnification provided by foreground lensing
clusters. This technique, first referred to as the ``gravitational telescope''
by Zwicky, has proven highly successful in identifying a large fraction
of the most distant galaxies known today thanks to magnifications
by typically 1--3 magnitudes (e.g.\  Franx et al.\ 1997, Ellis et al.\ 2001, Hu et al.\ 2002,
Kneib et al.\ 2004). 

We present a color-color selection similar to the Lyman-Break technique, used to
identify very high redshift objects using their specific spectrophotometric
properties. As first targets for our survey, we have chosen fields centered on
lensing clusters with well-constrained mass distributions, and already known to be
efficient gravitational telescopes. We use the sample of high-redshift
candidates selected in these fields to constrain the abundance of star-forming
galaxies up to $z\sim10$.

The plan of this paper is as follows: in Sect.\ \ref{criteria} we justify the
observing strategy and the photometric criteria adopted in this
project. Photometric observations and data reduction are described in detail
in Sect.\ \ref{photo} and \ref{reduc}. The construction and analysis of the
photometric catalogs is given in Sect.\ \ref{ana}.
The photometric selection of very high redshift candidates is presented in
Sect.\ \ref{selec}. The properties of the final list of candidates, including
spectral energy distributions (hereafter SEDs), photometric redshifts and
magnification estimates, are detailed in Sect.\ \ref{result}. 
In Sect.\ \ref{discuss} we discuss the intrinsic (lens-corrected) properties
of this sample, the number-density of star-forming galaxies at $z\sim6-10$ as
compared to simple model expectations, and the implications for the cosmic
SFR. We also summarize our ongoing spectroscopic survey in these fields. Conclusions are given in
Sect.\ \ref{conclusions}. 
In the appendix we provide more details on the improvement of the data 
reduction procedure, the completeness
and false-positive detection estimates in the different fields and filters, 
as well as additional tests performed on the reliability of optical dropouts.

Throughout this paper, we adopt the following cosmology: a flat $\Lambda$-dominated
Universe with the values $\Omega_{\Lambda}=0.7$, $\Omega_{m}=0.3$, 
$\Omega_{b}=0.045$, $H_{0}=70\ km\ s^{-1}\ Mpc^{-1}$, and $\sigma_8=0.9$. All
magnitudes given in the 
paper are quoted in the Vega system. Conversion values 
between Vega and AB systems for these filters are given in Table \ref{images}.

\section{\label{criteria}Simulations and observing strategy}

   Our project aims to search for $z \gtrsim 6$ galaxies, typically in the
$z \sim$ 7--10 domain. 
We have performed simulations to estimate the expected magnitudes of galaxies at
such redshifts, and to establish robust photometric criteria to select
high-redshift candidates behind lensing clusters. 
For this, we have used the evolutionary synthesis starburst models
by Schaerer (2002, 2003) for Population III and extremely metal deficient
galaxies, together with the usual templates available for normal galaxies. In
particular, we used the empirical SEDs compiled by Coleman, Wu and Weedman (1980)
to represent the local population of galaxies, with spectra extended to
wavelengths $\lambda \le 1400$\,\AA\ and $\lambda \ge 10000$\,\AA\ using the
equivalent spectra from the Bruzual \& Charlot GISSEL library for solar
metallicity (Bruzual \& Charlot 1993). We also
included the starbursts templates SB1 and SB2, from Kinney et
al.\ (1996), and the low metallicity galaxy SBS0335-052
(Izotov, 2001, private communication).

\begin{figure}[ht!]
\psfig{figure=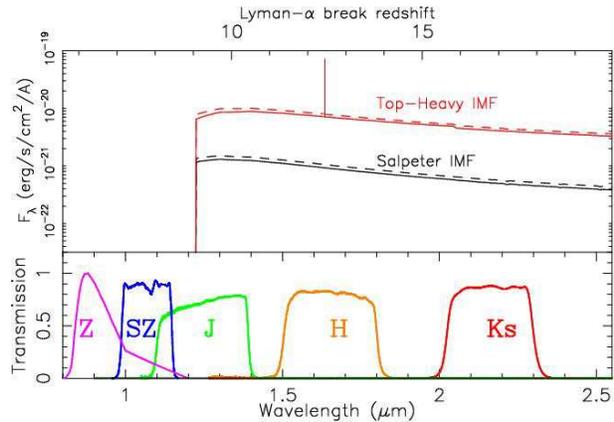,width=8cm,angle=270}
\caption{\label{filters}
Examples of SED for Pop III models used in
Fig.\ \ref{mags_heavy1} and Fig.\ \ref{mags_heavy2} (upper panel), for a
fiducial $10^7$ M$_{\odot}$ burst at $z=9$, with ages $10^4$ and $10^6$ yrs
(solid and dashed lines), and two IMF (normal Salpeter -black- and top-heavy
-red-), compared to the transmission curves of the FORS/ISAAC filters set used in this
survey (lower panel). The top axis gives the corresponding redshift at the
wavelength of the Lyman-$\alpha$ break. We overplot the location of the HeII
1640 emission line (without scaling its flux), which has a small impact on the 
k-correction as shown in Figs.\ \ref{mags_heavy1} and Fig.\
\ref{mags_heavy2}. 
Different assumptions were considered for the Lyman-$\alpha$ emission.
}
\end{figure}

   We consider a fiducial stellar mass halo of $10^7$ M$_{\odot}$,
corresponding to a collapsing DM halo of $2 \times 10^8$ M$_{\odot}$.
With a $\Lambda$CDM model and the cosmological parameters adopted here, we
expect $\sim 10$ of such DM haloes per Mpc$^3$ within the relevant redshift
range $z\sim5-10$, and $\sim 1\ $Mpc$^{-3}$ with a DM halo of $10^9$
M$_{\odot}$ (e.g. Loeb \& 
Barkana 2001). 
The virial
radius is of the order of a few kpc, and thus we consider that sources
are unresolved on a 0.3'' scale, with spherical symmetry. The reionization
redshift is assumed to be $z \sim 6$, but results
discussed below are independent of this precise value.  Lyman series
troughs (Haiman \& Loeb 1999), and Lyman forest absorptions following the
prescription of Madau (1995) are included. Although a detailed description of
Lyman-$\alpha$ emission is out of the scope of this paper, we have taken into
account the possible impact of the main emission lines on the integrated
colors using rather simple and extreme assumptions. Simulations accounting
for an extended Lyman-$\alpha$ halo (cf. Loeb \& Rybicki 1999) have been
computed, together with a simple parametrization of the fraction of
Lyman-$\alpha$ flux entering the integration aperture. 
Two extreme assumptions are considered here for the IMF, either a standard
Salpeter IMF, with stars forming between 1 and 100 M$_{\odot}$, or a top-heavy
IMF, with stellar masses ranging between 50 and 500 M$_{\odot}$.
Some preliminary versions of these simulations were presented in
Schaerer \& Pell\'o (2001) and Pell\'o \& Schaerer (2002).
Figure \ref{filters} shows, for each IMF, the expected SED of a
$10^7$ M$_{\odot}$ stellar mass halo, using two different assumptions for age.

\begin{figure}[ht!]
\psfig{figure=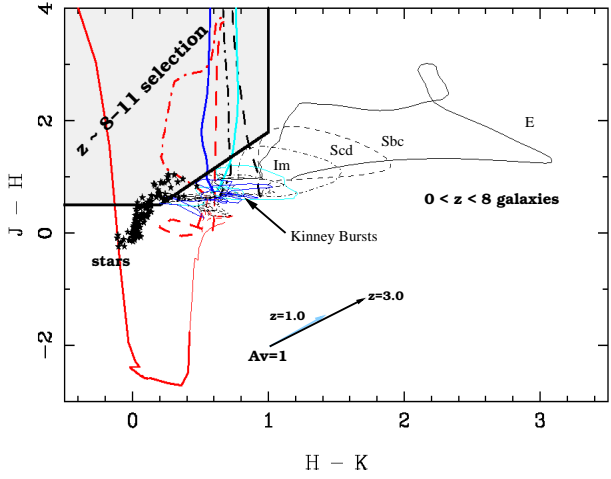,width=8cm}
\caption{\label{JHK_colors}
$J$ dropout selection in the $z \sim$ 8-11 domain.
$J-H$ versus $H-Ks$ color-color diagram (Vega system) showing the position
expected for different objects over the interval $z \sim$ 0 to 11.
The position of stars and normal galaxies up to $z\ \le$ 8
are shown, as well as the shift direction induced by $A_V=1$
magnitude extinction. Thin and thick lines display models below and above
$z=8$. 
Several models for Pop III starbursts are presented, for different fractions
of Lyman-$\alpha$ emission flux entering the integration aperture:
100\% (red solid line), 50\% (red dashed line) and 0\% (red dot-dashed line). 
The location of Kinney et al.\ (1996) starbursts templates is also given for
comparison (SB1(cyan) and SB2 (blue)).
All star-forming models enter the high-$z$ candidate region at $z\ge8$.
}
\end{figure}

\begin{figure}[ht!]
\psfig{figure=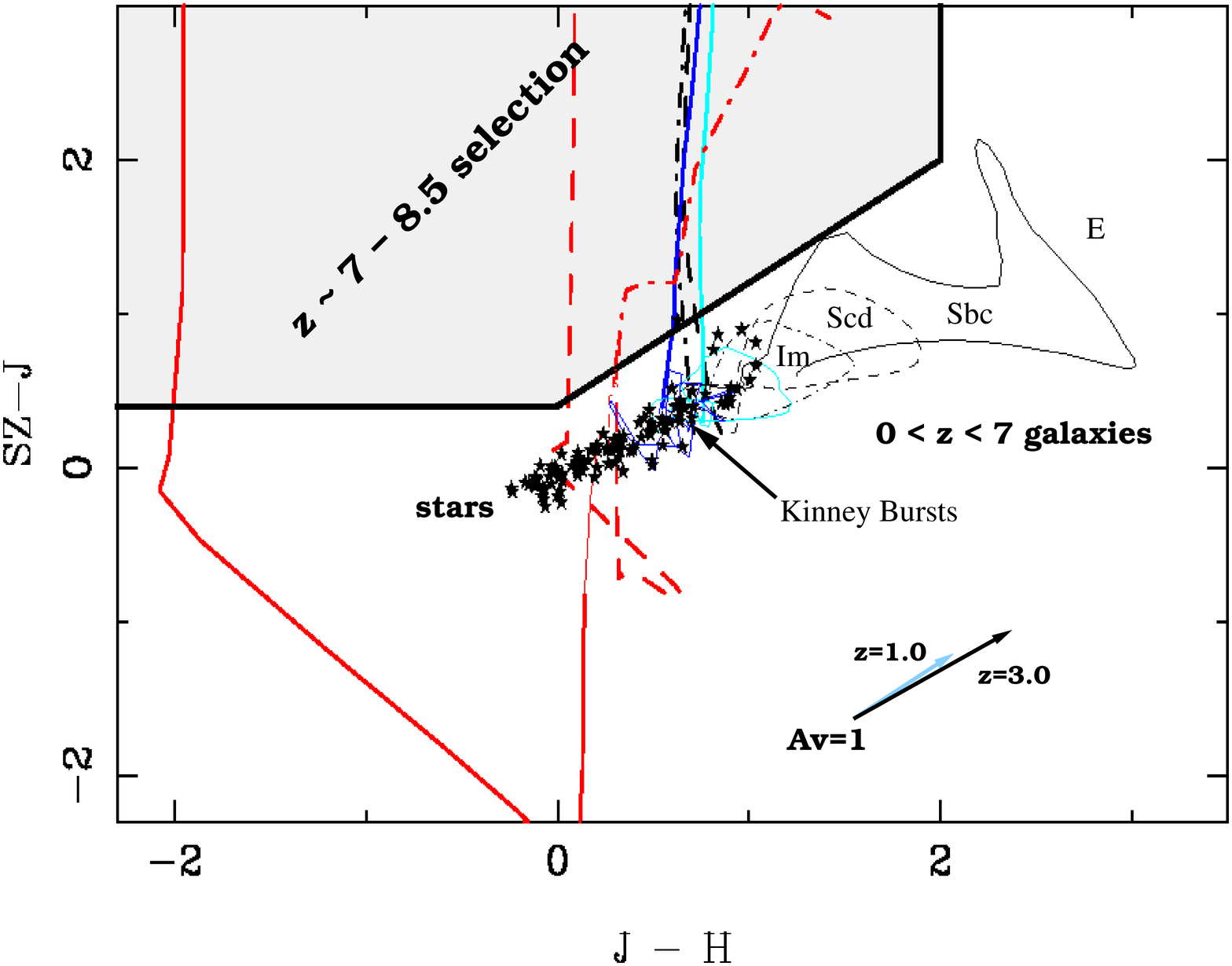,width=8cm}
\caption{\label{SZJH_colors}
$SZ$ dropout selection in the $z \sim$ 7-8.5 domain.
$SZ-J$ versus $J-H$ color-color diagram (Vega system) showing the position
expected for different objects over the interval of $z \sim$ 0 to 8.5.
Thin and thick lines display models below and above $z=7$ respectively. 
Models displayed and general comments are the same as in Fig.~\ref{JHK_colors}.
The position of stars and normal galaxies up to $z\ \le$ 7
are shown.
}
\end{figure}

\begin{figure}[ht!]
\psfig{figure=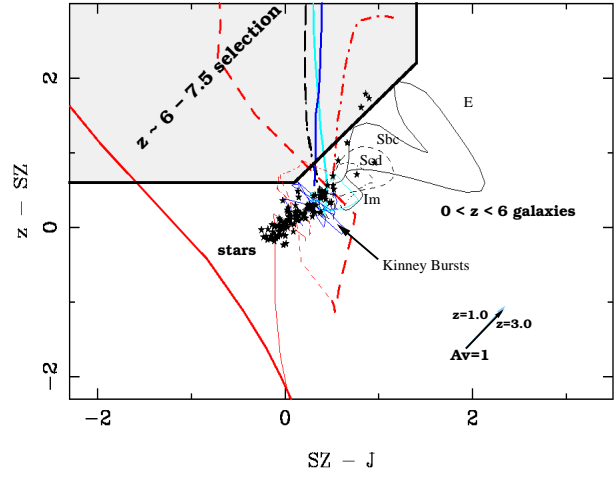,width=8cm}
\caption{\label{zSZJ_colors}
$z$ dropout selection in the $z \sim$ 6-7.5 domain.
$z-SZ$ versus $SZ-J$ color-color diagram (Vega system) showing the position
expected for different objects over the interval of $z \sim$ 0 to 7.5.
Thin and thick lines display models below and above $z=6$ respectively. 
Models displayed and general comments are the same as in Fig.~\ref{JHK_colors}.
The position of stars and normal galaxies up to $z\ \le$ 6
are shown.
}
\end{figure}

   Nebular continuous emission and strong emission
lines could have important effects on the integrated fluxes and colors of
such galaxies, although broad-band colors alone do not allow one to precisely
constrain the physical properties of these galaxies. 
The main signatures of genuine star-forming
sources at $z>6$ are common to all models: they are
optical dropouts, displaying a strong break and 
``red'' optical vs. IR colors, whereas they exhibit a ``blue'' SED 
longward of Ly-$\alpha$ provided reddening is sufficiently small.
Different redshift intervals can be defined using an appropriate set of
near-IR filters in combination with optical data. This particular 
application of the Lyman break technique as a function of redshift
is shown in Figs.\ ~\ref{JHK_colors} to
~\ref{zSZJ_colors}, for the different redshift intervals considered in this
paper. For clarity, only a reduced number of models is presented in these figures. 
Color shifts corresponding to an intrinsic extinction of $A_V=1$ (Calzetti et
al.\ 2000) are shown by arrows, for starbursts at $z=1$ and $z=3$. 
Stellar colors presented in these diagrams were computed from the library of
Pickles (1998).  

\begin{figure}[ht!]
\vbox{
\psfig{figure=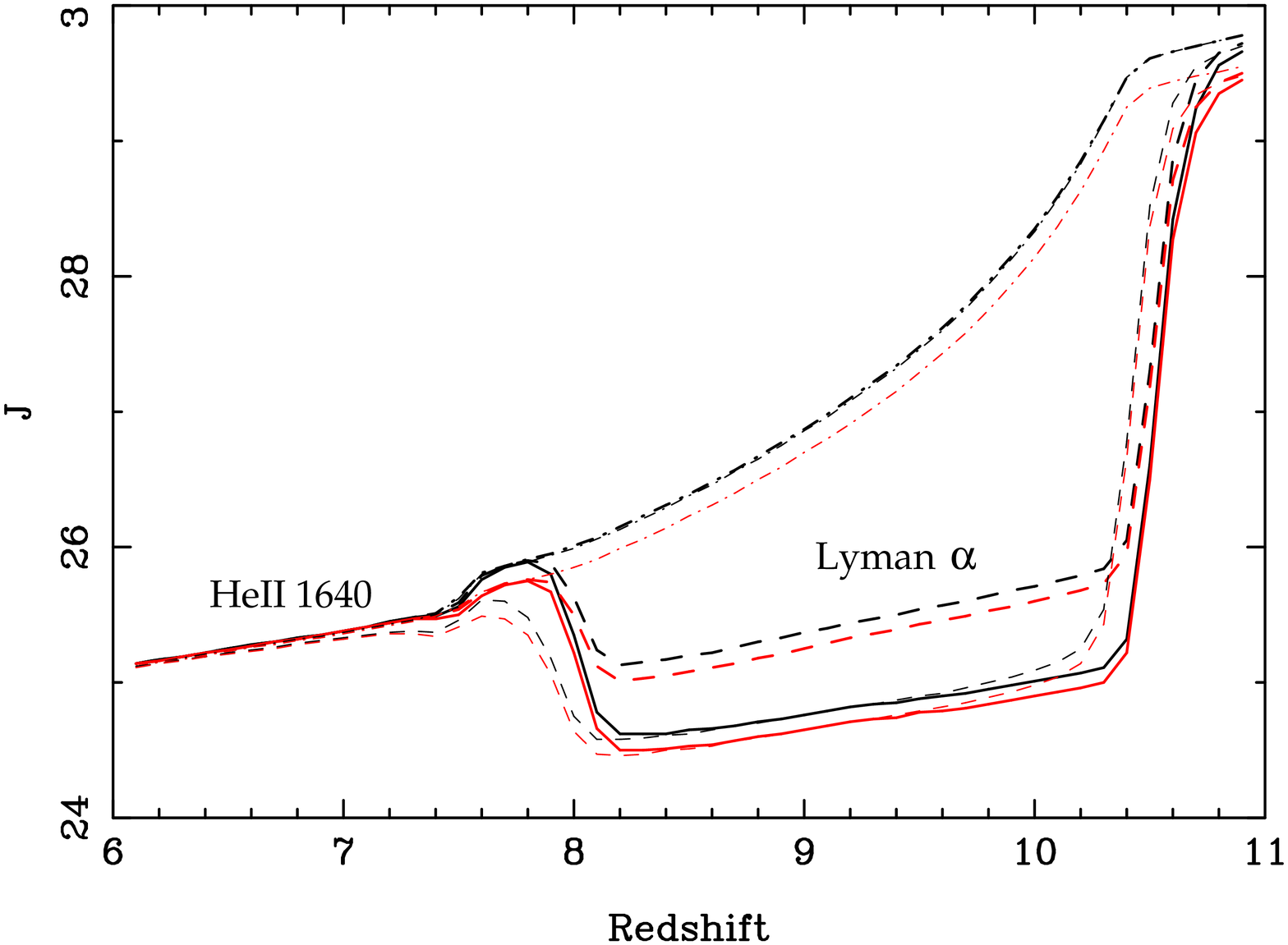,width=8cm}
\psfig{figure=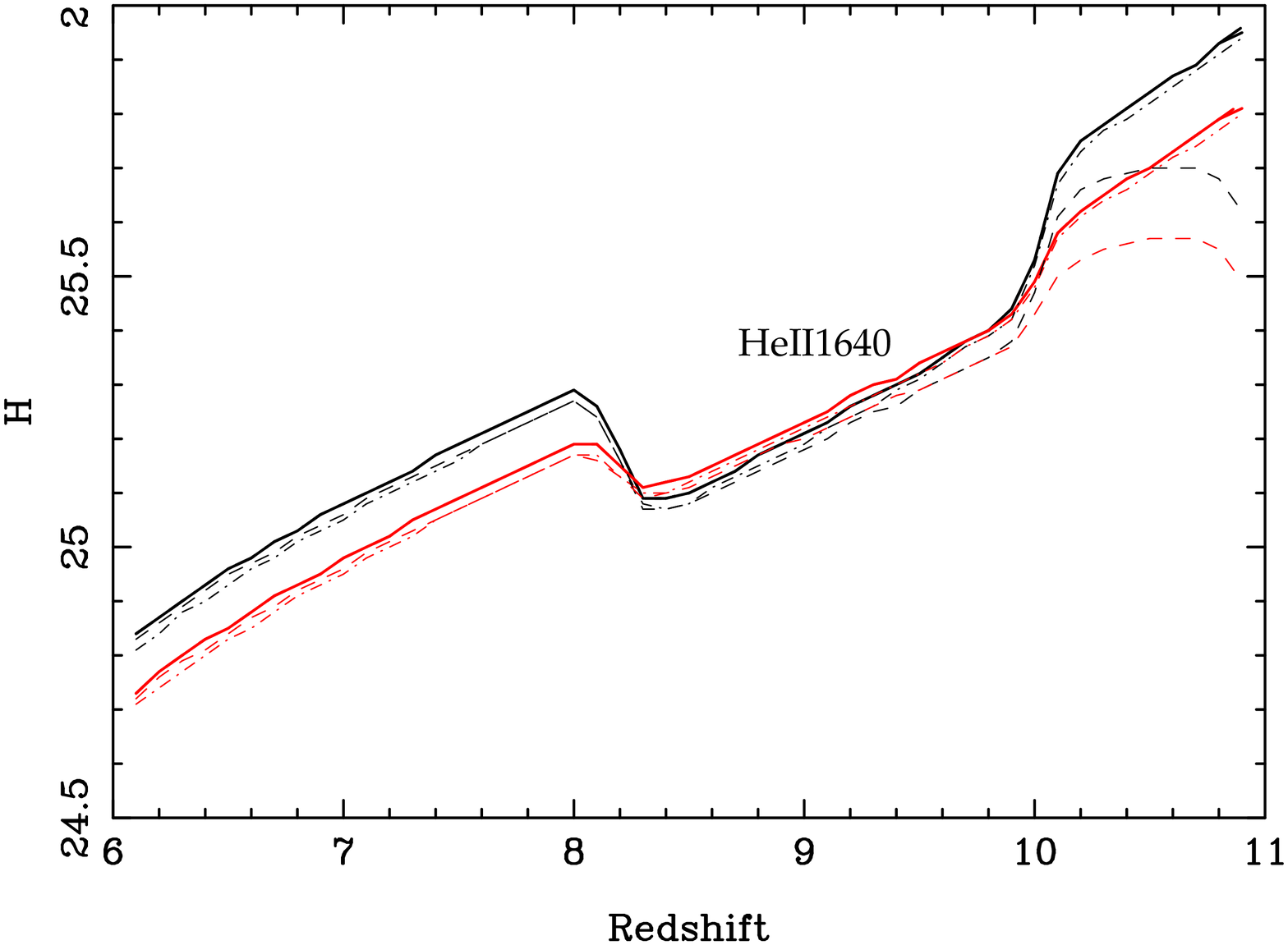,width=8cm}
\psfig{figure=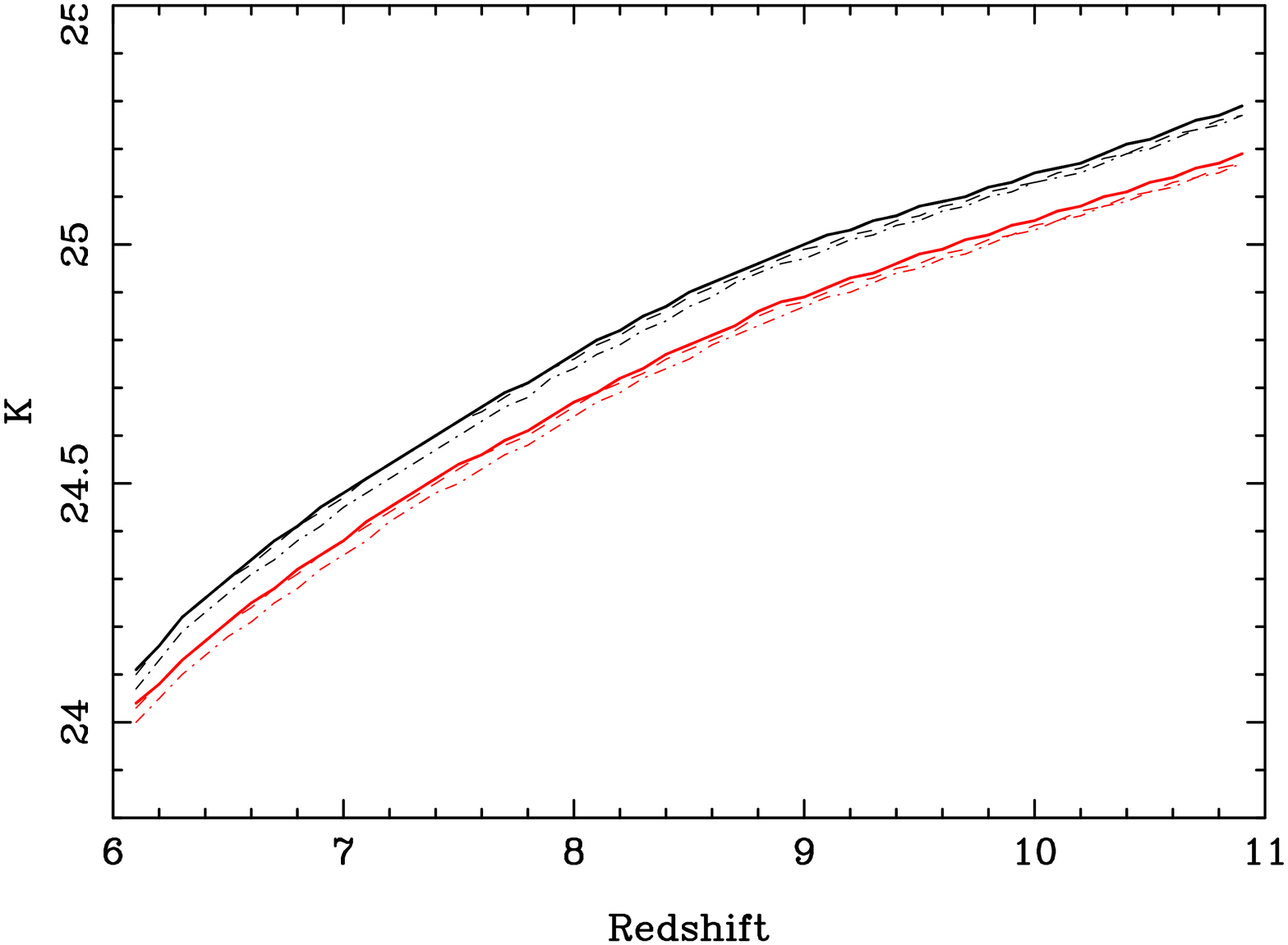,width=8cm}
}
\caption{\label{mags_heavy1}
From top to bottom, $J$, $H$ and $Ks$ magnitudes as a function of redshift
for a top-heavy IMF, for a fiducial stellar halo of $10^7$
M$_{\odot}$. The values corresponding to a normal Salpeter IMF are
about 2 magnitudes fainter over all the redshift interval.
Black and red lines correspond respectively to burst ages
$10^6$ and $10^4$ yrs. Various models for Pop III objects are
presented, for different fractions of the Lyman-$\alpha$ emission
entering the integration aperture: 0 \% (thick dot-dashed line),
50 \% (thick dashed line), and 100 \% (thick solid line). Thin
dot-dashed lines correspond to a self-consistent
extended Lyman-$\alpha$ halo emission (Loeb \& Rybicki 1999), whereas
thin dashed lines display the same model with 100 \% of Lyman-$\alpha$
emission entering the integration aperture.
}
\end{figure}

Figure ~\ref{JHK_colors} displays the $J-H$ versus $H-Ks$
color-color diagram for different extreme Population III starbursts within the
$5 \le z \le 11$ interval, compared to the location of stars and normal galaxies at
different redshifts. This diagram is particularly well suited to identify
galaxy candidates in the $8 \le z \le 11$ interval among the optical
dropouts. 
At redshifts above $\sim10$, galaxies are not detected in the $J$ band
(see also Fig. \ref{mags_heavy1}). For galaxies at
$6 \le z \le 9$, the same photometric selection can be performed
including the $z$ (0.9 $\mu$m) and $SZ$ (1.1 $\mu$m) filters 
(Fig. \ref{SZJH_colors} and \ref{zSZJ_colors}). 
The characteristics of these filters are
summarized in Table \ref{images} and Fig. \ref{filters}. 


   Once the color-selection is well established, an important issue is the
photometric depth needed to detect typical stellar haloes up to a given mass,
in order to derive statistically significant results on the nature and properties of
$z>6$ sources. 

\begin{figure}[ht!]
\vbox{
\psfig{figure=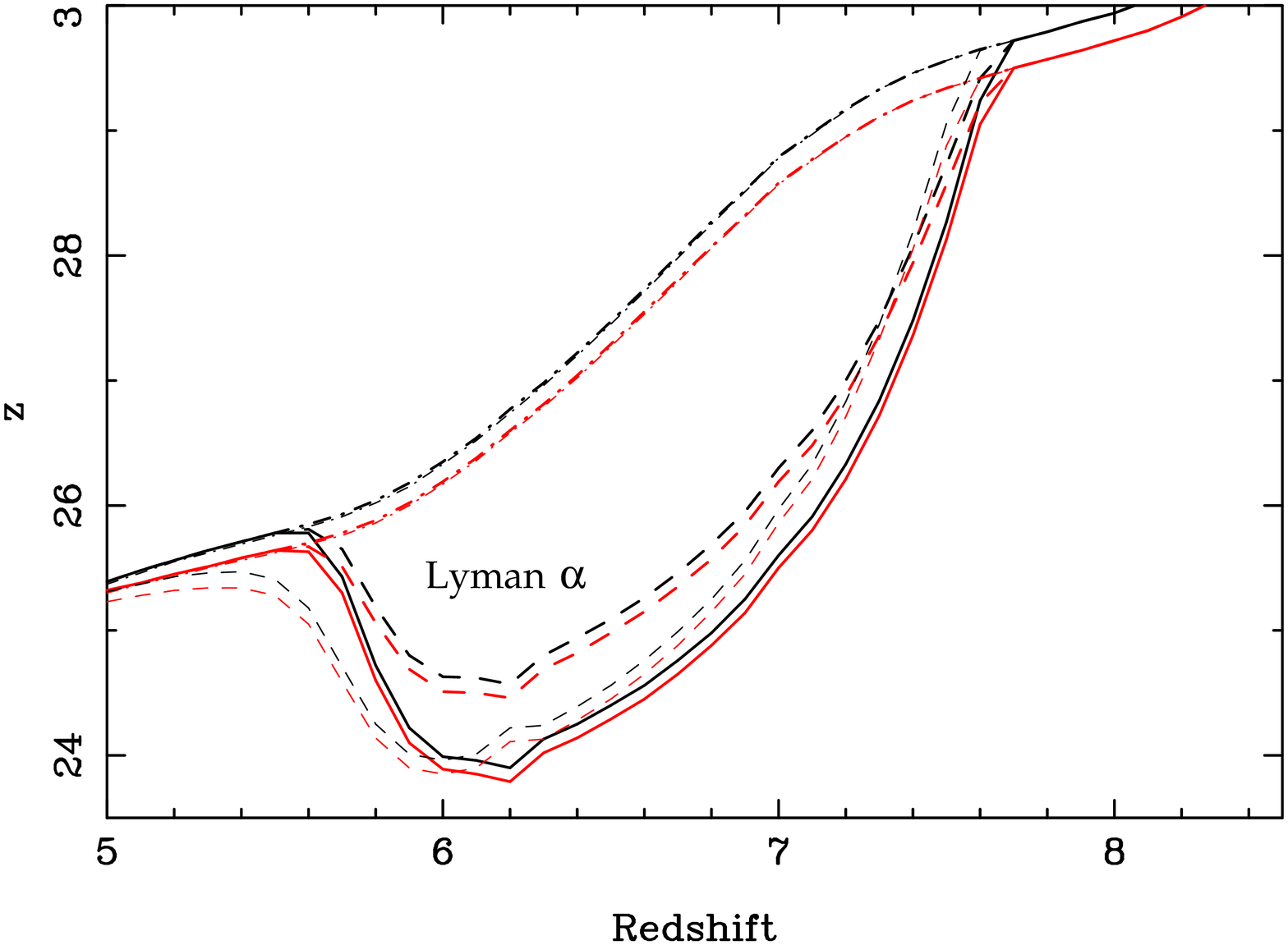,width=8cm}
\psfig{figure=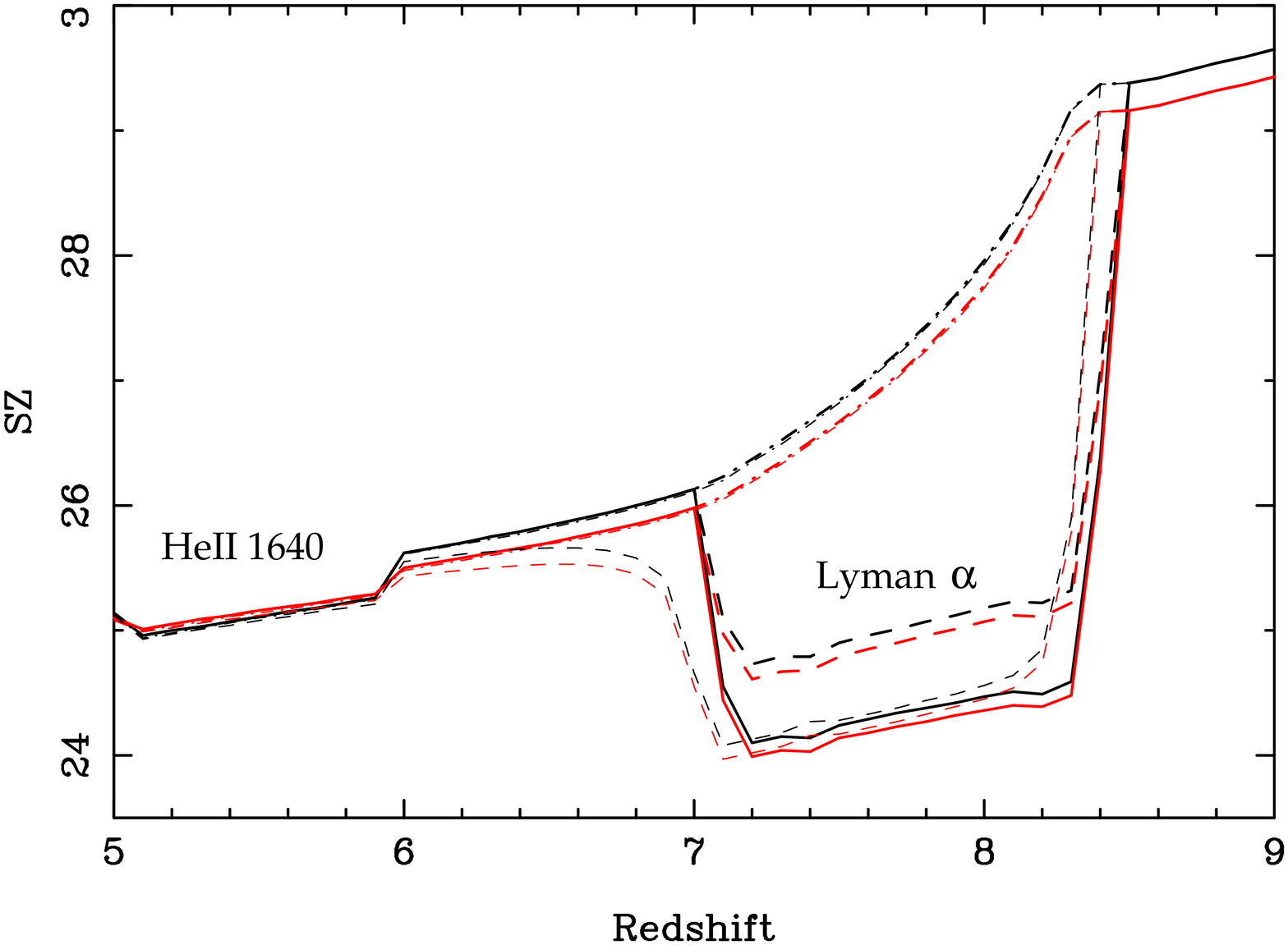,width=8cm}
}
\caption{\label{mags_heavy2}
From top to bottom, $z$ and $SZ$ magnitudes as a function of redshift,
for a fiducial stellar halo of $10^7$ M$_{\odot}$ and the same models
as in Fig.~\ref{mags_heavy1}.
}
\end{figure}

According to our simulations, the predicted magnitudes
in the Vega system for a reference stellar halo mass of $10^7$ M$_{\odot}$,
zero metallicity stars, a top-heavy Salpeter IMF, and a starburst younger than $10^6$ yrs,
typically range between
$\sim$ 24.5 and 26.0 in $J$ in the relevant redshift range ($z\le8$),
$\sim$ 24.5 and 25.5 in $H$, and
$\sim$ 24 to 25 in $Ks$ ($z \lesssim 10$), depending on models,
within the $z \sim 6-10$ interval (see Fig. \ref{mags_heavy1}).
For the same models, $SZ$ and $z$ range between $\sim 25-26$, for $z\le7$ and
$z\le6$ respectively, and sources become dropouts in these filters beyond
these redshifts (see Fig. \ref{mags_heavy2} and Fig. \ref{filters}).
For a standard Salpeter IMF (from 1--100 $M_\odot$) these values are $\sim 2$
magnitudes fainter than for a top-heavy IMF; increasing the metallicity for this IMF
implies a somewhat larger UV restframe flux (up to 0.5 mag brighter for
solar metallicity; cf.\ Schaerer 2003, Fig.\ 2).
Also, obviously these magnitudes scale with stellar mass.
In other words, a stellar halo with a standard IMF exhibits about the same magnitudes
as a top-heavy IMF which is a factor of 10 less massive, all the other
parameters being the same. This comment not only stands for PopIII models, but
also for solar metallicity starbursts and constant star-forming models with
standard IMF and metallicity.

Assuming a minimum gravitational magnification of $\sim$1 magnitude, 
if we intend to detect stellar haloes up to $10^8$ M$_{\odot}$ (or a few
$10^7$ M$_{\odot}$, depending on IMF), the photometric depth required is of
the order of $H\sim$ 24.0 and $Ks\sim$ 23.5 for a positive detection, and up
to $\sim$ 26.0 in $z$, $SZ$ and $J$ to identify significant dropout sources in
these filters. The number of sources expected can be roughly estimated as
follows. Taking into account the typical covolume surveyed in a lensing
cluster under these conditions
\footnote{a few $10^4\ $Mpc$^3$ between $z=6$ and 10, for a $\sim 2 \times 2$
arcmin$^2$ field of view, after correction for a typical magnification factor
of $\sim 2$},  
the relevant density of DM haloes assuming a conservative fraction of 10\% of
baryonic mass converted into stars before $z=6$ ($\sim 0.1\ $Mpc$^3$,
corresponding to $10^{10}$ M$_{\odot}$ DM halo), and the probability of detection
related to the visibility of the starburst within the relevant redshift
interval ($10^6$ yrs restframe, thus a {\it duty-cycle} factor $\sim$ 0.1 to a
few 0.01; see Sect.~\ref{discuss}), the number of sources expected ranges
between a few tens and a few hundreds. We still expect a few positive
detections at $z=6-10$ with a completeness level of the order of 10\% or even
lower. However, strong lensing effects have to be carefully taken into account
in this survey, as explained below.
A detailed comparison between the number of sources expected and the
number of sources actually detected in our lensing fields is provided in
Sect.~\ref{discuss}. 



\section{\label{photo}Photometric data}

Two lensing clusters were selected for this pilot study with the VLT:
AC114 ($\alpha$=22:58:48.26 $\delta$=$-$34:48:08.3 J2000, $z=0.312$) and Abell
1835 ($\alpha$=14:01:02.08 $\delta$=$+$02:52:42.9 J2000, $z=0.252$). AC114 is a
well-known ``gravitational telescope'', for which multiwavelength observations are
available. The corresponding lens model is well-constrained by
a large number of multiple-images spectroscopically identified at high redshift (Smail et al.\
1995, Natarajan et al.\ 1998, Campusano et al.\ 2001, Lemoine-Busserolle et al.\ 2003). Its
Einstein radius is about 30\arcsec\ for $z>6$.
Abell 1835 is the most X-ray luminous cluster in the $XBACS$ sample
({\it X-ray-Brightest Abell-type Clusters of galaxies}, 
Ebeling et al.\ 1998), thus potentially one of the most efficient gravitational
telescopes.
Indeed, strongly lensed features were identified in this cluster,
based on deep ground-based and $HST$ images, and it was used to search for 
bright submm galaxies with SCUBA (Smail et al.\ 1999, Ivison et al.\ 2000). 
The mass model used is similar to the one developed by Smith et al. (2005),
and gives an Einstein radius of $\sim 40$\arcsec\ at high $z$.

We observed these clusters with ISAAC and FORS in the near-infrared domain 
($\sim$ 0.9 to 2.2 \micron) between
September 2002 and April 2004, covering as far as possible the $z$, $SZ$, $J$,
$H$, and $K$ bands. 
Transmission curves for these filters are presented in 
figure \ref{filters}.
In addition, optical images at shorter wavelengths
(from $U$ to $I$ band) are available in our group from previous surveys, or
from archival data. 
Table \ref{images} summarizes the main characteristics of the photometric dataset.

Note that in the remainder of the paper we shall loosely use
the term ``near-IR'' for the $SZ$, $J$, $H$, and $K$ filters,
whereas ``optical'' refers to all filters shortwards of 1.0 \micron,
from $U$ to $I$ (or $z$-band when available).

\subsection{Near-IR photometric observations}

We obtained imaging data with the Infrared
Spectrometer And Array Camera (ISAAC, Moorwood 1997) located on
the Nasmyth-B focus of the 8.2m VLT-UT1 (Antu telescope), using
the Short-Wavelength channel of the instrument (Cuby et al.\ 2000).
The field of view of the camera is about 2.5 arcmin $\times$ 2.5 arcmin
with a pixel size of 0.148\arcsec. The data for AC114
were acquired during UT 2002 August 19-20-22 (period 69). Due to
technical problems with the instrument in this period, the usual
ISAAC detector was changed for the Aladdin 1024 $\times$ 1024 InSb array.
The data for Abell 1835 were acquired during UT 2003 January 14,
February 9-11-12-14-15 (period 70, $JHK$), and UT 2004 April 20 and May 15
(period 73, $SZ$) with the usual Hawaii Rockwell 1024 $\times$ 1024 Hg:Cd:Te Array.
Differences in efficiency between these two detectors 
have been reported, the Hawaii Rockwell detector providing better results in terms
of photometric accuracy. \footnote{For more details see Section 1.2.2 of the 
ISAAC user manual (Cuby et al. 2002)
}

Near infrared imaging is challenging because of the dominant and variable
 sky background. We used dithering of short
exposures with subintegration $\times$ integration times of 4 $\times$ 45 s in the $SZ$ and $J$ bands,
11 $\times$ 12 s in the $H$ band, and 6 $\times$ 15 s in the $Ks$ band, with a 30\arcsec \
jitter box. These values provided a good compromise between an optimal
photometric depth over a large fraction of the field of view, and 
good sky-subtraction in a crowded field.
For Abell 1835, the field center was chosen such as to prevent
contamination by a very bright star located at the north of the cluster.
Therefore the brightest cluster galaxy is not at the center of the field
(see Fig.\ \ref{cline_a1835}).

Calibration data were obtained in the usual way (detector darks, twilight
flats, ...). Standard stars from the LCO/Palomar NICMOS list (Persson et al.\
1998) were used for photometric calibration. 

\subsection{Optical and intermediate band images}

$z$ band observations of Abell 1835 were obtained during UT 2004 March 26 and April 10
with the FOcal Reducer/low dispersion Spectrograph (FORS2) at VLT. This instrument has a
0.252\arcsec\ pixel size and a field of view of 7.2 arcmin $\times$ 7.2 arcmin. We
used dithered individual exposures of 120 s.

To be able to distinguish between ``low'' ($z \la 6$) and ``high'' ($z > 6$) redshift objects,
we have compiled the available optical images for the two lensing clusters.
For AC114, we used the data from Campusano et al.\ (2001) covering $U$ to $I$ filters,
including a mosaic of deep HST/WFPC2-F702W ($R$) observations. Images in this
band were obtained in both \textit{high-sky} and \textit{low-sky} modes, with
different orientations on the sky, in such a way that the final composite
image in this filter exhibits three regions with different behavior (noise
properties, photometric depth) across the ISAAC field of view. Each
of them is presented separately in Table \ref{images}.

Optical data for Abell 1835 include $BVRI$ imaging obtained with the CFH12k camera
at CHFT, and HST/WFPC2-F702W ($R$) images acquired in \textit{low sky}
mode. Because of the field centering chosen for the near-infrared data, only half of
the ISAAC field of view is covered by the HST/WFPC2 image. 

For the two clusters, all optical data fully cover the relevant region studied
around the cluster center. The entire ISAAC field is covered in most cases.
The overlap fraction between the optical images and the near-IR data
is indicated in Table \ref{images},
where the references and main properties of the data set are summarized.
Overlap fractions refer to the ISAAC field. 

\begin{table*}[ht]
\caption{\label{images}Main properties of the photometric dataset used in this
paper: filter identification, total exposure time,
average seeing measured on the original images, pixel size,
1 $\sigma$ limiting magnitude inside a 1.5 \arcsec \ diameter aperture, filter effective wavelength,
AB correction, overlap fraction relative to the ISAAC frames (covering an area of 2.5 arcmin $\times$ 2.5 arcmin), and references. Exposure time values of three different subsets of the HST-$R_{702W}$ image, annotated $R_{702}^{1,2,3}$ are converted into \textit{low sky} mode for
comparison (see text for details). AB corrections ($C_{AB}$) correspond to $m_{AB}=m_{Vega}+C_{AB}$.}

\begin{tabular}{lrllcrrrl}\hline
Filter & $t_{exp}$ & seeing & pix & depth &  $\lambda_{eff}$& $C_{AB}$ & overlap & Reference \\
& [ksec] & [\arcsec] & [\arcsec] & [mag] & [nm] & [mag] & [\%] & \\\hline AC114&&&&&&&&\\\hline
$U$ & 20.00 & 1.3 & 0.36 & 29.1 & 365 & 0.693 & 43.1 & Barger et al.\ 1996\\
$B$& 9.00 & 1.2 & 0.39 & 29.0 & 443 &-0.064 & 100.0 & Couch et al.\ 2001\\
$V$& 21.60 & 1.1 & 0.47 & 28.5 & 547 & 0.022 & 76.6 & Smail et al.\ 1991\\
$R_{702}^{1}$ & $\geq$ 8.30 & 0.13 & 0.100 & $\geq$ 27.7 & 700 & 0.299 & 84.9 & Natarajan et al.\ 1998\\
$R_{702}^{2}$ & $\geq$ 24.90 & 0.13 & 0.100 & $\geq$ 28.4 & 700 & 0.299 & 41.4 & Natarajan et al.\ 1998\\
$R_{702}^{3}$ & 40.00 & 0.13 & 0.100 & 28.6 & 700 & 0.299 & 17.0 & Natarajan et al.\ 1998\\
$I_{814}$ & 20.70 & 0.3 & 0.100 & 26.8 & 801 & 0.439 & 77.6 & Smail et al.\ 1991\\
$J$ & 6.48 & 0.52 & 0.148 & 25.5 & 1259 & 0.945 & 100.0 & This work\\
$H$ & 13.86 & 0.40 & 0.148 & 24.7 & 1656 & 1.412 & 100.0 & This work\\
$Ks$ & 18.99 & 0.34 & 0.148 & 24.3 & 2167 & 1.873 & 100.0 & This work\\
\hline Abell 1835&&&&&&&\\\hline
$V$ & 3.75 & 0.76 & 0.206 & 28.1 & 543& 0.018 & 100.0 & Czoske et al.\ 2002\\
$R$ & 5.40 & 0.69 & 0.206 & 27.8 & 664& 0.246 & 100.0 & Czoske et al.\ 2002\\
$R_{702}$ & 7.50 & 0.12 & 0.100 & 27.7 & 700& 0.299 & 45.7 & Smith et al.\ 2005\\
$I$ & 4.50 & 0.78 & 0.206 & 26.7 & 817& 0.462 & 100.0 & Czoske et al.\ 2002\\
$z$ & 6.36 & 0.70 & 0.252 & 26.7 & 919& 0.554 & 100.0 & This work\\
$SZ$ & 21.96 & 0.54 & 0.148 & 26.9 & 1063& 0.691 & 100.0 & This work\\
$J$ & 6.48 & 0.65 & 0.148 & 25.6 & 1259& 0.945 & 100.0 & This work\\
$H$ & 13.86 & 0.50 & 0.148 & 24.7 & 1656& 1.412 & 100.0 & This work\\
$Ks$ & 18.99 & 0.38 & 0.148 & 24.7 & 2167& 1.873 & 100.0 & This work\\
\hline
\end{tabular}
\end{table*}

\section{\label{reduc}Data reduction and calibration}
Near infrared photometry of extremely faint sources requires a careful data reduction.
The general procedure described here was performed for all the ISAAC data ($SZ$, $J$, $H$ and $Ks$ bands). A number of specific improvements are given with more details in App. \ref{improvements}. 
For the FORS2 ($z$ band) data, we used a standard flat-field correction and combination of the
individual frames with bad-pixel rejection.

We reduced our data using {\sc IRAF} procedures and
according to the ISAAC Data Reduction Guide v.1.5\footnote{See {\tt
http://www.eso.org/instruments/isaac/\\
drg/html/drg.html}}. The
different steps are the following: photometric calibration, bias
subtraction, flat-fielding, sky subtraction, registration and
combination of the images. The reduction recipe we used, as well as some
of the improvements, were mostly inspired from the reduction of
near-IR observations on the HDFS field with the same instrument
(Labb\'e et al.\ 2003).

Photometric zero-points were derived from LCO/Palomar NICMOS standard stars
(Persson et al.\ 1998), observed each night using a five-point jitter pattern.
After subtracting from each image the median sky of all the pattern, we
measured the total counts in a 20 pixel radius circular aperture, and from
these integrated fluxes we derived the zero-points.
Airmass differences between science exposures were corrected 
to a reference value for each filter using a linear relation between
zero-point and airmass derived from standards stars observed at different
airmasses. 

After removing the instrumental ghost, substracting a median dark frame and
flat-fielding our data, we used the {\sc IRAF} package XDIMSUM\footnote
{XDIMSUM is a modified version by the {\sc IRAF} group of the Deep
Infrared Mosaicing Software package by P. Eisenhardt et al. See
ftp://iraf.noao.edu/extern-v212/xdimsum for details} to apply a
two-step sky-subtraction. During the first pass, each image is
sky-subtracted using the sky pattern obtained from a group of adjacent
frames and a bad pixel mask is created in the process.
The relative shifts between images are derived from the position of
several stars matched in each frame. Then, images are registered and combined
using integer shifts values to preserve the noise properties
and rejecting all bad pixels. Bright sources are 
detected in order to create an object mask, and a second
sky-subtraction is applied to the data, this time using the mask to reject
pixels located on objects in the evaluation of the sky. This
improves the quality of the final stacked image.

As a cross-check for our stacking procedure, we produced another
version of the final images using a slightly different reduction
recipe. After the usual ghost and dark removals, we flat-fielded
each image using a sky flat, created by evaluating the sky in a
group of adjacent frames, again masking the bright objects. Then
individual frames were registered and coadded in a standard way. The
resulting images are found to be similar, in terms of quality,
detection level and photometric depth, to the general procedure
described above. The main difference is an enhanced quality around the bright
galaxy haloes close to the cluster core, and thus we finally adopted the
two-step sky-subtraction procedure.

\section{\label{ana}Analysis of images}

Since we use imaging data acquired with very different filters and
instruments, we had to match them to a common reference when measuring
the required multi-band photometry. However, the registration and
seeing matching process generates the resampling of data, and therefore
modifies the noise properties of the background in the sense that the
error bars measured on these modified images by the standard means tend to be
underestimated. For this reason, we preferred to use the original
images to derive the error bars in each band, as explained below.

\subsection{\label{regis}Image registration and astrometry}

We registered the final $SZ$, $J$, $H$,and $Ks$ band images for
each cluster with a simple shift, except in the case of AC114,
where we corrected for a slight distortion that appeared in the
$J$ band image.

In order to measure relative photometry in the near-IR bands, we
matched all the images to a common seeing using a simple Gaussian
convolution, the worst case being the $J$ band for both clusters.

As all the photometric high-$z$ candidates are expected to be detected in the
$H$ band we have defined an $H$ band selected sample.
The detection images were created with the original $H$ band images,
weighted by the square root of the corresponding exposure time maps
in order to get a uniform background noise across the field.

The available optical images for each cluster were registered to
the ISAAC combined images, using standard IRAF procedures for
rotation, magnification and resampling of the data. These images
were mainly used to exclude well-detected low-$z$ sources, and also for the
first visual inspection of the optical dropouts.
However, resampling of data could produce both spurious detections and
false non-detections among the faint sources. For this reason, 
further discussed in Sect.\ \ref{check}, we have used the {\it original}
images instead to define our final sample of optical dropouts and we work on 
object coordinates.

We performed an astrometric calibration for all these images using
$\sim 30$ bright unsaturated objects present in the USNO catalog
(USNO A2.0, Monet et al.\ 1998). The error obtained in the absolute
astrometric calibration 
is typically $\sim 0.2$ \arcsec\ for a whole ISAAC field of view.

\subsection{Photometry}

We used the $SExtractor$ package version 2.2.2 (Bertin \& Arnouts 1996) to
detect objects and to compute magnitudes within our images. We optimized the
parameters to detect very faint unresolved sources. Magnitudes were measured
within identical circular apertures in all filters (from optical to
near-infrared), with the $SExtractor$ ``double-image'' mode, using the $H$ band
detection images described in Sect.\ \ref{regis}. 
Near infrared images were seeing-matched to the $J$ band (0.52\arcsec or 0.65\arcsec, 
see Table \ref{images}), but 
the seeing measured on ground-based optical bands is worse. Since optical
images were only used here for non-detection purposes (i.e., for the
identification of optical dropouts), we preferred to keep their original
seeing instead of degrading the quality of near-IR and HST images. 

$SExtractor$ detection parameters were the following: 4 connected pixels above
a threshold of $1\ \sigma$ {\it on the detection image}, which corresponds to
a central value of about $4.5\ \sigma$ for a seeing-limited source, where
$\sigma$ stands for the typical local background noise. 
Magnitudes were measured within a 1.5\arcsec-diameter aperture (i.e., 10
pixels on the ISAAC images). This value is an optimized compromise allowing us to
obtain a fair estimate of the total magnitude for 
point-like sources on near-IR images, while keeping a good S/N on the optical
images with the worst seeing ($\sim 1\arcsec$). 
We have also checked that magnitudes and error-bars measured within a 12
pixels (1.7\arcsec) diameter aperture are usually consistent within 1$\sigma$
error bars for the faintest sources considered in this study.

Since $SExtractor$ uses the registered, seeing-matched images to compute the photometric
errors in all bands, these values are systematically too optimistic,
thus leading to artificially high S/N determinations.
To get more realistic errors in our photometry, we preferred
an empirical method to derive them. We measured the typical RMS in the pixel distribution
within apertures of the same physical size as for flux measurements: we averaged the
pixel to pixel statistics in about 5000 non-overlapping apertures, randomly thrown
inside regions free of objects in each original (unregistered, unconvolved) image.
For each entry in the $SExtractor$ output catalog, we recomputed the photometric error
using the value of $\sigma$ derived from these simulations.
This photometric error measured in the original images was also used
to compute the limiting magnitude in each band, reported in
Table~\ref{images}. All S/N values reported
throughout the paper refer to these mock simulations.

The final catalogs include photometry within 1.5\arcsec\ aperture for 
all objects detected in the $H$ band; we were able to measure photometry of very faint
sources ($J \sim 24.4-24.8$, $H$ and $Ks$ $\sim$ 23.5) with a
relatively good accuracy (S/N$\ \gtrsim\ 3-4$).
The effective exposure time towards the edges of the field is smaller than in
the central region due to the dithering pattern used in near-IR images, thus
leading to brighter detection limits and an increasing number of spurious
detections at the edges of the frame. In this study, we use only the region of
the field for which the effective exposure time is above 50 \% of the (maximum) total
value. The overlap fractions relative to the ISAAC frames given in
Table \ref{images} refer to these central regions, corresponding to
6.34 arcmin$^2$ for Abell 1835 and 6.10 arcmin$^2$ for AC114. The images
shown in Fig.~\ref{cline_a1835} and ~\ref{cline_ac114} also refer to these
central regions. 

\subsection{\label{calccomp}Photometric completeness}

\begin{figure}[ht!]
\psfig{figure=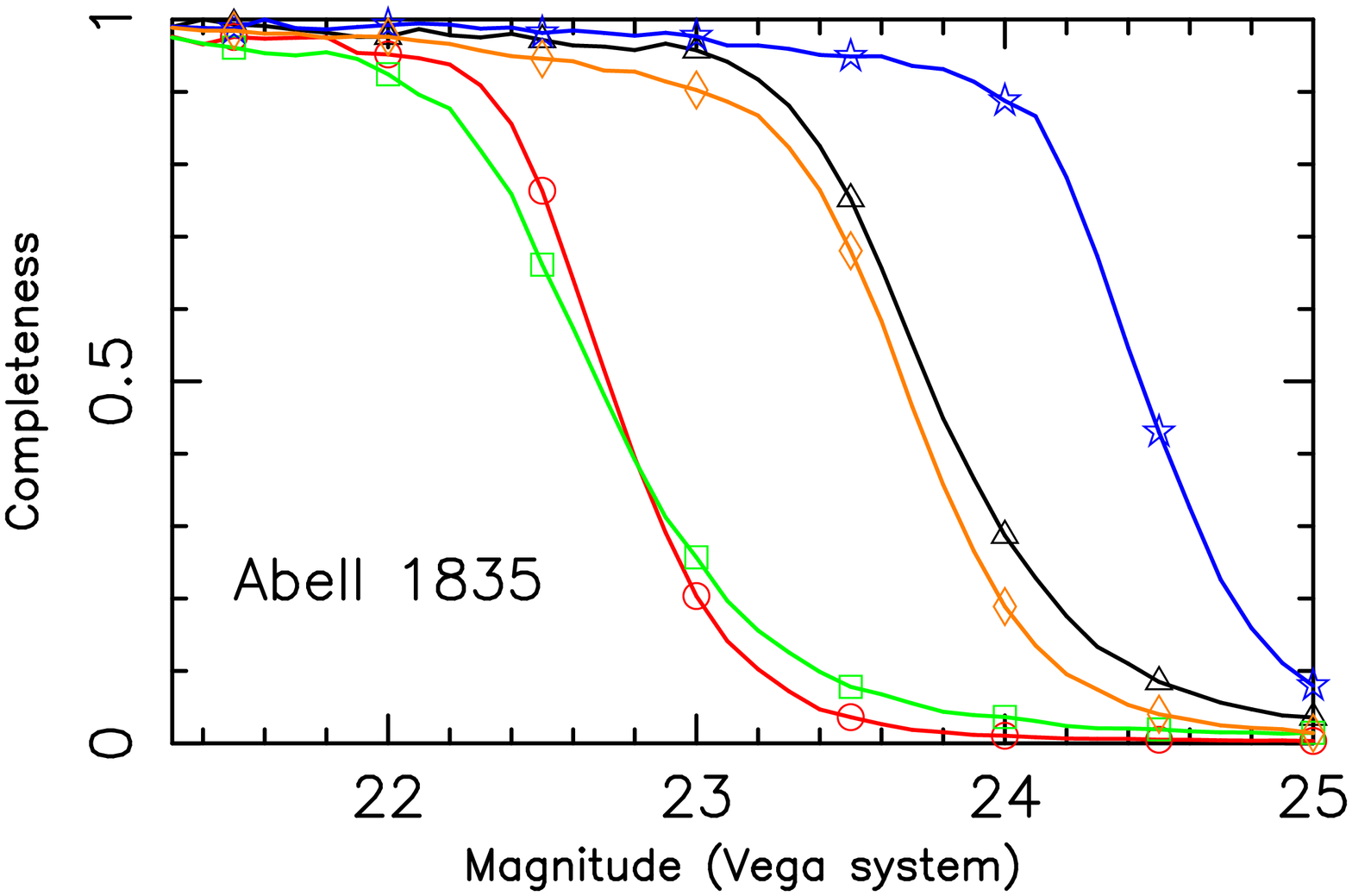,width=8cm,angle=270}
\psfig{figure=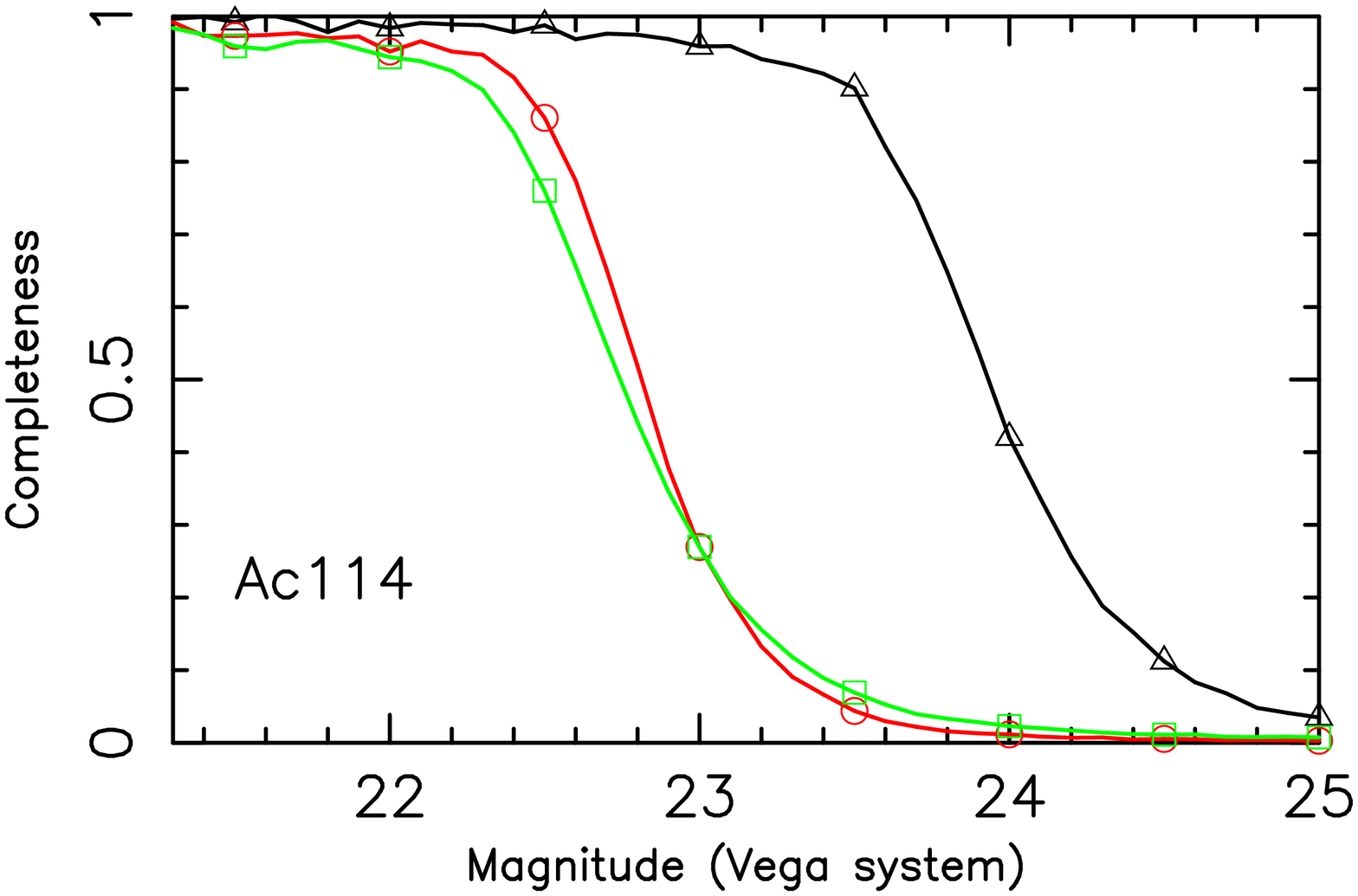,width=8cm,angle=270}
\caption{\label{compcurves}Completeness curves obtained from simulations in
each cluster, for the $z$ band (orange diamonds), $SZ$ band (blue stars), 
$J$ band (black triangles), $H$ band (red circles) and $Ks$ band (green
squares). 
}
\end{figure}

The characteristics of the final processed images are presented in Table
\ref{images}. The seeing was measured on the original co-added images.
We have computed the completeness values for point-sources, in each cluster and band,
for near-IR magnitudes within the relevant intervals. These limits were obtained from
direct simulations as follows. Artificial stars (i.e., seeing limited sources)
of fixed magnitude, ranging between 21 and 25, were  
added 1000 times at 30 different random locations
on our images, and then extracted using the same method for detection and photometry
as for astronomical sources (described above).
Only ``free'' sky regions were used for this exercise, with uniform
noise properties. This excludes in particular the cluster core and the
edges of the images, where the effective exposure time is less than 50 \%.
Completeness values are derived from the fraction of objects we recovered
in our images. The corresponding curves are shown in Fig.\ \ref{compcurves} and the
completeness levels are reported in Table \ref{compval}.
A completeness level of $\sim 20-30$\% is achieved for $H$ and $Ks$ $\sim$ 23,
thus in good agreement with the requirements given in Sect.~\ref{criteria} for
at least a few positive detections at $z \sim6-10$.

\begin{table}[ht]
\caption{\label{compval}90 \% and 50 \% completeness limits for each cluster
and band, corresponding to simulations shown in Fig.\ \ref{compcurves}
}
\begin{tabular}{lcccccc}\hline
Cluster & Band & 90 \% compl. & 50 \% compl. \\\hline
AC114 & $J$ & 23.5 & 23.9\\
AC114 & $H$ & 22.5 & 22.8\\
AC114 & $Ks$ & 22.4 & 22.8\\
Abell 1835 & $z$ & 23.0 & 23.6\\
Abell 1835 & $SZ$ & 24.0 & 24.4\\
Abell 1835 & $J$ & 23.3 & 23.6\\
Abell 1835 & $H$ & 22.3 & 22.7\\
Abell 1835 & $Ks$ & 22.1 & 22.7\\
\hline
\end{tabular}
\end{table}

Our near-IR survey has reached $SZ \sim$ 25.6, $J \sim$ 24.3-24.4, $H \sim$
23.5 and $Ks \sim$ 23.1 (AC114) to 23.5 (Abell 1835) (3 $\sigma$ detection level
within 1.5\arcsec\ aperture), i.e. AB$\sim$ 25-25.5 in $JHKs$ and AB$\sim$26.3 in $SZ$. 
The {\it minimum} magnification factor over the region covered by our near-IR
survey is $\sim$ 0.7 magnitudes, and at least $\sim$ 1 magnitude over
50 \% of the ISAAC field of view. Thus, the {\it effective} $3 \sigma$ limiting
magnitudes reached here are
close or similar to those attained in the HDFS (Labb\'e et
al. 2003) in $JHKs$ (respectively AB$\sim$ 26.8, 26.2 and 26.2). 
Note that the limiting Vega magnitudes given in
Table \ref{images} correspond to $1 \sigma$ values. Our $3 \sigma$ 
limiting magnitudes in the $H$ band 
also comeclose to the magnitudes
of the $z\sim7-8$ $z$-dropouts detected by Bouwens et al.\ (2004b) in the Hubble
Ultra Deep Field, with $H_{160}(AB) \sim$ 26.0 to 27.3, after correction for a 
typical magnification factor of at least $\sim$ 1 magnitude. 

We determine below the additional correction for incompleteness
generated by our sample selection procedure. Lensing models were used 
to derive the effective completeness of our sample as a function of
redshift and magnitude, as compared to blank fields.

\section{\label{selec}Selection of high-$z$ photometric candidates}
This Section presents the procedure adopted 
to select the sample of high-redshift candidates. As described in the
observing strategy (Sect.\ \ref{criteria}), color-color diagrams 
of optical dropouts 
have been used to select high-redshift candidates from our deep near-infrared images. 
For a subsample of them, individual probability distributions and photometric
redshifts can be reliably derived from their photometric SEDs, as discussed in
Sect.~\ref{result}.
The location of the photometric candidates with
respect to the critical lines, and thus the typical magnification factors
reached by our sample, are also briefly described.

\subsection{Near-IR color-color diagrams of bright objects}

\begin{figure}[htb]
\psfig{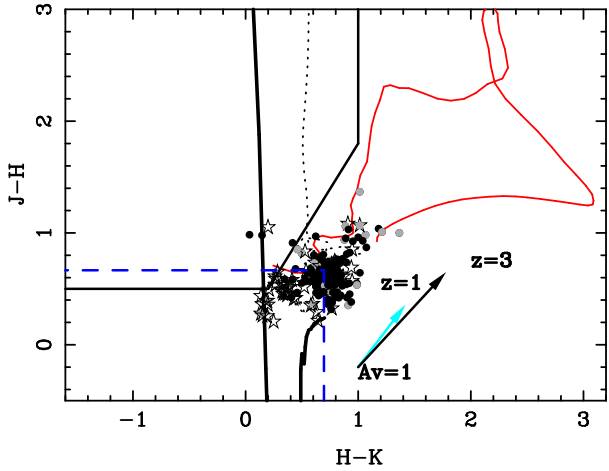}
\caption{\label{diagtotal}Location of objects brighter than $H=22.5$
in the $J-H$ vs $H-Ks$ color-color diagram for Abell 1835 (black
dots). Special symbols are used for $SExtractor$ stars
($flag_{*}>0.95$, open stars) and EROs (grey dots).
Predicted colors for an elliptical cluster galaxy are shown with dashed lines, and the
direction of reddening is indicated by an arrow. 
Evolutionary tracks are overplotted as in Fig.~\ref{JHK_colors},
to facilitate the comparison: E galaxy (red solid line), a
low-metallicity starburst (thick solid black line) and 
a local starburst from Kinney et al.\ 1996 (dotted line). Error bars are
typically $\le 0.1$ mags for these objects. 
}
\end{figure}

As an additional test of the photometric selection of sources, we have checked that
magnitudes and colors of bright sources ($H<22.5$, i.e. S/N$\ \gtrsim\ 8-10$)
are in good agreement with expectations. We have already secured the colors of cluster 
members during the photometric calibration described in Appendix~\ref{improvements}.
Stars morphologically identified by $SExtractor$
($flag_{*}>0.95$) are located at the expected position in these
diagrams. A representative example is given in Fig.\ \ref{diagtotal}
for Abell 1835. This diagram is to be compared to the theoretical
expectations displayed in Fig. \ref{JHK_colors}.

If we adopt the $(R-Ks>5.6)$ definition
from Daddi et al.\ (2000) to select Extremely Red Objects (hereafter EROs)
from this sample of bright sources, we find that a few of them 
lie inside the low-redshift region of the diagram for each cluster, 
a location mainly compatible with dust-reddened starbursts.
This diagram also shows that only very few {\it bright} objects
correspond to our color-selection criteria
for high redshift galaxies. All these objects are detected in the
optical bands. We have carefully inspected the morphology of these
sources on the HST images and found two cases: either they clearly
correspond to two blended objects, or they are point-like sources. In
the first case, the two objects merge in a single source on the
near-IR images, in which case the photometric measurements are
contaminated. In the latter case, we found stars close to 
saturation in one or several filters. 
Thus, except for these understandable cases, no bright
objects ($H<22.5$) are found to fulfill our two color selection criteria.

\subsection{\label{check}Catalog of optical dropouts}

Optical dropouts are defined as objects non-detected in all the
available optical images, from the $U$ band to the $z$ band. 
A source is considered as non-detected in a given band when its
magnitude corresponds to a flux below the $1 \sigma$ detection level within
a 1.5\arcsec\ aperture. These are the magnitude limits reported in
Table~\ref{images}, also used in all tables and figures through the 
paper to derive limits in color. 

In the first step, we used $SExtractor$ in ``double-image'' mode, with the $H$ band
detection images described above as a reference (see Sect.\
\ref{regis}). We cross-checked the detection (or non-detection) of
sources on their original images, before any geometric correction or resampling. 
Positions for all the sources detected in the $H$ band were computed
in the original images using the appropriate geometrical
transformations. An input catalog was created with these
coordinates, used by $SExtractor$ as input for the detection
(or non-detection) of each object on the original image. 

The automatic procedure described above provided a first catalog of
optical dropouts containing 122 and 38 objects up to
$H=24.0$, in Abell 1835 and AC114 respectively. 
Since we detected sources with $SExtractor$ at very faint limits in
flux, all the optical dropouts were carefully examined to reject both
spurious detections in the near-IR bands and false non-detections in
the optical bands, using the original images. 

Some objects in the first catalog were rejected because they were found to
be located in noisy 
regions, close to the limits of the deep images or close to the haloes of
bright galaxies. Some of them were contaminated because they lie too close to
bright objects, in particular towards the cluster center. All these
objects were removed from the dropout catalog either because they
were more likely false detections, or because their photometry was
highly contaminated. A mask was created to remove the remaining noisy regions (bright
galaxies, galaxy haloes, ...) from the subsequent analysis. The region
masked typically corresponds to $\sim 20$ \% of the surface in both
clusters.  At the end of the visual inspection, only $\sim$ 20 \% of 
the original sample remains in the list. 

Several dropout sources were found to be brighter in the $Ks$ band image 
than in the $H$ band. Their centroid determined by $SExtractor$ on the
detection $H$ band image is, in principle, less accurate than in the
$Ks$ band, thus potentially leading to less optimal colors. There are 7 objects of
this kind in Abell 1835 and 1 in AC114.
We corrected the photometry and centroid
positions for these objects by running $SExtractor$ with the $Ks$ band
image as detection frame, keeping all other parameters unchanged.

Since photometry was obtained in \textit{double mode}, 
magnitudes measured by $SExtractor$ in 
the infrared bands could be incorrect due to flux contributions at the
limits of the aperture whereas no object is clearly seen in the
center. For 5 objects in Abell 1835 and 2 in AC114, a non-detection
was forced after visual inspection. 

We have considered that sources detected
in at least two near-IR bands had more significance, since the probability
of false-positive detections in two different bands decreases strongly 
compared to our 
estimates only based on the detection band (Sect.\ \ref{calccomp}). 
These objects constitute our ``second-category''
sample. Among them, we define a ``first-category'' subsample including 
only the best-detected sources (having $\Delta m_H<0.4$, equivalent to
2.5$\sigma$ detection within the aperture). Objects clearly
detected in the reference $H$ band, after 
visual inspection by two different persons, but not detected in
another filter, constitute the ``third category'' sample. The
remaining ones (only detected in $H$ band, and dubious after visual
inspection), are considered as a ``fourth category'' sample, which is
not discussed hereafter. 
Tables~\ref{table_a1835} for Abell 1835 and \ref{table_ac114} for AC114 provide
the coordinates and photometric properties for all optical-dropouts from the first, second
and third-category samples. Identification numbers increase with measured 
$H$ band magnitudes for a given cluster. The number of first/second/third-category
dropouts for Abell 1835 and AC114 is 11/7/5 and 4/4/2 respectively.
In the case of Abell 1835, the best limit for the optical non-detection
is provided by the $z$ band ($z_{AB} \gtrsim $ 27.3, cf.\ Table \ref{images}). 
For AC114 the strongest non-detection criterion is in $R/HST$ with 
$R_{AB} \gtrsim $ 28.--28.9, depending on the source location.

Figures ~\ref{trombino_a1835} and ~\ref{trombino_ac114} display the thumbnail
images of the relevant optical dropouts in Abell 1835 and AC114
respectively. For each source, the available near-IR images 
($SZ$ $J$ $H$ $Ks$ for Abell 1835 and $J$ $H$ $Ks$ for AC114), together with
the strongest non-detection band ($z$ for Abell 1835 and $R/HST$ for AC114)
are shown.

In order to derive global properties of the different types of 
candidates, we carefully estimated our sample completeness and the number of 
false-positive detections, as detailed in App. \ref{sampcomp}. 
According to our results, 
\textit{false-positive detections} are not expected up to
$H=23.0$, and they account for less than 30 \% (typically 12\% in A1835 and
25-33\% in AC114) for first and second-category candidates brighter than
$H=23.3$, depending on the set of filters where an object is being detected
(see Table ~\ref{spurious}). 

We also performed a number of additional tests on the reliability of optical dropouts, reported 
in App. \ref{addtests}.

\subsection{Third and Fourth category candidates}

The manual classification of objects detected only in the reference $H$ band
into ``third'' and ``fourth'' category dropouts seems arbitrary at this
point. We have used rather conservative criteria in this study to avoid the sample
being dominated by false-positive detections, increasing with
magnitude. However, a fraction of these rejected sources is actually real,
although difficult to quantify with present data. A
good example is A1835\#35, a source kept as a secondary target for
spectroscopy, for which we have obtained a spectroscopic confirmation
of $z=1.68$, using H$\beta$ and [O{\sc iii}]$\lambda,\lambda$ 4959,5007 lines 
detected in the $J$ band with ISAAC (see Richard et al.\ 2003 and discussion in
Sect.~\ref{spectro}). This object should have been removed from our
present sample: it is marginally detected in $H$ and $K$ bands, and is
fainter than any other object reported in Tables~\ref{table_a1835} and
\ref{table_ac114}. \\

\subsection{Crude redshift estimate of high-$z-$ candidates from near-IR colors}

\begin{figure*}[ht]
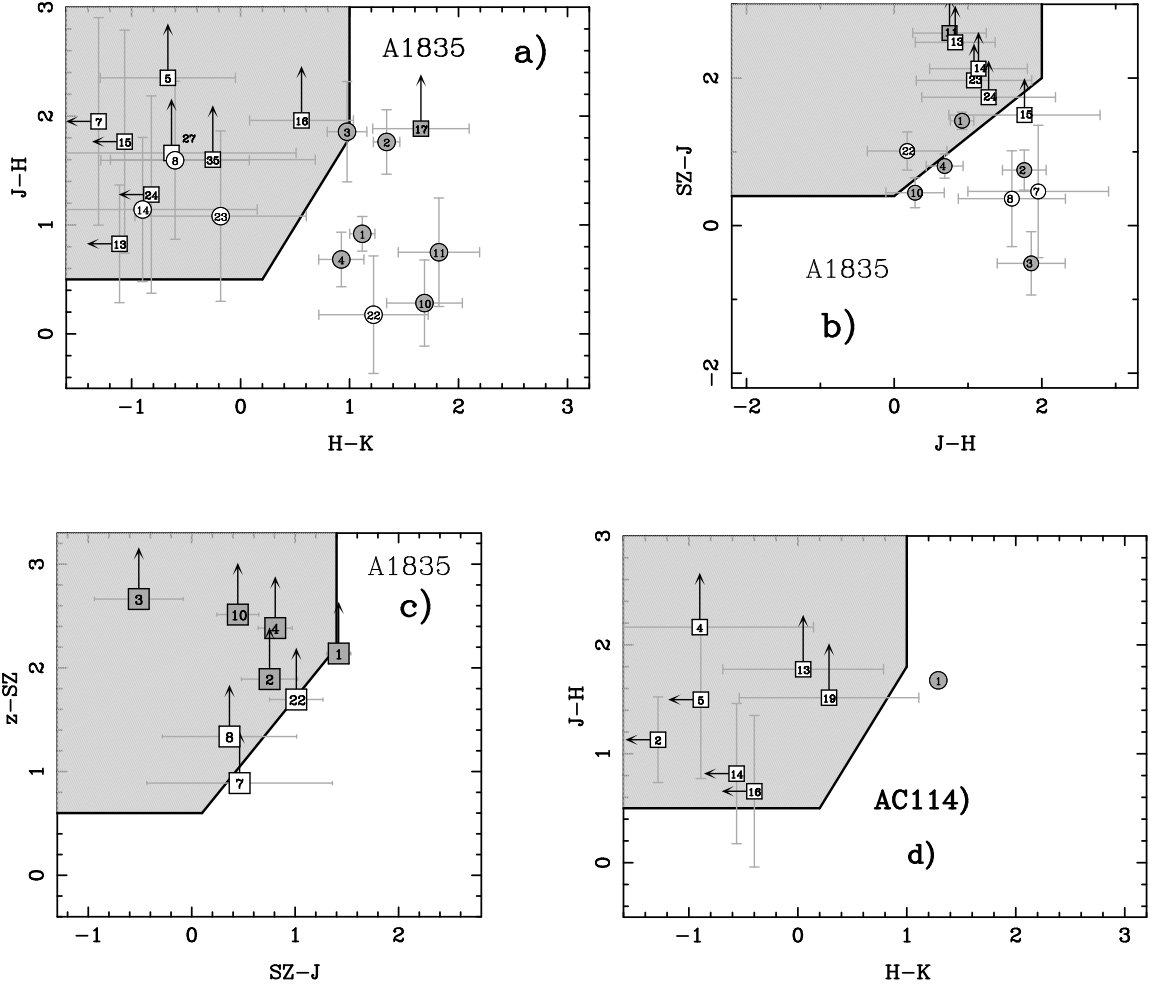

\centerline{\mbox{\psfig{figure=a1835_JHK.ps,height=6cm,angle=270}
\hspace{1cm}\psfig{figure=a1835_SZJH.ps,height=6cm,angle=270}}}
\vspace{1cm}
\centerline{\mbox{\psfig{figure=a1835_zSZJ.ps,height=6cm,angle=270}
\hspace{1cm}\psfig{figure=ac114_JHK.ps,height=6cm,angle=270}}}
\caption{\label{cc}Color-color diagrams showing the location of all optical-dropouts
detected in Abell 1835 (a-b-c), AC114 (d), and the delimitation
of the selection region used in the different 
redshift domains: $JHKs$ ($z \sim 8-10$), $SZJH$ ($z \sim 7-8.5$),
and $zSZJ$ ($z \sim 6-7.5$) (see text). The identification numbers are indicated 
according to Tables \ref{table_a1835} and \ref{table_ac114}. For a
given diagram, circles and squares correspond to objects detected in
three and two filters respectively; upper limits are displayed by an
arrow. Optical dropouts fulfilling the EROs definition are shown in
grey. Dropouts are all detected in the $H$ band, and non-detected in
the $z$ band. Sources are presented in diagrams (a) and (d) if they are 
detected either in $J$ or $K$, and in diagrams (b) and (c) if detected
in $J$ or $SZ$. 
}
\end{figure*}

  We have produced a first list of potential ``high redshift''
candidates by applying the color-color selection criteria described in
Sect.\ \ref{criteria} to the catalog of optical dropouts (see Fig.\ \ref
{cc}). 
We select objects with a fairly red color at wavelengths
close to the Lyman-$\alpha$ break/Gunn-Peterson trough
(``dropout''), and blue colors longward of it, indicative of a blue
UV restframe spectrum.

In practice, depending on the available near-IR photometry, the candidates
can be classified in three approximate redshift bins between 6 and 10.
From the $JHK$ color-color diagram, available for both clusters, 
we selected a sample of candidates
in the range $8\lesssim z \lesssim 10$. The selection region we used is defined by :
\begin{itemize}
\item[]{$(H-K)<1.0$}
\item[and]{$(J-H)>0.5$}
\item[and]{$(J-H)>1.625\ (H-K)+0.175$.}
\end{itemize}
As shown in Fig.~\ref{cc}, the majority of optical dropouts in both clusters
fulfill the high-$z$ requirements. 
The majority of those in the remaining part of this diagram
fulfill the
EROs selection criterion of $R-K>5.6$, and thus they are possible intermediate-redshift
reddened starbursts.

To further distinguish the objects at $z \la 8$
we use, where available, the $SZJH$ color-color diagram to select candidates in the 
range $7\lesssim z \lesssim 8.5$ and the $zSZJ$ diagram in the range
$6\lesssim z \lesssim 7.5$.
The selection in the $SZJH$ diagram is defined by 
\begin{itemize}
\item[]{$(J-H)<2.0$}
\item[and]{$(SZ-J)>0.4$}
\item[and]{$(SZ-J)>0.8\ (J-H)+0.4$}
\end{itemize}
In the $zSZJ$ diagram it is defined by 
\begin{itemize}
\item[]{$(SZ-J)<1.4$}
\item[and]{$(z-SZ)>0.6$}
\item[and]{$(z-SZ)>1.23\ (SZ-J)+0.477$}
\end{itemize}

Figure ~\ref{cc} presents the color-color diagrams for 
the 18 (8) first and second-category dropouts detected in Abell 1835 (AC114). 
We used the location in these diagrams to
distribute the candidates within the different redshift ranges.
For a subsample of these optical dropouts, individual photometric
redshifts were derived from their photometric SEDs. The
attribution of a redshift to each candidate is discussed in
Sect.~\ref{result}. 

\section{\label{result}Results}

We shall now present the results concerning the magnification of the 
high-$z$ candidates, their SEDs and photometric redshifts. Some individual
objects deserve particular discussion. We also provide some elements to
understand the differences found between the two clusters. 

\subsection{Magnification of the high-$z$ candidates}
\label{magn}

High-$z$ candidates were selected based {\it only} on their photometric
properties. Their positions with respect to the critical lines were 
not considered as a selection criterion. However, objects located
close to the high-$z$ critical lines are of greater interest, because
of the larger magnification.

\begin{figure*}[ht]
\centerline{\fbox{\psfig{figure=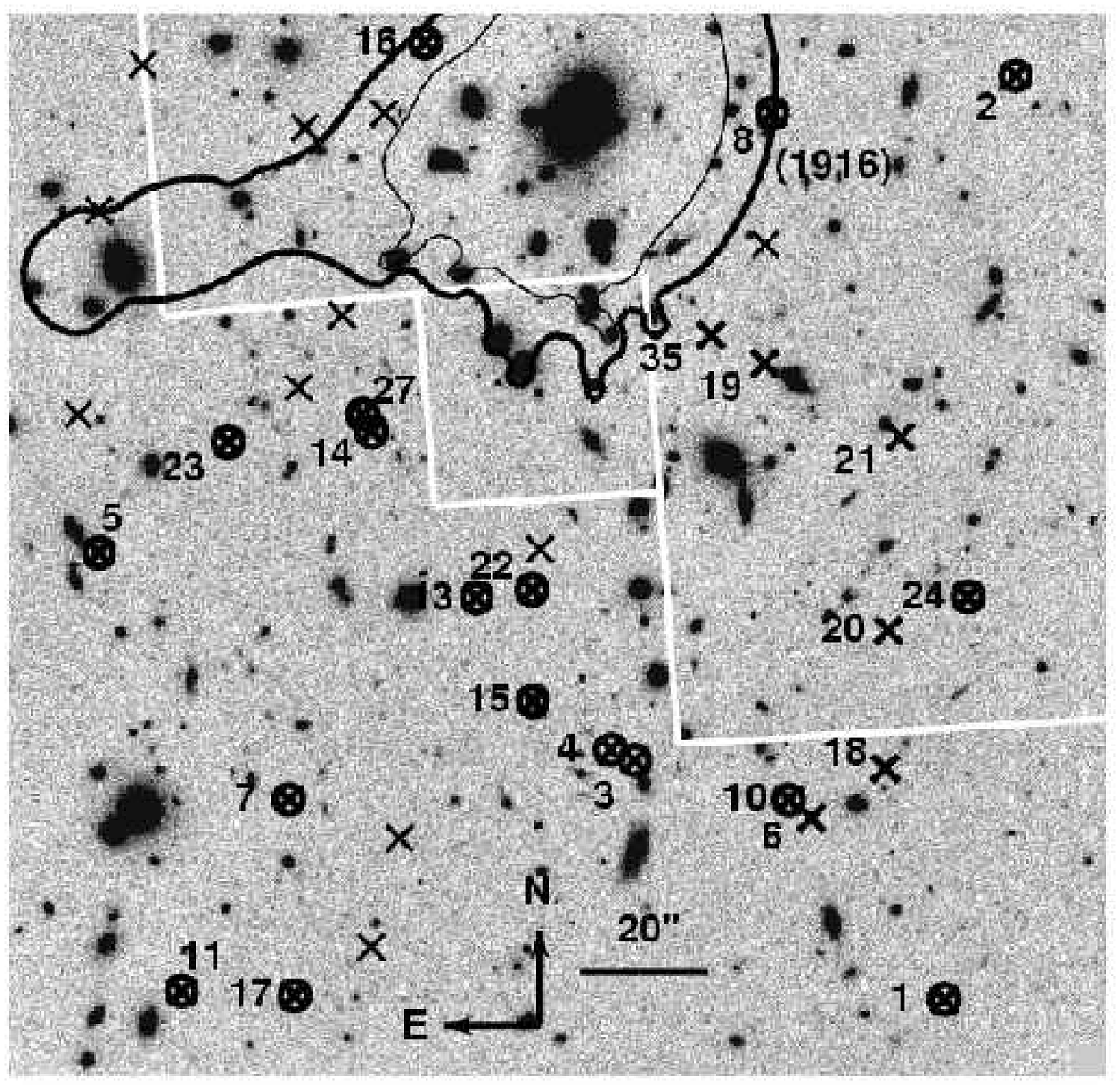,width=7cm}\psfig{figure=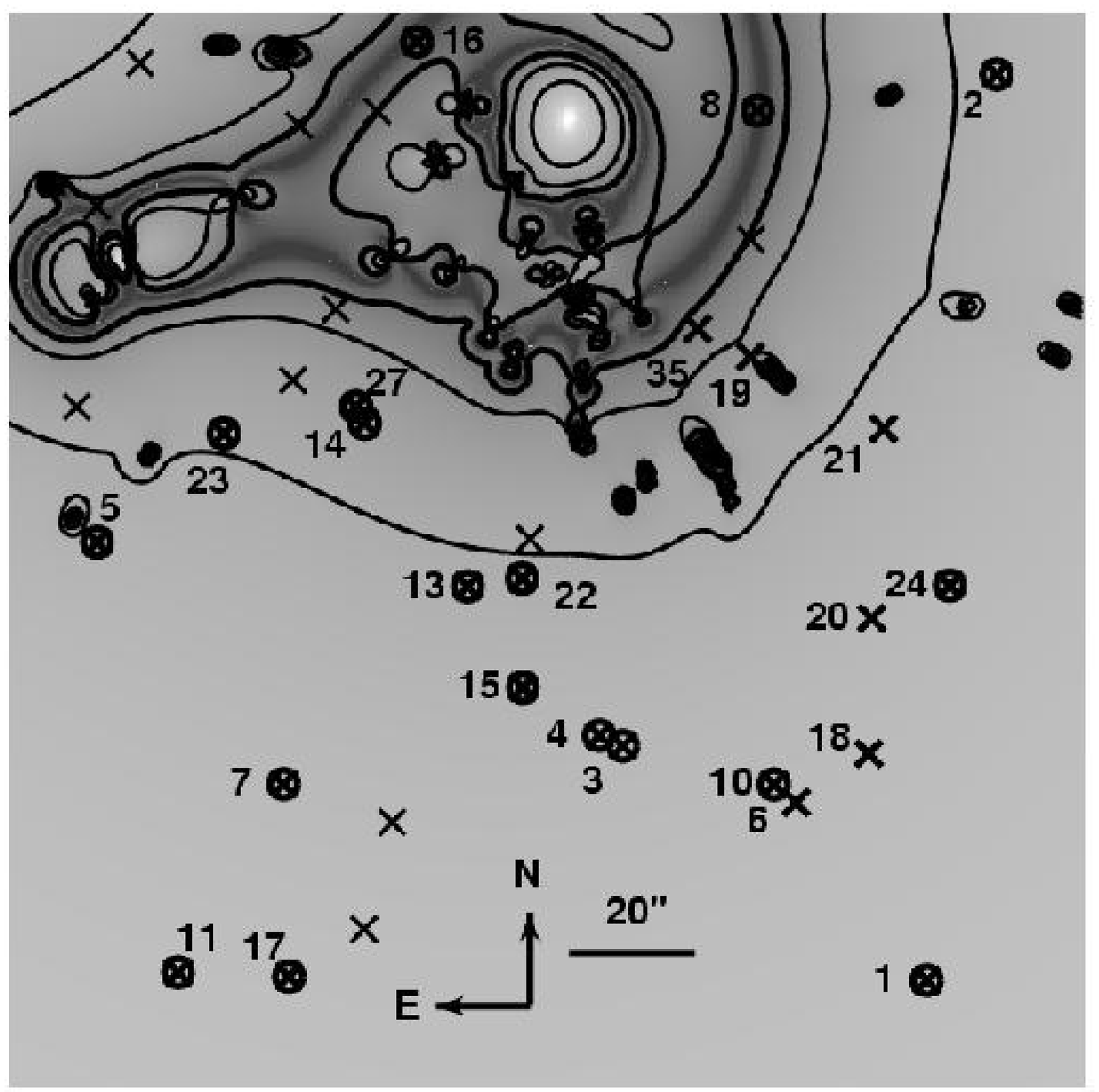,width=7cm}}}     
\caption{\label{cline_a1835}Left : $H$ band image of the lensing cluster Abell 1835
showing the location
of the critical lines at $z=1.5$ (thin solid curve) and $z=10$ (thick
solid curve). All candidates are shown with
crosses; identification numbers are the same as in Table \ref{table_a1835}.
First/Second-category dropouts are circled and
fourth-category objects are not labeled. Right: location of the same objects
relative to the magnification across the field. Contours are overplotted for
magnification values of 1, 2 and 3 magnitudes, computed assuming sources at $z=9$,
although the position of these lines is weakly sensitive to source
redshift within the relevant range $z \sim 6$ to 10.
White lines delimit the footprint of the $R_{702}$ WFPC image, covering $\sim
46\%$ of the whole ISAAC field of view.
}
\end{figure*}

\begin{figure*}[ht]
\centerline{\fbox{\psfig{figure=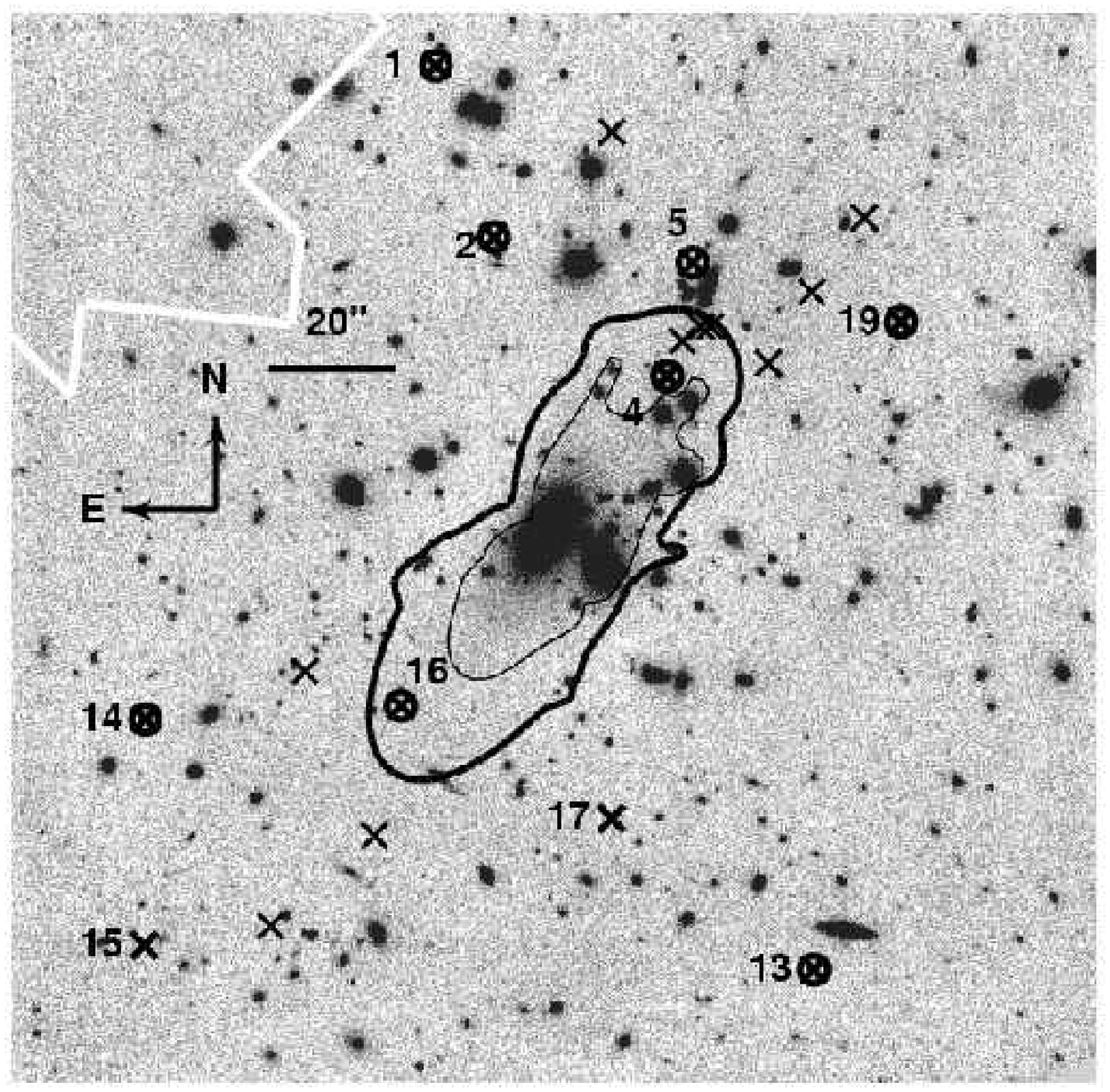,width=7cm}\psfig{figure=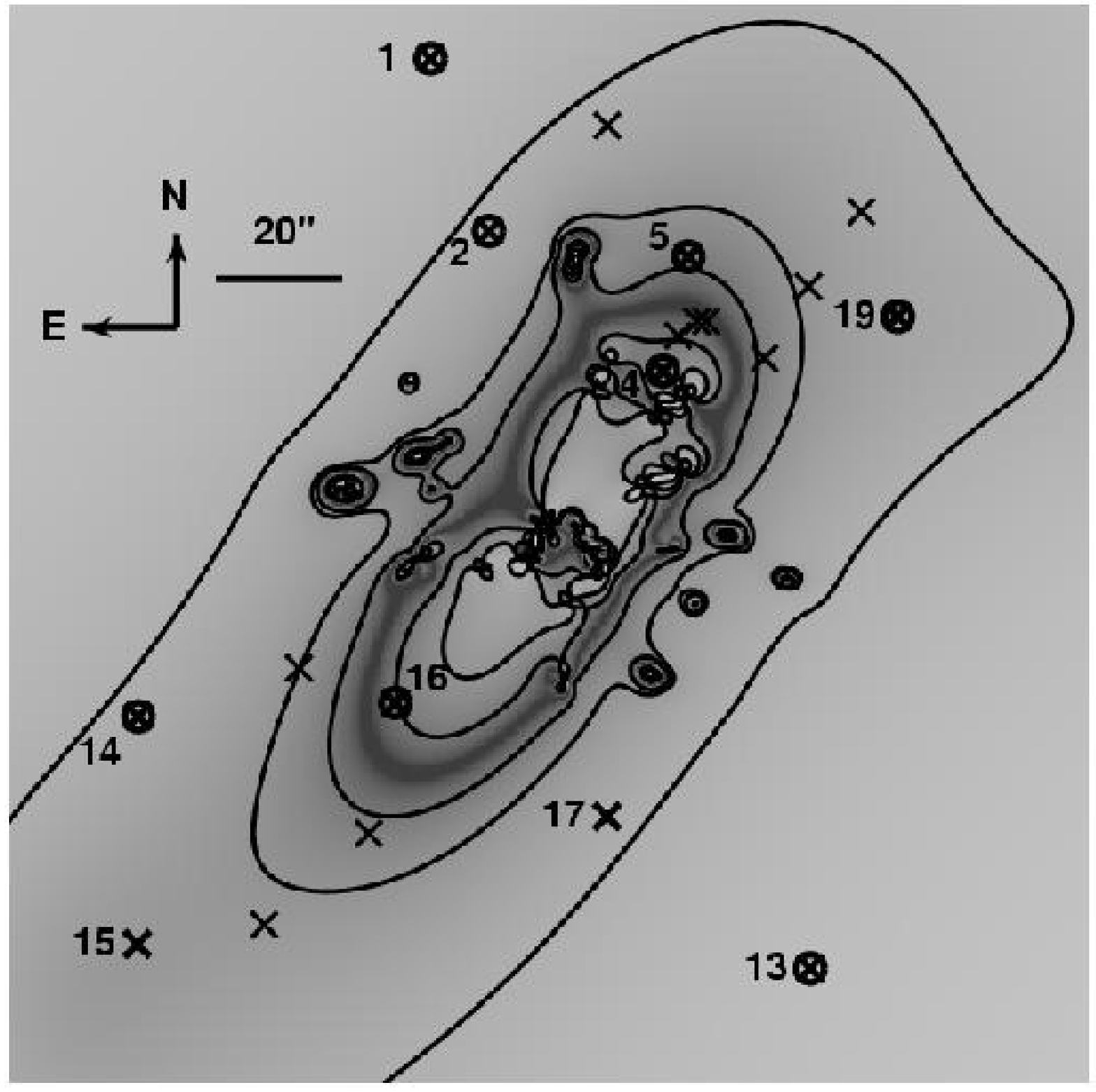,width=7cm}}}
\caption{\label{cline_ac114}Same caption as Fig. ~\ref{cline_a1835} for the
cluster AC114. Identification numbers are the same as in Table \ref{table_ac114}. 
White lines delimit the footprint of the $R_{702}$ WFPC image, covering $\sim 85\%$ 
 of the whole ISAAC field of view.}
\end{figure*}

Figures \ref{cline_a1835} and \ref{cline_ac114} show the final-processed
$H$ band images used for the object detection with $SExtractor$, together
with the location of our candidates in both clusters.
Also plotted 
are the critical lines at $z=1.5$ and $z=10$, and contours of iso-magnification
assuming a source redshift of $z=9.$ computed from the 
lensing models for Abell 1835 (similar to Smith et al.\ 2005) and AC114
(Natarajan et al.\ 1998, Campusano et al.\ 2001). 
The position of these lines is weakly sensitive to source redshift within 
the relevant range $z \sim 6$ to 10. 
For a given redshift estimate, the location of the high-$z$ candidates
on the field allows us to derive their magnification factors 
(see Tables \ref{table_a1835} and \ref{table_ac114}).
$\mu_6$ and $\mu_{10}$ give the magnification factors assuming a source
redshift of 6 and 10 respectively. Where applicable, the adopted magnification 
$\tilde{\mu}$ is computed assuming the ``adopted'' redshift  $\tilde{z}$
described in Sect.\ \ref{hyperz}.

The uncertainty in the magnification factor associated
with the uncertainty on the redshift value is usually smaller than 10\%, except
for a few objects exhibiting the largest magnification factors ($\mu > 10$),
i.e.\ located  within a few arcsecs of the critical lines.
Also the magnification factor at a given position on the image plane varies slowly with
redshift for sources located more than $\sim 10''$ away from the
critical lines. 
Since lensing models are mainly based on the identification of multiple images with 
secure spectroscopic / photometric redshifts, the uncertainty in the derived
magnification factor is usually smaller close to these regions. However, it
could be larger in the case of Abell 1835, because the model is based only on one
multiple image system. Because of the relative insensitivity to source
redshift and position on the image plane, a refined version of lensing models
will not change our present results and conclusions. 

For objects located close to the critical lines at high redshift,
we used the same lensing models to look for possible multiple images
which could affect our analysis (number counts, etc.)
or allow us to better constrain the 
position of the critical lines at $z\gtrsim 6$. With the present
data, we could not find any pair of
objects that would be a fair multiple image candidate, although this
possibility cannot be ruled out because of completeness
considerations (see Sect.\ \ref{calccomp}).

\subsection{\label{hyperz}SED properties and photometric redshift estimates}

  The position of optical dropouts on the different color-color diagrams
provides a first estimate of their photometric redshifts, and an objective
criterion allowing us to classify them into different redshift intervals. 
This criterion can be refined for about 30$\%$ of our candidates presented
below, for which the S/N is sufficient to derive individual probability
distributions in redshift. These sources are among the brightest candidates in
the $H$ band, or have been detected in other filters with S/N$\sim 4-10$. Of
particular interest are the optical dropouts which can be unambiguously
excluded from the $z\gtrsim 6$ sample using photometric redshift
considerations. 

Probability distributions and photometric redshifts were derived 
for our candidates from broad-band photometry over a wide wavelength
interval, using an adapted version of the public photometric redshift software $Hyperz$
(Bolzonella et al.\ 2000). Best-fit redshifts and redshift probability
distributions between $z=0$ and 12 were computed through a standard SED fitting
procedure. We used a variety of template models: starbursts (Kinney et
al.\ 1996; SBS0335-052, Izotov\ 2001), evolutionary models from the
GISSEL library (Bruzual \& Charlot 1993), empirical templates from
Coleman et al.\ (1980), and theoretical templates for zero metallicity
(PopIII) and low metallicity starbursts (Schaerer 2003). Intrinsic
reddening was considered as a free parameter ranging between $A_V=0$
and 3 magnitudes, according to the Calzetti et al.\ (2000) extinction
law. The Lyman forest blanketing is included following the
prescription of Madau (1995). The non-detection in the optical bands
was used as a constraint when computing photometric redshifts.

For further discussion on the candidates, we shall adopt a redshift value
$\tilde z$ combining all the constraints obtained from the color-color
diagrams and the photometric redshift determinations. Redshift values $\tilde
z$ and redshift intervals adopted for each optical dropout are reported in
Tables ~\ref{table_a1835} and ~\ref{table_ac114}. Depending on its 
photometric redshift probability distribution $P(z)$,
we attribute a redshift quality to each candidate ($\phi_{z}$ in
Tables~\ref{table_a1835} and \ref{table_ac114}) as follows:
 \begin{itemize}
\item[(A)]{
Objects displaying a unique solution in their probability
distribution in redshift ($P(z)$), irrespective of the redshift, 
with a good SED fitting for this unique solution (i.e. absolute probability
higher than 90\% in most cases), with no other secondary solutions with reduced
$\chi^2$ better than $\chi^2$(best fit) $+$ 1. 
The best fit redshift in this case corresponds to the $\tilde z$,}
and $z1$,$z2$ given in the tables correspond to $1\sigma$ redshift intervals (68\%
confidence intervals).
\item[(B)]{Objects showing a degenerate solution 
between $z \gtrsim 6-12$  and a lower redshift solution, 
in general within $z\sim$ 1.5--2.5. 
In this case, the two solutions are equally significant, with a good SED
fitting in both cases (reduced $\chi^2\la 1$).
We adopt in this case the
higher redshift solution in a similar way as for (A).}
\item[(C)]{Objects for which no reliable {\it individual} photometric redshift solution could
be obtained, 
either because their $P(z)$ is basically ``flat'', without a significant
solution, or because multiple degenerate solutions exist, all of them
providing a poor fit in terms of absolute probability.
In this case, the redshift limits [$z1$, $z2$] and $\tilde z$ 
in the Tables correspond to the mean redshift 
and the redshift intervals defined by the color-color diagram selection:}
\item[]\centerline{[6.0-7.5] and 6.75 for $zSZJ$}
\item[]\centerline{[7.0-8.5] and 7.75 for $SZJH$}
\item[]\centerline{[8.0-10.0] and 9.0 for $JHK$}
\end{itemize}
For objects displaying a peak at high-$z$ in their
probability distributions $P(z)$ (quality types A and B), the best-fit $\tilde
z$ is always found as expected within the redshift interval defined by the
color-color diagrams. $SZ$ band photometry is not available for AC114, and
thus we only considered the [6.0-8.0] and [8.0-10.0] redshift intervals for
this cluster. 

Figures ~\ref{trombino_a1835} and ~\ref{trombino_ac114} also display the SEDs
and best-fit models for optical dropouts, only for first and well-detected second-category sources.
Several of these objects seem too bright to be high-$z$ sources. 
We discuss below these and other peculiar objects individually.
A more detailed description of individual objects will be presented
elsewhere.
%
\subsection{Contamination by mid-$z$ interlopers}
\label{midz}

   Because of the large photometric errors of the selected optical dropouts in
the near-infrared bands (up to $\sim$ 0.6 magnitudes in some cases) our high
redshift sample is susceptible to contamination by low redshift interlopers
falling in our color-color selection regions. We derive a basic idea of this
contamination factor based on the spectroscopic catalog from the HDFS (Noll et
al.\ 2004) for which
the near-infrared photometry was obtained with the same instrument and filters
(Labb\'e et al.\ 2003).

  Photometric errors were introduced in the HDFS catalog
following a Gaussian distribution of fixed $\sigma$ in each $J$, $H$ and $K_s$
band. Different values of $\sigma$ were used to mimic the typical S/N
in the sample of optical dropouts, and different redshift intervals were
considered for the HDFS sample. The worst contamination level by mid-$z$
interlopers on the ($J-H$) vs ($H-K$) color-color selection diagram ranges
between 20 and 25\%, depending on the redshift intervals, when applying a 0.6
mag. photometric scatter to all filters, i.e. error bars sensibly larger than
those reported in Tables~\ref{table_a1835} and \ref{table_ac114}. None of the
contaminant sources would have been selected as an optical dropout. Therefore,
this source of contamination should represent a second order correction for our
sample.

%
\subsection{Individual objects in Abell 1835}
\label{result_ind}

Optical dropouts in this cluster are distributed as follows 
(cf.\ Table \ref{table_a1835}, Figure \ref{trombino_a1835}). 
We have detected 7 objects satisfying the ERO criterion. Among them, two
(A1835\#2 and A1835\#17) are unambiguously identified as low-$z$ sources, two
display a non-standard behavior in their SED (A1835\#10 and A1835\#11), and
one is clearly variable in (at least) the $SZ$ band (A1835\#4). In general,
optical dropouts displaying anomalous SEDs as compared to young
starbursts are either variable sources (A1835\#4) or sources whose nature could
not be determined with the present data (A1835\#8, A1835\#10, A1835\#11).
These objects,  marked as ``Ex.'' in Tables \ref{table_a1835} and \ref{table_ac114}, 
are not considered as true high-$z$ candidates and are therefore excluded from 
the high-$z$ sample.
Similar arguments can be used to remove the other two ``bright'' EROs
(A1835\#1, A1835\#3) from the sample.

 A general comment concerning the
``bright'' EROs is needed here. As mentioned above, only 2 of these in A1835 and 1
in AC114 could be identified unambiguously as intermediate-$z$ galaxies using
photometric redshift considerations. All the others are difficult to reconcile
with normal ``mid-$z$'' galaxies because of their relatively blue continuum in
the near-IR $JHK$ domain, yielding solutions at $z \sim$ 6--8 which are equally
likely or even better (see Fig. ~\ref{trombino_a1835} and
Table \ref{table_a1835}). However, they are extremely luminous if at such
high-$z$, typically M$_B$ $\sim$ $-25.0$ to $-26.5$ for the brightest ones.
Although ``bright'' high-$z$ sources of this kind
may exist, up to $50 L_{*}$, as recently proposed by Mobasher et al.\
(2005) for their massive post-starburst galaxy at z$\sim$6.5, we exclude these sources from the high-$z$ sample, and consider the
low-$z$ solution as more plausible at this stage. Spectroscopic information is
needed to further characterise these faint lensed EROs, and determine their nature by a measurement of their redshifts. 

Some sources in A1835 deserve specific comments.
\begin{itemize}

\item{A1835\#2 : This source corresponds to J5, the near-IR
counterpart of the SCUBA-selected galaxy SMMJ14009+0252 (Ivison et
al.\ 2000, Smail et al.\ 2002). The near-IR photometry reported here is compatible
with the recent results by Frayer et al.\ (2004). It satisfies the ERO
criterion, and it is likely a low-$z$ source as suggested by our SED
fitting result: $z$=1.34 (with 1$\sigma$ ranging between $z$=1.18 and
1.64).  This solution is in good agreement with the redshift constraints 
$0.7\lesssim z\lesssim 1.8$ derived from the radio-submm spectral 
index $\alpha^{850}_{1.4}$, but inconsistent with the range 
$3\lesssim z\lesssim 5$ suggested by the submm colors (Ivison et al.\ 2000).}

\item{A1835\#4 : As mentioned in Appendix~\ref{addtests}, this source
displays a difference of 0.4 mags in $SZ$ (more than 3
$\sigma$) between the two series of images obtained on the 19 April and
the 15 May, which seems to indicate an intrinsically variable source.
}

\item{A1835\#8 : This object (previously named A1835\#1916, also known as the $z\sim 10$ candidate) was
studied in detail in Pell\'o et al.\ (2004ab).
The photometry presented here is an improved version; the present and earlier
magnitudes in the common filters are compatible within 1$\sigma$ error-bars.
The field around A1835\#1916 has been reobserved
between 30 May and 6 June 2004 by Bremer et al.\ (2004) with NIRI/GEMINI in the $H$ band.
Surprisingly, the object is not redetected in these images, which
are at least $\sim$ 0.5 mag deeper than the ISAAC images taken
approx.\ 15 months earlier. The reality of our initial photometric detections
is not questioned by Bremer et al.\, who reconfirm it using our data, although
the photometric properties of this source are still a matter of debate 
(Lehnert et al.\ 2005, Smith et al.\ 2006). 
The detection in 3 bands where the object is
re-detected, including our new $SZ$ images (see Table~\ref{table_a1835}), makes a spurious event highly
unlikely (about 12\% probability, from our estimates given in Table
\ref{spurious}). When we consider the present results together with  
our previous findings (the source was virtually non-detected in our $J$
images), and the non-detection by Bremer et al.\ in the $H$ band with new
independent data,
this source could be intrinsically variable.
Its nature (and hence also its redshift) presents a puzzle, and will be
discussed elsewhere. Hereafter we do not consider this source within the 
high-$z$ sample. }

\item{A1835\#35 : This is a third-category candidate (previously
named A1835\#2582) and a rather unusual emission line galaxy
already studied in detail by Richard et al.\ (2003). 
As for \#8, the present and earlier magnitudes in the common filters 
are compatible within 1$\sigma$ error-bars. The marginal detection of this
object in the $I$ and $J$ bands reported by Richard et al.\ (2003) is found 
to be non-significant with our new and more conservative error estimates.
Thanks to [O{\sc iii}]$\lambda\lambda$4959,5007 and H$\beta$ detected in the $J$ band,
\#35 has been identified as an extremely faint galaxy at $z=1.68$, with $M_B
\sim -16.4$ and a gravitational magnification of $\sim 2$ magnitudes.
This object has been removed from the photometric sample of
high-$z$ candidates.} 
\end{itemize}

\subsection{Individual objects in AC114}

From the 10 optical dropouts selected in AC114 
(cf.\ Table \ref{table_ac114}, Figure \ref{trombino_ac114}), 
8 are in the first and second-category samples.
Among those, only one object satisfies the ERO criterion (AC114\#1, cf.\ below). 
The lack of $z$ and $SZ$ photometric data for this cluster precludes
a further classification into redshift bins between $z \sim 6-8$.

\begin{itemize}
\item{AC114\#1 : This source satisfies the ERO criterion, and it is
likely a low-$z$ galaxy according to our SED fitting result: $z=1.62$
(with 1$\sigma$ ranging between $z=1.58$ and $1.89$).  
}
\end{itemize}

\subsection{Differences between the two lensing fields}
\label{cluster-to-cluster}

The total number of remaining first and second-category high-$z$ candidates in AC114 as
compared to Abell 1835 is found to be in a ratio of 7/10
(9/15 whith third-category
candidates), after excluding 
EROs in both two clusters.
Several reasons could explain this difference, in addition to field-to-field variance:

\begin{itemize}

\item The depth of the near-IR images is less in AC114 as compared to Abell
1835, by 0.1 and 0.4 magnitudes in $J$ and $Ks$ respectively. Also the overall
detection image in $H$ is noisier for AC114, as shown in Table ~\ref{spurious},
rendering the identification and visual inspection of near-IR
detections more difficult. We expect a larger fraction of (blind) false positive detections
in AC114 than in Abell 1835 (Table \ref{spurious}). Thus, if the sample was dominated by such
detections, the number of candidates should be smaller in Abell 1835 than in
AC114. However, the opposite trend is observed, which
means an efficient (manual) control of the sample. 

\item The optical images used to identify dropouts are not identical
in the two cases. In Abell 1835, the main constraint comes from a deep $z$
band image (limiting magnitude $z_{AB} \gtrsim $ 27.3, whereas it is $I_{AB}
\gtrsim $ 27.2 in AC114). In AC114, the strongest non-detection criterion 
is set by the $R/HST$ image (limiting magnitude $R_{AB} \gtrsim 28.7-28.9$ in
the relevant region of the field, whereas it is $R_{AB} \gtrsim $ 28.0 in
Abell 1835). Star-forming {\it and} highly-reddened intermediate-$z$ sources such as
A1835\#35 could survive more easily in the Abell 1835 sample than in AC114
because of the difference in the $R$ band.

\item Geometrical considerations coming from lensing are also to be taken into
account. Abell 1835 is not centered on the CD galaxy, and the two clusters have different
redshifts ($z=0.252$ and $0.312$). As shown in Fig. ~\ref{cline_a1835} and
~\ref{cline_ac114}, the magnification factors across the ISAAC field of view
are different thus leading to different {\it effective} surveys, as discussed
in detail in Sect.~\ref{discuss}. In summary, within the redshift domain
considered here ($6 \lesssim z \lesssim 10$), the two clusters cover about 
the same area {\it on the sky} close to the critical lines for the largest 
magnifications factors beyond 5 (24 and 21\% of the total surface respectively), and 
thus we expect (and observe) the same number of such sources in
the two fields. On the contrary, for magnification factors below 5,
the two clusters exhibit a 
different behavior:  31 and 47\% of the total surface respectively for Abell
1835 and  AC114, for magnifications factors between 2 and 5, and
45 and 32\% of the total surface respectively for magnifications factors below
2 (0.75 magnitudes). In other words, the sample observed in
Abell 1835 is expected to be dominated by sources with magnification factors
below 2 (up to $\sim$50-60\%, assuming sources with a uniform distribution in
$z$ and same luminosities, after completeness correction), whereas the
majority of candidates detected in AC114 are expected to exhibit magnification 
factors above 2 under the same conditions. This is the qualitative behavior
of candidates actually observed. For magnifications factors above 2, and taking
into account that we are dealing with small number statistics,
the relative number of candidates detected in both clusters is in good agreement
with expectations. On the contrary, the simple considerations given here
cannot explain the excess in the number of candidates detected in Abell
1835 with respect to AC114, for magnification factors below 2.

\end{itemize}

\section{\label{discuss}Discussion}

   In this section we discuss the implications of the present
results in terms of abundance of star-forming galaxies at $6 \lesssim
z \lesssim 10$. We present and discuss the intrinsic properties of candidates
actually detected, after correction for lensing magnification, and we
compare the observed counts with
order-of-magnitude expectations obtained from simple modeling. The observed
number densities of candidates, as seen through gravitational
lenses, are translated into {\it effective} number densities through 
a careful modeling of lensing effects, easily comparable with blank
field studies. The luminosity function and the cosmic star formation
rate derived from our sample of $6 \lesssim z \lesssim 10$
candidates is also presented and discussed. 
The large correction factors applied to this
sample make the determination of integrated quantities, such as luminosity
functions and SFR densities extremely difficult.
In the final section, we
briefly describe the preliminary results obtained on the spectroscopic
follow-up of the photometric candidates. 

\subsection{\label{intrin}Intrinsic properties}

The typical magnification values of our candidates range between 1.5 ($\sim$
0.44 mags) and 10 (2.5 mags). For some objects very close to the critical
lines, we found magnification values $\mu>25$. However, because of the
underlying error in the precise location of the critical lines from the
models, we prefer to adopt a more conservative lower limit of $\mu=25$ for
these objects. The average(median) magnification values among the first-priority
high-$z$ candidates are 6.5 (2.3) in Abell 1835 and 7.9 (3.5) in AC114.

We derived the unlensed $L_{1500}$ luminosity, at 1500 \AA\ in the restframe, for all high-$z$ candidates,
using the adopted value $\tilde z$ for the redshift and a flux estimate from
the photometry in the band closest to this restframe wavelength ($SZ$, $J$ or $H$).
Deriving the intrinsic properties of $6\lesssim z\lesssim 10$ galaxies can be challenging due to the significant level of contamination.

After correction by the lensing magnification affecting each object (see Sect.\ \ref{magn}) the
unlensed $L_{1500}$ luminosities were converted into Star Formation Rate (SFR) through the
usual calibration from Kennicutt (1998):

$$SFR\ ({\rm M}_{\odot}\ {\rm yr}^{-1})=1.05\ 10^{-40}\ L_{1500}\ ({\rm ergs}\ {\rm s}^{-1}\ {\rm \AA}^{-1})$$

These physical properties are summarized for each candidate in Table
\ref{table_a1835} and \ref{table_ac114}. The typical SFR obtained for
objects included in the final sample (excluding EROs and anomalous SEDs) 
is about $\sim$ 10 M$_{\odot}\ $yr$^{-1}$, with extreme values ranging between a
few units and $\sim$ 20 M$_{\odot}$ yr$^{-1}$. The
conversion of $L_{1500}$ into SFR assumes a constant star formation at
equilibrium, and such conditions are not necessarily reached in these
objects. 

Interestingly, although the selection criteria are only based on near-IR colors 
irrespective of magnitude, {\it almost all} the photometric candidates fulfilling 
our selection criteria are {\it fainter} than $H = 23.0$ (AB $\sim
24.5$). Only three exceptions are found in Abell 1835 among the possible
low-$z$ EROs, as described above. After correction for magnification accross
these fields, the lack of ``bright'' sources means that we have not detected
young starbursts at $z \sim6-10$ more luminous than typically $L_{1500}$ $= 3\times 
10^{-41}\ {\rm ergs}\ {\rm s}^{-1}$, i.e. more massive than typically a few
10$^8$ M$_{\odot}$ (for starbursts younger than 10$^6$ yr, under standard
assumptions for the IMF). 

Also, a direct comparison between low-$z$ and high-$z$ SEDs shows that
our high-$z$ candidates tend to be very blue in $H-K$. 
For $z \ge 6$ candidates, $H-K$ colors provide an estimate of the
restframe UV slope $\beta$, assuming $3\sigma$ detection levels in $K$ when
the source is not detected in this band, or when the S/N in this band is lower
than $3\sigma$, with large uncertainties due to photometric error bars. 
The UV slope $\beta$ usually ranges from $-2.5$ to $-3.5$
for the first category dropouts, with two sources (A1835\#7 and AC114\#2)
reaching $\sim -3.9$ at 1$\sigma$, and between -0.7 and $-3.0$ for the second
category dropouts. This systematic trend towards extremely blue colors was
also reported by Bouwens et al.\ (2004b) for their sample of $z\sim7-8$
candidates. 

Although the optical dropouts we found are stretched by the magnification
factor $\mu$, they appear as point-like sources 
in our ground-based images. If we assume a minimum magnification of 0.4 mags
for all the field, the physical size of these objects at $z>7$ is likely to be
smaller than 1.7 kpc, in good agreement with size calculations of Kneib et al. (2004) 
and Bouwens et al. (2004b).

\subsection{\label{Nzmodels} Observed number counts vs model expectations}

 The {\it efficiency} of using lensing clusters as gravitational telescopes to find
high-$z$ galaxies can be evaluated with model expectations and simple
assumptions. 

We first use
a simple model to estimate the expected number counts, both in blank
fields and lensing clusters, taking into account our photometric selection
criteria. We use semi-analytical models for dark-matter halo formation
(the Press-Schechter formalism, Press \& Schechter 1974). Starburst models
presented in Sect.~\ref{criteria} were used to scale the SED, assuming that the
fraction of the baryonic mass converted into stars is fixed to $f_{*}=0.1$
within the redshift interval $6\lesssim z\lesssim 11$. 
We focus on two different ``extreme'' assumptions for the IMF:
a standard Salpeter IMF from 1 to 100  M$_{\odot}$ and a top-heavy IMF (with stars ranging
from 50 to 500 M$_{\odot}$). 

To correct these estimations for the relative \textit{visibility time} of the bursts, which are 
typically of $t^{*}=10^6$ yr timescales from our simulations, we multiply the
number density of objects expected at a given redshift $z$
by the \textit{duty-cycle} factor : $t_{*}\ (1+z)/(t_H(z)-t_H(17))$, 
where $t_H(z)$ is the age of the Universe at redshift $z$. This corresponds to the 
probability for a burst to be visible at redshift $z$, assuming that all
haloes convert a constant fraction of their baryonic mass into stars, at some time
between redshifts 17 and $z$.

Lensing introduces two opposite trends in the observed sample compared to
blank fields: 1) gravitational magnification by a factor $\mu$, increasing the 
number of faint sources and thus the total number of sources, and 2) reduction of
the effective surface by the same factor 
thus leading to a dilution in observed counts.  
We explicitly compute the expected number counts with these models in clusters 
by a pixel-to-pixel integration of the magnification maps as a function of
redshift, using the lensing models, after masking all the
pixels lying in the mask of bright objects described in Sect.~\ref{check}.
The expected number counts up to $H\le24$ are very similar for the two clusters
(less than 10\% difference over the redshift interval), and thus we averaged
together both results into a unique ``strong lensing cluster''
prediction. These results hold for intermediate
redshift clusters ($z\sim0.2-0.3$) and should be revised for lensing clusters
at higher redshifts and/or a different field of view, although the trends
remain the same.  

\begin{figure}[ht!]
\psfig{figure=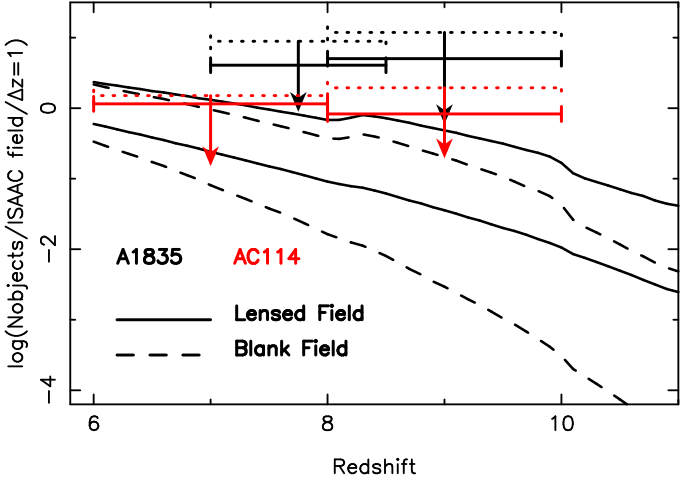,width=9cm,angle=270}
\caption{\label{ncounts} Comparison between the expected number counts of
galaxies in the field of ISAAC, up to $H\le$24, per redshift bin
$\Delta z=1$, in a blank field and in the field of a
strong lensing cluster (see text for details).
Expected counts are obtained with the simple model discussed in
Sect.~\ref{Nzmodels}, for two extreme IMF: 
a standard Salpeter IMF (lower curves) and a top-heavy IMF (upper
curves). The differences between a blank-field (dashed line) and lensing
fields (solid lines) are more pronounced at higher redshift. 
Observed counts are displayed for the two lensing clusters, corrected for
incompleteness effects. Solid lines 
display the results for first-category sources only, whereas dotted lines correspond to first and second-category candidates. 
}
\end{figure}

 The comparison between expected and observed number counts of
galaxies in the field of ISAAC, up to $H\le24$, per redshift bin
$\Delta z=1$, in a blank field and in the field of a
strong lensing cluster are presented in Fig.~\ref{ncounts}. 
Blank field number counts are not corrected for bright-object masking, and
thus they correspond to an upper limit. 
Observed number counts in the two lensing clusters have been corrected for
photometric incompleteness (typically a factor of $0.05-0.1$) using the simulations
reported in Sect.~\ref{calccomp}, sample incompleteness (by a constant factor
$C_{sample}\sim0.42$ as detailed in Appendix~\ref{sampcomp}) 
and for the expected fraction of
false-positive detections, from our estimates given in Table
\ref{spurious}. Results are presented in Fig.~\ref{ncounts}.

For each redshift bin, we corrected the number counts for both first and second-priority 
candidates (using $\tilde{z}$ as their redshift estimate), and the results are
directly plotted on the model 
expectations, without any normalization. Observed number counts are 
upper limits, as our sample is likely contaminated by low-$z$ interlopers. 

As shown in Fig.~\ref{ncounts}, strong lensing fields are a factor of
$\sim 5-10$ more efficient than blank fields of the same size in the 
$z\sim 7-11 $ domain, all the other conditions being the same. 
Observed number counts of candidates at $z\sim6-8$ are in good
agreement with these order of magnitude estimates, 
in particular with $\sim$10\% of the baryonic
mass converted into stars at $z\ge6$. At $z\ge8$, the observed number counts
are more consistent with a top-heavy IMF, with a large
cluster-to-cluster variance. 

\subsection{Lens-corrected number densities of high-$z$ sources}

We have used the lensing models to derive the {\it effective} areas and
corresponding volumes surveyed in the different source planes. The aim is to
translate the observed number densities of candidates into {\it effective}
number densities easily comparable with blank field studies. 
We also correct our observed sample of candidates for incompleteness 
and false-postive detections.

Magnification and dilution effects by the lensing field are carefully
taken into account to compute number densities and derived quantities. 
The average magnification value over a whole ISAAC field
is about 2, thus leading to a dilution close to 50\% over the whole field. 
However, a careful modeling is needed to properly take into account the
intrinsic incompleteness of the sample as a function of redshift and
position on the field.
For each candidate in the field, with observed magnitude $H_o$,
we compute the magnification factor $M(\Omega,z)$ as a function of its
position and redshift $z$, as well as the lens-corrected magnitude $H_e$
(referred hereafter as {\it effective magnitude}), using the lensing models
presented in Sect.\ \ref{magn}.  
The \textit{effective completeness} $\eta\ (H_e,z)$ gives the ratio between the observed
number counts in the lensing field, $N_{o}(H_e,z)$, and the equivalent value
measured on a blank field of the same observed surface, $N(H_e,z)$, complete
up to the magnitude $H_e$:

$$
\eta\ (H_e,z)=\frac{N_{o}(H_e,z)}{N(H_e,z)}
$$

This quantity can be written as a function of the solid angle surveyed
on the sky, $\Delta\Omega$, 

$$
\eta\ (H_e,z)=\frac{\displaystyle \int_{\Delta\Omega}\frac{\displaystyle N(H_e,z)}{\displaystyle M(\Omega,z)}C(H_o)d\Omega}{\displaystyle \int_{\Delta\Omega} N(H_e,z)d\Omega}=
$$
$$
=\frac{1}{\Delta\Omega}\int_{\Delta\Omega}\frac{C_{sample}\ C(H_e-2.5\ log_{10}M(\Omega,z))}{M(\Omega,z)}d\Omega
$$

where $C(H_o)$ stands for the photometric incompleteness correction for an {\it observed}
$H$ band magnitude $H_o$ (plotted in Fig.~\ref{compcurves}) and $C_{sample}=0.42$ 
is the additional incompleteness factor for the sample, as given in Appendix~\ref{sampcomp}.

\begin{figure*}[ht]
\begin{minipage}{9cm}
\centerline{\mbox{\psfig{figure=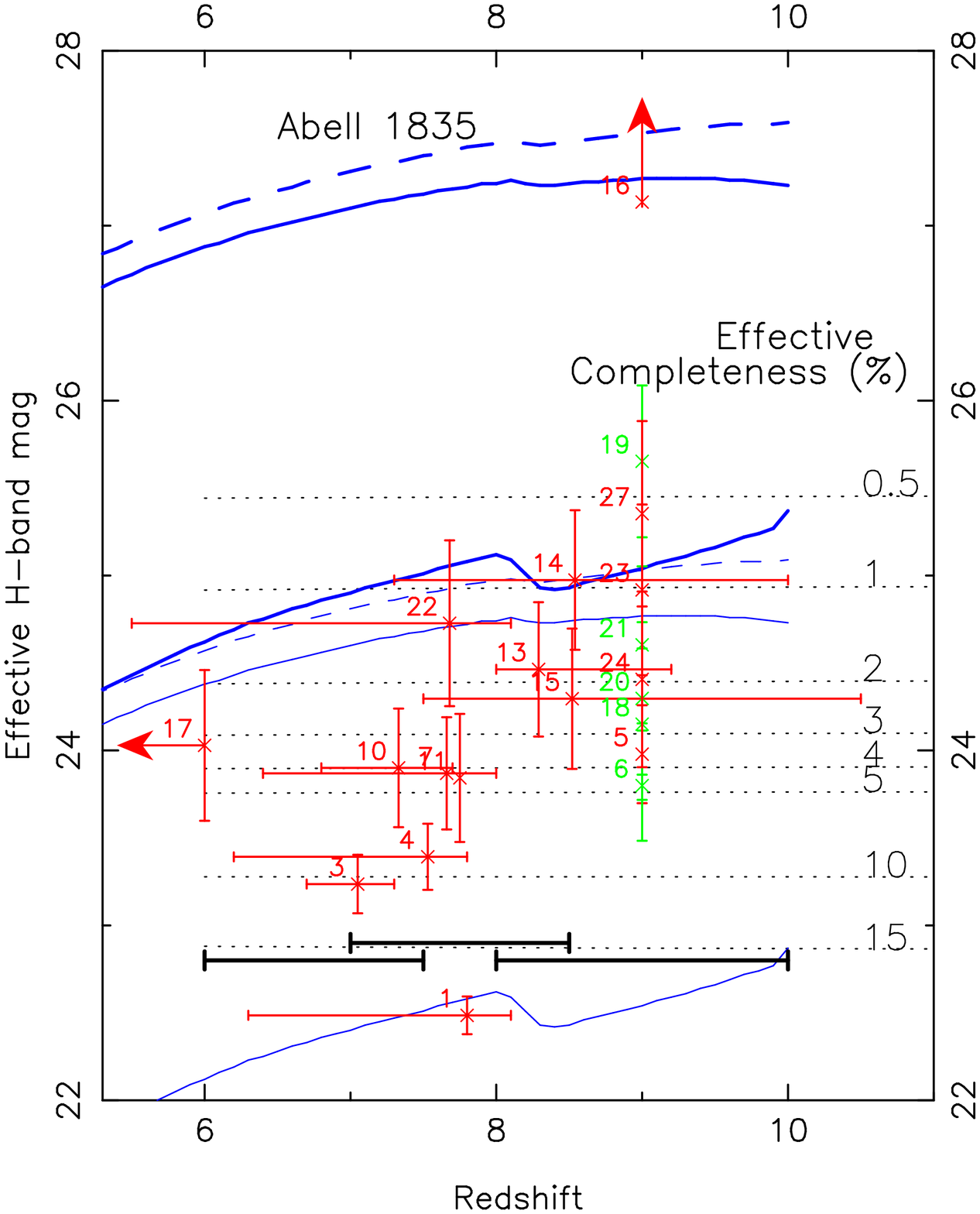,width=8.cm}}}
\end{minipage}
\begin{minipage}{9cm}
\centerline{\mbox{\psfig{figure=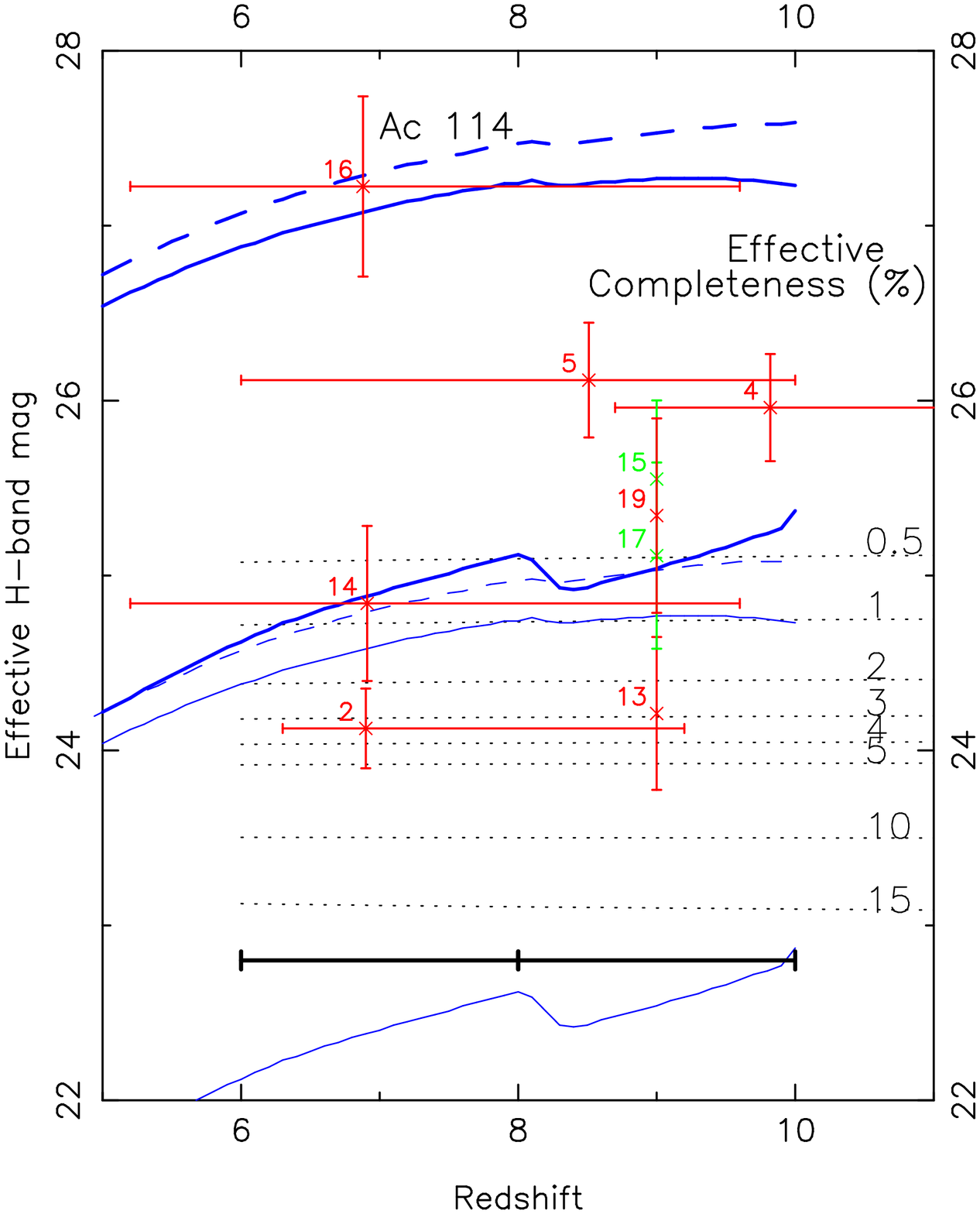,width=8.cm}}}
\end{minipage}
\caption{\label{comp_heff}Effective magnitudes $H_e$, corrected for lensing, 
as a function of $z$ for the two lensing clusters. Values of effective completeness 
$\eta(H_e,z)$ with respect to a blank field (for number counts complete up to
$H_e$) are overplotted as dotted lines for the 0.5 to 40\% levels (see text
for details). 
First and second-category (red) and third-category (green) candidates are positioned at their
adopted redshift $\tilde{z}$, with errors bars in $z$ corresponding to $[z1-z2]$ range (see
text for details). Redshift intervals considered for third-category
candidates are given at the bottom of the figure.
Photometric errors in $H_e$ coincide with photometric
errors observed in $H$. 
Solid lines display the predicted magnitudes versus redshift for 4 different
starbursts models used in Sect.~\ref{criteria}, with a stellar masses scaling
to $10^7$ M$_{\odot}$ (thick line) and $10^8$ M$_{\odot}$ (thin line).
From top to bottom, these models correspond to single bursts with Salpeter IMF
and stellar masses ranging from (1) 1 and 100 M$_\odot$, 
and (2) 50 and 500 M$_\odot$. For comparison, the dashed line
corresponds to a constant star-forming model with age $10^8$ yr.
}
\end{figure*}

   In practice, we computed the values of $\eta$ as a function of $H_e$ and
redshift using the magnification maps over the field, after masking all the
pixels lying in the mask of bright objects described in Sect.~\ref{check}. 
Figure \ref{comp_heff} shows, for each cluster, the location of
the candidates in the ($z$,$H_e$) plane. Overplotted are different
models of high-$z$ starbursts, and completeness levels $\eta$.
Excluding the 3 brightest EROs in Abell 1835 (which are possibly low-$z$
sources), all our candidates range from
$H_e$=24.0 to 27.2 (AB $\sim $ 25.5 to 28.7). 
From these diagrams, it appears that our sample of candidates is sensitive to
stellar mass scales in the range $10^7-10^8$ M$_{\odot}$, and that our typical
correction levels range from 1 to 15\% (including both lensing dilution and
photometric incompleteness).
Candidates with the smallest $\eta$ factors in these diagrams have the largest
weights in the number densities and derived quantities.

\subsection{Luminosity Function}
\label{lf_sec}

A fair estimate of the luminosity function (LF) at 1500 \AA\ can be derived from the 
intrinsic luminosities of our candidates presented in Sect.~\ref{intrin}, 
using the same approach as in the previous section to compute
number densities corrected for incompleteness and spurious detections.
As discussed in the previous sections, the correction factors applied to
this sample are relatively large, thus leading to large error bars in the LF
determination.
We discuss here two redshift intervals for which we have enough sources for
this exercice: $6\lesssim z\lesssim 10$ and $8\lesssim z\lesssim 10$.
In practice, we derive the number density of objects in the co-volume surveyed,
with $\Delta log_{10}(L)=1$, after correction for the individual $\eta$ factors. 
Error bars are estimated as a combination of two independent sources of noise:
the $1\sigma$ confidence levels for a Poisson distribution and the  
uncertainty in luminosity introduced by the $\eta$ factor (typically a factor
of 3). The latter is a combination of the photometric error bars, and
uncertainties in the incompleteness corrections and lensing modeling. 
When no object was detected in a luminosity interval, we corrected the
Poisson-noise upper limit by the typical effective completeness $\eta$ for this
luminosity. 
We also corrected our data points for the fraction of false-positive 
detections expected from Table \ref{spurious}.

\begin{figure}
\psfig{figure=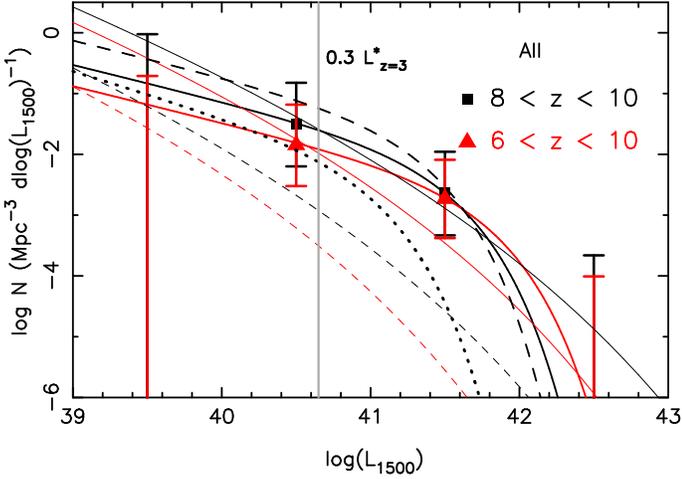,width=9cm,angle=270}
\caption{\label{lf} 
Combined $L_{1500}$ LFs for the two fields, 
for two redshift intervals: 
$6\lesssim z\lesssim 10$ (red) and $8\lesssim z\lesssim 10$
(black). 
Data points are corrected for spurious sources, and 
error bars combine Poisson noise statistics and uncertainty in the 
effective completeness. Note the large correction factors applied to this
sample, which translate into large error bars.
The STY fits to the LF data are represented by thick solid lines.
For comparison, the LF fit by Steidel et al. for LBGs at
$z\sim 3$ is also overplotted (thick dashed line), 
as well as the $z\sim 6$ fit from Bouwens et al.\ (2005) (thick dotted line),
{\it without any additional renormalization to fit the data}. 
Also the $L_{1500}$ LF derived from the simple models presented in
Sect.~\ref{Nzmodels} is shown, in the redshift intervals 
$6\lesssim z\lesssim 10$ (red thin lines) and $8\lesssim z\lesssim 10$
(black thin lines), for the classical salpeter IMF (dashed lines) and the top-heavy
IMF (solid lines). The cosmic SFR value has been derived 
by integrating this LF down to $0.3 L^*_{z=3}$ (shown as a vertical grey line).
See text for more details. 
}
\end{figure}

The combined $L_{1500}$ Luminosity Functions for both clusters, with the 
corresponding error-bars, are given in Fig.~\ref{lf}.
Only first and second-priority candidates have been considered, but the difference
obtained when using the full sample is within the $1\sigma$ error bars.

The observed LFs have been fitted by the STY method (Sandage, Tammann \&
Yahil\ 1979), a maximum likelihood fit of data points to the analytical
Schechter function $\phi(L)$ (Schechter et al.\ 1976):
$$
\phi(L) dL  =  \phi^*_{1500} \left(\frac{L}{L^*}\right)^\alpha
\exp \left(-\frac{L}{L^*}\right)\, d\!\left(\frac{L}{L^*}\right)
$$
assuming that this function provides a good representation of the data. 
Due to a lack of information towards the faint end, a strong degeneracy is
expected between $L^*$ and $\alpha$, which we do not discuss here. To avoid this problem, we assumed a fixed value of $\alpha=1.6$,
corresponding to the Steidel et al.\ (1999) determination for Lyman Break
Galaxies (LBGs) at $z\sim 4$, and leave the other parameters free.
The typical values found for $L^{*}$ are $10^{41.5}$ \ergs\ s$^{-1}$\ \AA\ $^{-1}$, and these 
results are not affected by the way we binned the data points. The STY fits to the 
data presented in Fig.~\ref{lf}. 

Without the blind correction for false-positive detections, 
the data points increase by $\sim$0.4-0.6 dex, which is 
a relatively small effect compared to the error bars. 

For comparison, we overplot in Fig.~\ref{lf} the LF fit
found by Steidel et al. for LBGs at $z\sim 3$, after correction for
differences in the respective cosmological parameters. 
It is shown as a thick dashed line in Fig.~\ref{lf} {\it
without any renormalization} to fit the data points. This LF for LBGs at
$z\sim 3$ seems to be slightly higher (by $\sim 0.5$ dex) than the LF for our
candidates, but compatible within the $1\sigma$ error bars. 
Our results are also fully compatible with the LF derived by Bouwens et al. (2005)
(presented as a dotted line in Fig.~\ref{lf}) 
for their sample of $z\sim 6$ candidates in the UDF, UDF-Ps and GOODS fields
in the low-luminosity regime, i.e. for $L_{1500}\lesssim 0.3 L^*_{z=3}$, but
we do not see the turnover observed by these authors towards the bright end
relative to the $z\sim 3$ LF. 

We also display in Fig.~\ref{lf} the LF derived from the simple models
presented in Sect.~\ref{Nzmodels}, in the redshift intervals 
$6\lesssim z\lesssim 10$ (red lines) and $8\lesssim z\lesssim 10$
(black lines), for the standard Salpeter IMF (thin dashed lines) and the top-heavy
IMF (thin solid lines). 
A better overall fit to data points is obtained with the top-heavy IMF.

Considering only first-priority or first$+$second priority
candidates does not change the results substantially. 
Also, including or removing the brightest EROs
in Abell 1835 does not change these conclusions. 

\subsection{Cosmic Star Formation Rate at $z>6$}

   We use the individual properties computed in Sect.\ \ref{intrin} to derive
the Cosmic SFR value for each redshift bin considered in our previous
analysis. The large correction factors applied to this sample, which
dominate the error bars on LF measurements, make the determination of 
SFR densities challenging. We use different approaches here to derive an
estimate for this important quantity.
On the other hand, since these objects are only photometric candidates, the obtained
values are to be considered as upper limits. However, all values derived here 
neglect possible extinction corrections.

The first estimate is obtained in a very simple way.
We compute the total SFR in a redshift bin by summing all individual contributions to the
SFR within this bin, after correcting each object by its $\eta$ value
and its expected probability of being a false-positive detection (Table ~\ref{spurious}).
We divide this result by the total covolume surveyed 
in this redshift bin accross the ISAAC field area, assuming a blank field
(because magnification/dilution effects are already included in the $\eta$
factor). This corresponds to \\
$2.9\ 10^4\ $Mpc$^3$ for $z\in [6-7.5]$, $2.6\ 10^4\ $Mpc$^3$ for $z\in [7-8.5]$, 
$3.8\ 10^4\ $Mpc$^3$ for $z\in [6-8]$, $3.1\ 10^4\ $Mpc$^3$ for $z\in [8-10]$, for the size of an ISAAC field.
The resulting SFR densities $\rho_\star$ obtained are relatively high, even when the sample is restricted
to first and second-category candidates: $\rho_\star=$ 3.31 10$^{-3}$ M$_{\odot}$ yr$^{-1}$ Mpc$^{-3}$
for $z\in [6-7.5]$, 2.46 10$^{-2}$ M$_{\odot}$ yr$^{-1}$ Mpc$^{-3}$ for $z\in [7-8.5]$,
and 1.20 10$^{-1}$ for $z\in [8-10]$. 

  Another estimate of the cosmic SFR, allowing us to compare results to previous
findings, can be obtained by integrating the Luminosity Function fit found in
Sect. \ref{lf_sec} down to $0.3\ L^{*}_{z=3}$, the same limit used by Bouwens
et al.\ (2004b). In this case, we use the same redshift bins defined in
Sect. \ref{lf_sec}, 
and we obtain a somewhat lower value of the SFR for z in [6-10] if we consider the first and second category candidates.
Considering only the first category candidates, these values are
  lower by a factor $\sim 3$, as can be seen from Table \ref{sfrd}, summarising
  our different cosmic SFR density estimates.

\begin{figure}[ht!]
\psfig{figure=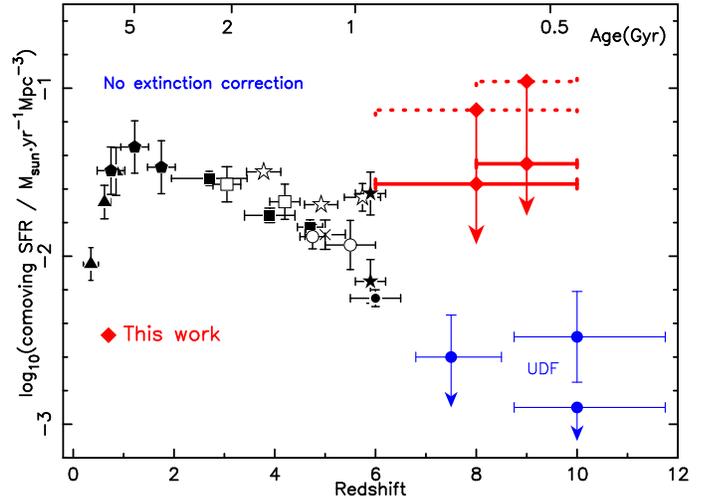,width=9cm,angle=270}
\caption{Evolution of the comoving Star Formation Rate density as a function
of redshift. Different approaches are used to derive an order of
magnitude estimate for this quantity, reported in Table \ref{sfrd}.
Results from other surveys, uncorrected for extinction, 
are compared to our upper limits taken at face value. Data
are compiled from the CFRS (Lilly et al.\ 1995, filled triangles), Connolly et 
al.\ 1997 (filled pentagons), LBG work from Steidel et al.\ 1999 (open
squares), Fontana et al.\ 2003 (open circles), Iwata et al.\ 2003 (cross), 
Bouwens et al.\ 2003a (filled squares), GOODS (Giavalisco et al.\ 2004, open
stars), different ACS estimates from Bouwens et al.\ 2003b (filled stars) and 
Bouwens et al.\ 2005b (filled circle). We also report the value derived by 
Bouwens et al.\ (2004b) and Bouwens et al.\ (2005a) in the Ultra-Deep Field (``UDF'' labels). 
Our results (filled red diamonds) are presented  for both clusters in the redshift 
ranges [$6-10$] and [$8-10$] : these values are obtained by integrating the 
Luminosity Function fit down to $L_{1500}=0.3\ L^{*}_{z=3}$. 
Solid lines refer to the first-category candidates only
($\Delta m_H<0.4$), whereas dotted lines correspond to first and
second-category sources. 
} 

\label{SFR}
\end{figure}

The final results are shown in Fig.\ \ref{SFR}, 
applying the correction for false-positive detections, for comparison 
with the cosmic SFR obtained in other surveys carried out on blank fields,
without applying any extinction correction. Our error bars were computed using
Poisson noise statistics in the number of objects within each redshift bin. 
The numerical values of the cosmic SFR are also summarized in Table \ref{sfrd}.

  When considering only the first-category candidates
(i.e\ with a $1\sigma$ error $\Delta m_H<0.4$), 
our results in the $z \sim 6-10$
domain are compatible with previous findings. 
However, our estimate of the comoving SFR density at $z \sim 8-10$
seems to be larger than all values derived at $z \sim 4-6$, although compatible within
the error-bars.
Taken at face value, our findings seem to be in good agreement with some theoretical
cosmic SFR density models previously published; e.g.\ with the model of Barkana \& Loeb 
(2001, their Fig. 29) for a reionization redshift between 6 and 8,
 recent hydrodynamical models of Nagamine et al.\ (2005),
and with the self-consistent reionization models of Choudhury \& Ferrara (2005).

However, compared to recent studies in the Hubble UDF at similar redshifts
(Bouwens et al. 2004b, 2005a) our SFR density, or upper limits thereof,
is larger by roughly 1 dex.
This difference is related to the bright end of the LF,
i.e. $L_{1500}\gtrsim 0.3 L^*_{z=3}$.  In all cases the sources are photometric candidates therefore 
providing upper limits to the actual UV flux densities.
The
effective fields surveyed are small in all cases, thus leading to a strong field-to-field
variation of $\sim 20-30\%$ in the number of sources. Cluster-to-cluster
variations already discussed in Sect.~\ref{cluster-to-cluster} are clearly
seen in our sample, although lensing and photometric considerations could
account for most of them. 
Recent spectroscopic results obtained by Le F\`evre
et al. (2005) on the VVDS, for an I-band flux-limited sample of galaxies
up to $z\sim5$, indicate that the Universe contains more star-forming
galaxies towards the bright end of the LF than previously reported using
color-color selection techniques, suggesting an active star formation
activity in the redshift domain covered by the present survey.

  There are several ways to reconcile our SFR measurements with
Bouwens et al.'s values, in addition to field-to-field variance. Residual
contamination by false sources combined with lensing effects are able to
affect our results in different ways.

On the one hand, residual spurious sources constitute a potential source of
contamination if the adopted corrections are underestimated. The extremely
blue $H$-$K$ colors obtained for the stacked 
images argue for a significant residual contamination for the
faint third category dropouts, as discussed in detail in
Appendix~\ref{addtests}. For this reason, third category dropouts were not used
to derive LFs and SFR densities. 
The blind corrections applied to the first and
second-category samples to obtain the above results are in good
agreement with the contamination levels derived from the stacked images and
the UV slope.  

On the other hand, {\it observed} number counts could be affected by residual positive magnification
bias, producing a systematic trend compared to blank fields.
This trend is
indeed expected under simple considerations, as shown in Sect.~\ref{Nzmodels}
and Fig.~\ref{ncounts}. A positive magnification bias is expected when the 
slope of number counts (with the approximation $\alpha = -d(log\ n)/d(log\ L)$)
is $\alpha \lesssim -1$ in the magnitude and redshift domains
considered, i.e. $N_{lensed}(\ >L)= N(\ >L) \times \mu^{\alpha-1}$ (see
e.g. Broadhurst et al.\ 1995). This is indeed the case within our $H$ band
limited sample, according to the simple assumptions given in
Sect.~\ref{Nzmodels}, and the shape of the observed LF. 
We have directly accounted for magnification biases using the lensing 
models. However, an additional magnification bias
could remain in our sample due to a systematic trend: up to a given limiting
magnitude, we tend to detect the sources with the largest magnification
factors, instead of (or in addition to) the intrinsically brighter sources, as
it happens in blank fields. This systematic trend could slightly modify the
slope of the LF derived for high-$z$ sources. It is
difficult to correct for without a complete mock simulation, assuming a shape for
the LF of background sources, and then statistically correcting for this
additional bias. Given the error bars obtained for the LF in
Sect.~\ref{lf_sec}, this residual magnification bias should be a second order
correction for high-$z$ sources. 
However, the same trend could exist for very faint intermediate-redshift
interlopers, such as the extremely-faint source A1835\#35 (Richard et al.\
2003). This effect is presently uncorrected in our sample for obvious reasons,
but it could be responsible for part of the discrepancy, 
because intermediate-redshift interlopers should mainly affect the bright end
of the LF.

   The standard calibration used to convert the
$L_{1500}$ luminosity into SFR, which assumes a standard Salpeter IMF under
equilibrium conditions (i.e.\ constant SF over timescales $\gtrsim 10^{8-9}$ yr), 
is not necessarily appropriate for objects at such early epochs. 

\begin{table}[ht!]
\caption{\label{sfrd}Summary of results obtained for the comoving Star Formation Rate density (in M$_{\odot}$ yr$^{-1}$ Mpc$^{-3}$) 
by integrating the Luminosity Function down to 0.3 $L^{*}_{z=3}$ in different redshift ranges, also presented in Fig.~\ref{SFR}, 
correcting for lensing, incompleteness effects and expected fraction of spurious sources.
}
\begin{tabular}{ll}
Assumption & SFR density\\
\hline
$[6-10]$ First-category & 2.7\ 10$^{-2}$ \\
$[6-10]$ First and second-category & 7.4\ 10$^{-2}$\\
$[8-10]$ First-category & 3.5\ 10$^{-2}$\\
$[8-10]$ First and second-category  & 1.1\ 10$^{-1}$\\
\end{tabular}
\end{table}

\subsection{Spectroscopic follow-up}
\label{spectro}

We have started the spectroscopic follow-up of our sample of high-$z$
candidates with ISAAC/VLT. This survey is presently ongoing, and the final conclusions
will be presented in a further paper. Results on these observing runs have
been (partly) published in Richard et al.\ (2003) and Pell\'o et al.\ (2004a),
as well as a first preliminary summary in Pell\'o et al. (2004b).

   To search for faint emission lines, we have systematically explored
the 0.9-1.40 $\mu$m domain ($SZ$ and $J$ bands of ISAAC), where
\lya\ should be located for objects within the $7 < z < 10.5$ redshift
interval. We intended to detect emission lines with intensities ranging between
$10^{-17}$ and a few $10^{-18}$ \erg,
with a spectral resolution for the sky lines of
$R=3100$ corresponding to the instrumental 1\arcsec\ slit width.
The fraction of spectral band lost due to strong OH sky emission
lines is of the order of 30\%. Slit configurations were set to optimize 
the acquisition of a maximum number of targets per night, with 
priority given to first-category dropouts. Secondary targets were
only observed when aligned with first priority targets. 

Up to now, our spectroscopic survey with ISAAC has targeted 2 priority
candidates in AC114, and 7 in Abell 1835 (4 ``first priority'' targets and 3
secondary ones). From this sample of 9 targets, 2/3 of the objects
observed display emission lines. 5 sources have clear
emission lines detected, and another one is still to be
confirmed. The distribution between first and second priority targets
for spectroscopy does not fully coincide with the present classification in
Tables~\ref{table_a1835} and ~\ref{table_ac114}, because it was based on an
earlier version of the image reduction and analysis. For instance, $z$ and
$SZ$ images were not available for Abell 1835 at this epoch. 
In summary, from 6 first priority spectroscopic targets observed in the two
clusters, we have clearly 
confirmed one candidate (A1835\#8, originally named A1835-1916, see
Sect.~\ref{result_ind}), which is found to be a puzzling 
source; two $z\ge7$ candidates show emission lines still to be reconfirmed; 
one candidate is found to be a low-$z$ contamination, and two of them do not
show emission lines. From 
the 2/3 secondary targets displaying emission lines, one is a possible $z\ge7$
source to be confirmed, whereas the other one is a faint low-$z$
galaxy (A1835\#35, $z=1.67$). 
According to these preliminary numbers, the efficiency of our survey
could range between $\sim 30$ and 50\%, with interesting low-$z$
by-products such as A1835\#35 (Richard et al.\ 2003). A large majority of our
high-$z$ candidates still need to be confirmed, either by a redetection of a faint
emission line, or by the non-detection of other lines expected at low-$z$.


\section{Summary and conclusions}
\label{conclusions}

We have obtained deep $JHK$ near-IR photometry of two well-known lensing clusters, A1835
and AC114, plus $z$ and $SZ$ imaging of A1835 with FORS and ISAAC at the VLT. 
Our photometric depth reached $SZ\sim 25.6$, $J\sim 24.4$, $H\sim 23.5$ and
$K_s\sim 23.3$ (Vega system), in addition to a minimum magnification factor of
1 magnitude over half of the ISAAC field of view.
These images, combined with existing data in various optical bands including HST images,
have been used to select galaxy candidates at very high redshift ($z \sim 6-10$).
The candidates have been selected with the dropout technique and two-color selection
criteria appropriate for high-$z$ galaxies.

 From our $H$ band selected sample we have identified 18 (8) ``first and second-category''
optical dropouts in A1835 (AC114) up to $H_{\rm Vega} \sim 23.9$ (AB $\sim 25.3$,
uncorrected for lensing). Second category is defined here as objects
detected in $\ge 2$ near-IR bands, the best-detected sources being defined as first priority.
Among them, 8(5)
exhibit homogeneous SEDs compatible with star-forming galaxies at $z\gtrsim
7$, and 5(1) are more likely intermediate-redshift EROs.
In both fields we have also identified a few additional dropouts detected only in the $H$ band
(``third category'' objects), which satisfy our photometric selection
criteria.
We have estimated the fraction of spurious sources expected in the
different filter combinations, and corrected all the relevant derived
quantities for this effect.

Typically our candidates are magnified by a factor of 1.5 ($\sim$
0.44 mags) to 10 (2.5 mags), with average (median) values of the order
of $6.5-7.9$ ($2.3-3.5$) for the two clusters.
All high-$z$ candidates turn out to be fainter than $H_{\rm Vega} \sim$ 23,
with typical effective (i.e.\ lensing corrected) magnitudes $H_{\rm Vega} \sim  24-25$
($H_{\rm AB} \sim$ $25.4-26.4$) and fainter in some cases.
Assuming standard SFR(UV) conversion factors, which may however be questionable for
galaxies of such presumably young age, the SFR is found to be between
few units and $\sim$ 20 \msun\ yr$^{-1}$.
Their UV restframe spectrum, measured by the $H-K$ color, seems to be very blue
-- a trend also reported for other high-$z$ galaxy samples
(e.g.\ Papovich et al.\ 2004, Bouwens et al.\ 2004b).

Taking into account the gravitational lensing effects, sample
incompleteness and expected spurious detections,
a first attempt was made to constrain the density
of star-forming galaxies present at $z \sim 6-10$ based on
lensing data. Integrated quantities, such as LFs and SFR densities, are affected by 
non-negligible uncertainties due to the large correction factors applied to this
sample.
The LF measured for LBGs at $z\simeq 3$ seems to be slightly
higher (by $\sim 0.5$ dex) but still consistent with the LF derived for our
sample. The turnover observed by Bouwens et al. (2005) towards the bright end
relative to the $z\sim 3$ LF is not observed in this sample.

We have also estimated an upper limit for the cosmic SFR density from these data. 
Our values in the $z \sim 6-10$ domain are higher than the estimates obtained
in the NICMOS UDF, even
when the most conservative assumptions and corrections are applied to the
data. This difference is related to the bright end of the UV LF for our candidates,
i.e. $L_{1500}\gtrsim 0.3 L^*_{z=3}$.
This systematic trend with respect to blank fields could be due to
to field-to-field variance, a positive
magnification bias from intermediate-redshift EROs,
and/or residual contamination by spurious sources.
Given the error bars, residual magnification bias should be negligible 
for high-$z$ sources, but a population of faint intermediate-redshift interlopers
affected by positive magnification bias cannot be excluded.
At least one of such intermediate sources was spectroscopically
identified in the field of A1835 (Richard et al. 2003). 

According to our simulations, in agreement with the first photometric results
presented in this paper, the use of lensing clusters as gravitational
telescopes seems to significantly improve the survey efficiency of $z\gtrsim
6$ galaxies compared to blank fields.

Given the uncertainties involved in the candidate-selection process, and the
faint fluxes observed for our photometric candidates, the present results are  
to be considered as a first attempt to constraint the population 
of $6 \lesssim z \lesssim 10$ star-forming galaxies using lensing clusters.
The present results and conclusions have to be confirmed and improved.
Spectroscopic follow-ups are underway to determine the efficiency of our selection
technique, and the contamination level by intermediate-redshift interlopers.
Additional deep photometry in various bands are being secured with HST, IRAC/Spitzer,
and from the ground to improve the SEDs characterization of the high-$z$
candidates.
Increasing the number of lensing fields with ultra-deep
near-IR photometry is essential to obtain more accurate constraints on the abundance
and physical properties of $z\gtrsim 7$ starburst galaxies.


\acknowledgements

 We are grateful to T. Contini, G.P. Smith, E. Egami, F. Lamareille, A.
 Hempel, C. Donzelli for useful comments and discussions, and to the
 anonymous referee for his/her helpful and constructive comments.
 We thank the ESO Director General for a
 generous allocation of Director's Discretionary Time for ISAAC
 spectroscopy (DDT 271.A-5013 and 272.A-5049).
 JR and JPK are grateful to Caltech, and in particular to R.S. Ellis, for
 their support.
 This paper is also based on
 observations collected at the European Space Observatory, Chile
 (069.A-0508,070.A-0355,073.A-0471), the NASA/ESA Hubble Space
 Telescope operated by the Association of Universities for Research in
 Astronomy, Inc., and the Canada-France-Hawaii Telescope operated by
 the National Research Council of Canada, the French Centre National
 de la Recherche Scientifique (CNRS) and the University of Hawaii.
 Part of this work was supported by the
 French {\it Centre National de la Recherche Scientifique},
 the French {\it Programme National de
 Cosmologie} (PNC) and {\it Programme National de Galaxies} (PNG),
 by the {\em International Space Science Institute} (ISSI), as well
as by the Swiss National Science Foundation.


\Online
\appendix

\section{Further improvements in the data reduction}
\label{improvements}
The following steps were introduced in addition to the standard scheme
to improve the data reduction:

\begin{itemize}
\item{Object masks created by XDIMSUM were not well-suited for our fields
centered on lensing clusters, because they did not correctly take 
the bright extended haloes in the cluster core into account. We improved the sky-subtraction
in these regions by applying a simple threshold above the sky background in
our images to create the object masks. This procedure greatly reduced the 
contamination close to the bright objects.}

\item{About 35 \% of the images taken with the Hawaii-Rockwell array suffered
from bias residuals, that appear more pronounced at the bottom
and middle region of the detector. Before combining the individual
sky-subtracted frames, we removed these residuals by subtracting from each line its
median, with a rejection of pixels flagged in the object mask.
}

\item{In the case of AC114 cluster, about 20\% of the individual $H$ band frames
presented strong low-frequency background variations, due to imperfect
sky-subtraction and possible contamination due to the proximity of the moon. We
corrected part of these residuals using a bidimensional fit of the large-scale
background for each
frame, after rejection of all pixels flagged in our object mask.
}

\item{ Before combining the frames into a final stack, we applied weight values accounting
for slight variations in quality during observations, in such a way that the
best-quality images will have the highest weight.
Weighting was optimized in order to improve the detectability of faint
point-like sources ; we computed individual weights using the following
relation :
$$weight \propto (ZP\times var\times {s}^2)^{-1}$$
where we computed the individual zero-point $ZP$ and seeing $s$ from the
magnitudes and FWHM of the 5 brightest unsaturated stars located in our
field. The local sky pixel-to-pixel variance $var$ was derived through
background statistics inside a small region free of objects.
}

\item{To check for the final accuracy of our absolute photometric
calibration, we compared the theoretical and observed colors of
several cluster elliptical galaxies for which we had
available spectroscopy, after having reduced the images as described
below, and seeing-matched them to the worst value. The empirical SED template
compiled by Coleman, Wu and Weedman (1980) was used to derive the expected colors for
elliptical galaxies at the cluster redshifts. In the single case of the $J$ band image 
in Abell 1835, we corrected a 0.1 mag offset in the zero-point. We also checked
that the optical to near-IR colors of the brightest elliptical
galaxies were consistent with the theoretical expectations.
We found that our final absolute photometry is accurate to about 0.05
mag throughout the wavelength domain.}

\end{itemize}

\section{Sample completeness and false-positive detections}
\label{sampcomp}
In addition to the purely photometric completeness effect in the detection of near-infrared 
sources, another incompleteness factor comes directly from our non-detection
criterion in the optical bands. We statistically expect, by integrating a
normal distribution function above this level, to measure a 1$\sigma$ flux for  
16 \% of all dropouts in each optical band, which we did not include in our current 
selection technique. Because there are 5 such non-detection filters for each cluster, 
this gives a statistical completeness factor of $C_{sample}=(1-0.16)^5\sim 0.42$, which is unrelated to the
observed NIR magnitude of the objects. We applied this additionnal
correction to the final sample of optical dropouts. 

Our detection scheme was optimized to identify faint sources which are only
detected on the near-IR bands, i.e. a subsample of the images.
To evaluate the fraction of spurious detections expected in our
photometric catalogs, we constructed a special  \textit{noise image}
for each cluster and each near-infrared band,
where all astronomical sources were removed while keeping the same noise
properties: we subtracted by pairs sequential images obtained
with similar seeing conditions, and then coadded them using the same procedure
described above. The result is an image with
the same noise properties compared to the final stacks, affected only by small
residuals at the location of the brightest sources. We used $SExtractor$ to
detect objects inside these noise images,
with the ``double-image'' mode and the same detection parameters
as for astronomical sources. After masking the regions around the brightest
objects and galaxy haloes to prevent any detection of source residuals (as for
the astronomical images), the number of objects detected in these noise images
was compared with the number of high-$z$ candidates \textit{blindly}
selected as optical-dropouts within the same region on the
astronomical images, for different ranges of magnitude and categories
of optical dropouts as defined in Sect. \ref{check}. $H$-band magnitude bins 
have been defined to include a similar number of spurious sources in each bin. 
A source was conservatively included as positive (spurious) detection in this
table when Sextractor in the double-image mode was able to measure a magnitude
for this source above the same detection level as defined for our candidates in Sect.\ \ref{check}.
Results are summarized in
Table~\ref{spurious} for each cluster. 

The fraction of false objects obtained with this
procedure is overestimated compared to real catalogs, since no attempt was
made to manually correct for obvious spurious detections, whereas all the
dropout candidates included in the final catalogs were visually inspected by
at least two different persons.
Their false-detection probability is therefore reduced compared to Table\ \ref{spurious}.

\begin{figure*}
\centerline{\mbox{\psfig{figure=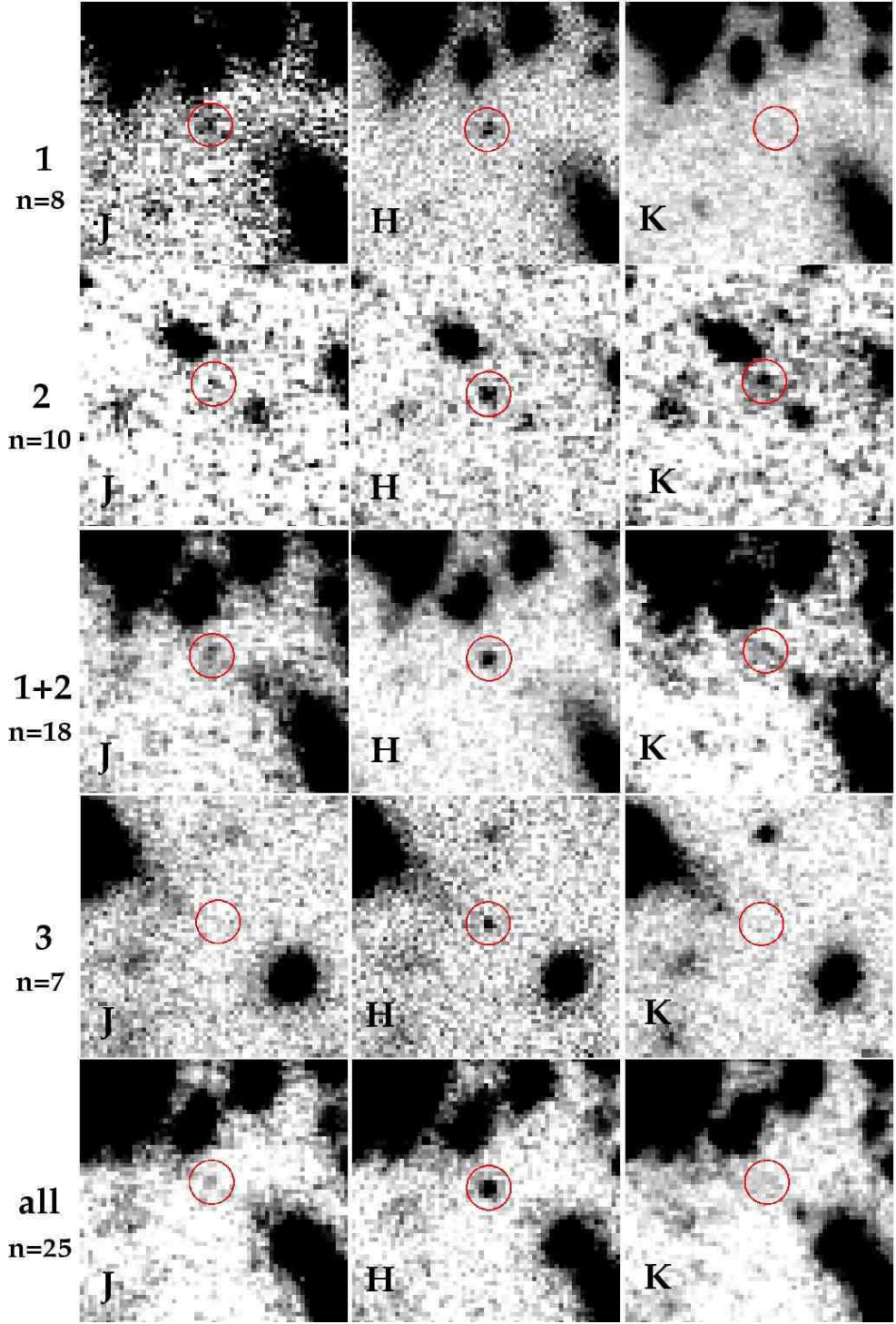,width=13cm,angle=0}}}
\caption{\label{stacked}
From left to right, stacked $J$, $H$ and $Ks$ images
for first, second, third, first $+$ second-category and all $z \ge 6$
candidate sources in Abell 1835 and AC114, excluding EROs. Images display a
10\arcsec $\times$ 10\arcsec region 
around the composite source. 
}
\end{figure*}

\begin{table*}[ht]
\caption{\label{spurious}Expected percentage of false-positive detections in
our samples of third-category (only detected
in $H$-band) and  second/first-category (detected in $H$- and at least another
near-IR band) dropouts for each cluster,
as a function of the detection filters and the $H$-band magnitude. The typical errors
in these values are $\sim 15$ and $\sim 30$ for Abell 1835 and AC114, respectively.
}
\begin{tabular}{ccrrrrrrrr}\hline
 & 3rd Cat.  &  \multicolumn{7}{c}{First / Second Category dropouts}\\
$H$ range &  $H$ & $SZ$+$H$ & $J$+$H$ & $H$+$K$ & $SZ$+$J$+$H$ & $J$+$H$+$K$ & $SZ$+$H$+$K$ & $SZ$+$J$+$H$+$K$ \\
$[$mags$]$ & \multicolumn{8}{c}{\% of spurious detections in Abell 1835 (AC114)}\\
\hline
22.75 - 23.00 &   0 (  0)&  0&  0 (  0)&  0 (  0)&  0&  0 (  0)&  0 &  0 \\
23.00 - 23.30 &  35 ( 57)& 27& 12 ( 25)& 25 ( 33)& 12& 12 ( 33)& 12 & 12 \\
23.30 - 23.75 &  98 (100)& 66& 65 ( 88)& 61 ( 61)& 28& 27 ( 38)& 30 & 12 \\
23.75 - 24.00 & 100 (100)& 56& 96 (100)& 73 (100)& 27& 28 (100)& 40 & 11 \\
\hline
\end{tabular}
\end{table*}


\section{\label{addtests}Additional tests on the reliability of optical dropouts}

In addition to the simulations presented in 
Sect.~\ref{calccomp} on
the completeness and spurious detections expected on the reference $H$ band
image and on the other near-IR images, 
we have performed additional tests on the reliability of optical-dropouts
identified on the near-IR images. 

\subsection{Pseudo-$\chi^2$ images}

 Detection pseudo-$\chi^2$ images were created from 
individual $J$, $H$ and $Ks$ band images in the following way : each image was normalized by the
noise 1$\sigma$ image, weighted by the square root of the corresponding
exposure-time maps, and then all images were registered and averaged
together. We applied the same $SExtractor$ detection parameters to these
images, and compared the detection results. 
This procedure has some obvious limitations, because the final stacks
are not independent from the original $H$-band detection images. 
For Abell 1835, $\sim$ 89\%  of the first and second-category sources presented in
Table~\ref{table_a1835} are re-detected in the pseudo-$\chi^2$ image, with positions
less than 1 pixel off with respect to original centroids, except \#13 and
\#15. For AC114, 75\% of the first and second-category sources are re-detected in the
pseudo-$\chi^2$ image (all sources except \#2 and \#5). 
By definition, third-category sources are clearly detected only in the $H$ band
image. However, sources \#21 and \#35 in Abell 1835 are also re-detected in the
corresponding pseudo-$\chi^2$ images.
On the other hand, 60(86)\% of the first and second-category high-z sources in 
Abell 1835 (AC114) (Tables~\ref{table_a1835} and \ref{table_ac114}) are
re-detected in at least another near-IR filter with a mean S/N$\ge2$ within a
1.5\arcsec-diameter aperture. 
Thus, a majority of our optical dropouts, at least the first and second-category ones,
have a significant signal detected in several near-IR bands. 

\subsection{Transient objects}

We have estimated the contamination level expected in our images
due to known transients, such as TNOs or supernovae, taking into
account the combination schemes used to build the final stack, which 
typically reject $10-20$\% of the brightest pixels at this stage. The
typical motions observed and expected for TNOs range
between $\sim (1-10)\arcsec /h$ (Trujillo et al.\ 2001). 
In the detection $H$ band
a TNO moves at least between 3 and 6 pixels in 10\% of
the exposure time, depending on overhead details and the distribution of
exposures during the observing period, thus making a TNO 
selection highly unlikely. 
On the contrary, supernovae events within the cluster, and
particularly a type I supernova in the halo of a cluster member, cannot be
removed with the usual scheme. We do not expect more than 1-2 events per
cluster in these deep observations, according to previous findings (e.g.,
Gal-Yam et al.\ 2002). Very rare and unique events, such as 
lensed supernovae (e.g.\ Marri \& Ferrara 1998, Marri et al.\ 2000),  
or a tidal disruption of a star by a BH (cf.\ Stern et
al.\ 2004) cannot be excluded either, but again such events are not expected
to dominate the sample.  

\subsection{Photometric stability}

The Abell 1835 data in the $SZ$ band were obtained at two different epochs: 19 April
2004 ($\sim 4h$ exposure, 77 images) and 15 May 2004 ($\sim 2h$ exposure, 45 images). 
Both series of images have identical seeing and photometric zeropoints. We used
these data to check the compatibility of our magnitude measurements at
different epochs, and particularly to set constraints on the possible
variability of faint optical dropouts in these bands. Among the 9
sources detected in the $SZ$ band, 7 of them show magnitudes and
1$\sigma$ error-bars (with $\sigma_{SZ} \sim $ 0.1 to 0.5 mag)
fully consistent between the two epochs and with the final stack.
A bright and clearly variable source (A1835\#4) and a particular case (A1835\#8), 
both discussed in Sect.~\ref{result_ind}, are clear exceptions.

\subsection{Stacked images}

 We have generated stacked $J$, $H$ and $Ks$ images for all first to third
priority candidates, excluding EROs. A 10\arcsec $\times$ 10\arcsec\ region has
been selected around the $H$-band centroid for each $z \ge 6$ candidate in
Tables~\ref{table_a1835} and \ref{table_ac114}, for all $JHKs$ images.
Multiplicative corrections have been applied to properly combine images
coming from the two clusters with different photometric zero-points. After
background subtraction, images have been averaged using IRAF routines
and different pixel rejection schemes in order to obtain a ``clean'' region
around the stacked source, although the final photometry does not
strongly depend on the combination procedure. Representative results are
displayed in Fig.~\ref{stacked}, for first to third-category sources combined
in different ways.
We used Sextractor to measure the corresponding fluxes, best magnitudes and
errors of these composite sources in the different bands 
and samples. Colors are obtained using the same apertures as for
individual sources.
Results are summarized in Table~\ref{composite}. The S/N of the 
composite sources in all filters increases with the number of stacked images, 
as expected if a significant signal was present in a majority of sub-images
and bands. 
Although these optical dropouts are likely to constitute a non-homogeneous
sample, all the stacked series display a break between $J$ and $H$ (typically
$J-H\sim1.5$ to 2.50 in the Vega system), and relatively blue $H-Ks$ colors, thus
a photometric SED corresponding to a dominant population of $z\ge8$ sources
(see Fig.\ ~\ref{JHK_colors}), or a noticeable contamination by spurious sources.
The flux detected in the $J$ band is clearly
higher for the first and first $+$ second-category dropouts, i.e. the
brightest sources in the $H$ band, as expected if a fraction of these sources
are at $z\le8$. This result is the same when combining all dropouts at $z\le8$
(6 sources, all of them first or second-category), but the final S/N in this band is
higher when blindly combining all first $+$ second-category dropouts. 
The same comments stand for the $Ks$ band, for which the best S/N is achieved
for the first and second-category dropouts.
The profiles obtained for the composite sources in the
different filters are all compatible with the seeing values
in the $H$ band ($\sim0.5$\arcsec), and slightly broader in $J$ and $Ks$
($\sim0.55$\arcsec and 0.45\arcsec respectively).

False or transient sources stacked in these images
will tend to enhance the trend towards artificially 
``blue'' $H-Ks$ and ``red'' $J-H$ colors, thus providing an independent method
to estimate the contamination levels reported in Table\ \ref{spurious}, and
applied throughout the paper. 
For $z \ge 6$ candidates stacked here, $H-K$ colors provide a rough
estimate of the restframe UV slope $\beta$. For young starbursts, $\beta$ is
found to be $\sim -2.5$ in the local universe (Heckman et al.\ 1998),
$\beta\sim -3$ for the $z\sim7-8$ candidates reported by Bouwens et al.\
(2004b), and it could reach values up to $\beta \sim -3$ for young
starbursts at very low metallicity (see Fig. 1 from Schaerer \&
Pell\'o \ 2005). As discussed in Sect.~\ref{intrin}, our high-$z$ candidates
tend to be extremely blue in $H-K$, but none of them could be excluded from  
the sample to a $3\sigma$ level on the basis of a clear $\beta \le -3.5$,
although two first category dropouts are close to this limit (A1835\#7 and
AC114\#2, see also Sect.~\ref{intrin}). 
Note that the contamination here refers to the {\it integrated light} in the $H$
band, whereas Table\ \ref{spurious} reports percentages in the number of
sources. The results are the following:

\begin{itemize}

\item The extremely blue colors of the third-category sample argue for a
significant contamination, between 60 and 65\% of the total light for $\beta$
ranging between $-3.5$ and $-2.5$ respectively. The faint population of
third category dropouts could be dominated by spurious sources, as already
expected from Table~\ref{spurious}. 

\item The second-category sample is compatible with a very low
contamination level by spurious sources. It is negligible for $\beta \sim
-3.5$, and it could reach up to 10\% for $\beta\sim-2.5$. 

\item The combined first-category image is clearly dominated by the 
brightest sources, and among them the two dropouts reach close to $\beta
\le -3.5$ at 2-3$\sigma$ level (Sect.~\ref{intrin}). This gives a
contamination level ranging between 62 and 70\% for the whole sample, whereas
it is $\sim$10\% when these two sources are removed. Note that these sources
are detected in several bands. 

\item The combined first $+$ second category (with all sources included) is
qualitatively in good agreement with expectations in Table~\ref{spurious} 
(i.e. 33 to 50\% maximum contamination on the whole sample). 

\end{itemize}

   Overall, the contamination levels based on UV slope
considerations are optimistic in general as compared to the blind corrections
applied according to Table~\ref{spurious}. 

\subsection{Spatial distribution of optical dropouts}

We carried out a Kolmogorov-Smirnov (hereafter K-S) monodimensionnal
test (Peacock, 1983) on the observed radial distance of the sources with
respect to the cluster central galaxy, 
because high redshift images are expected to be preferentially found around
the critical lines for positive magnification bias (Broadhurst et al. 1995), 
and thus inconsistent with a homogeneous distribution. Such positive bias is 
expected to be the case when probing the steep part of the luminosity function 
(see Sect. \ref{Nzmodels} and Fig.~\ref{ncounts}).
When comparing our candidates with a
uniform distribution (after masking the usual regions of the image), we obtain
an average to high resulting probability ($\sim40$\% for Abell 1835, $\sim
2$\% for AC114). 
Thus, the K-S test indicates that the spatial distribution of our candidates
has a low probability to be drawn from a homogeneous distribution.
However, applying a similar K-S test for the spurious 
sources used in Sect.~\ref{calccomp} gives us much lower probabilities 
($<0.01$\%) for the false-positive detections to be drawn from a homogeneous
distribution. Such a high clustering level is expected for spurious
detections, usually concentrated in specific regions of the image having a
high noise level. 
When comparing the distribution of candidates and false-positive detections in the
same K-S test, we find that these two samples are likely to have a different distribution
(with probability $<0.01$\% to be drawn from the same parent distribution).  

\begin{table*}[ht]
\caption{\label{composite}
Photometric magnitudes and colors derived for the composite $z \ge 6$ source
candidates in Abell 1835 and AC114 (see text for details). 
Upper limits correspond to 3 $\sigma$.
}
\begin{tabular}{lccccc}\hline
Combination & $J$ & $H$ & $Ks$ & $J-H$ & $H-Ks$ \\\hline
1st category dropouts & 25.28 $\pm$ 0.20 & 23.68 $\pm$ 0.09 & $\ge$24.7 & 1.71
& $\le -1.0$ \\
2nd category dropouts & 25.69 $\pm$ 0.23 & 24.02 $\pm$ 0.08 & 23.80 $\pm$ 0.13 & 2.20
& 0.48 \\
3rd category dropouts & $\ge$25.6 & 23.79 $\pm$  0.10  & $\ge$24.7 & $\ge$
1.82 & $\le$ $-0.92$ \\
1st $+$ 2nd category dropouts & 25.38 $\pm$ 0.17 & 23.81 $\pm$ 0.08 & 24.22 $\pm$
0.15& 1.53 & $-0.32$ \\
1st $+$ 2nd $+$ 3rd category dropouts & 26.18 $\pm$ 0.31 & 23.84 $\pm$ 0.07 &
24.76 $\pm$ 0.24 & 2.45 & $-0.86$ \\
\hline
\end{tabular}
\end{table*}

\clearpage
\onecolumn
\begin{figure}[ht!]
\psfig{figure=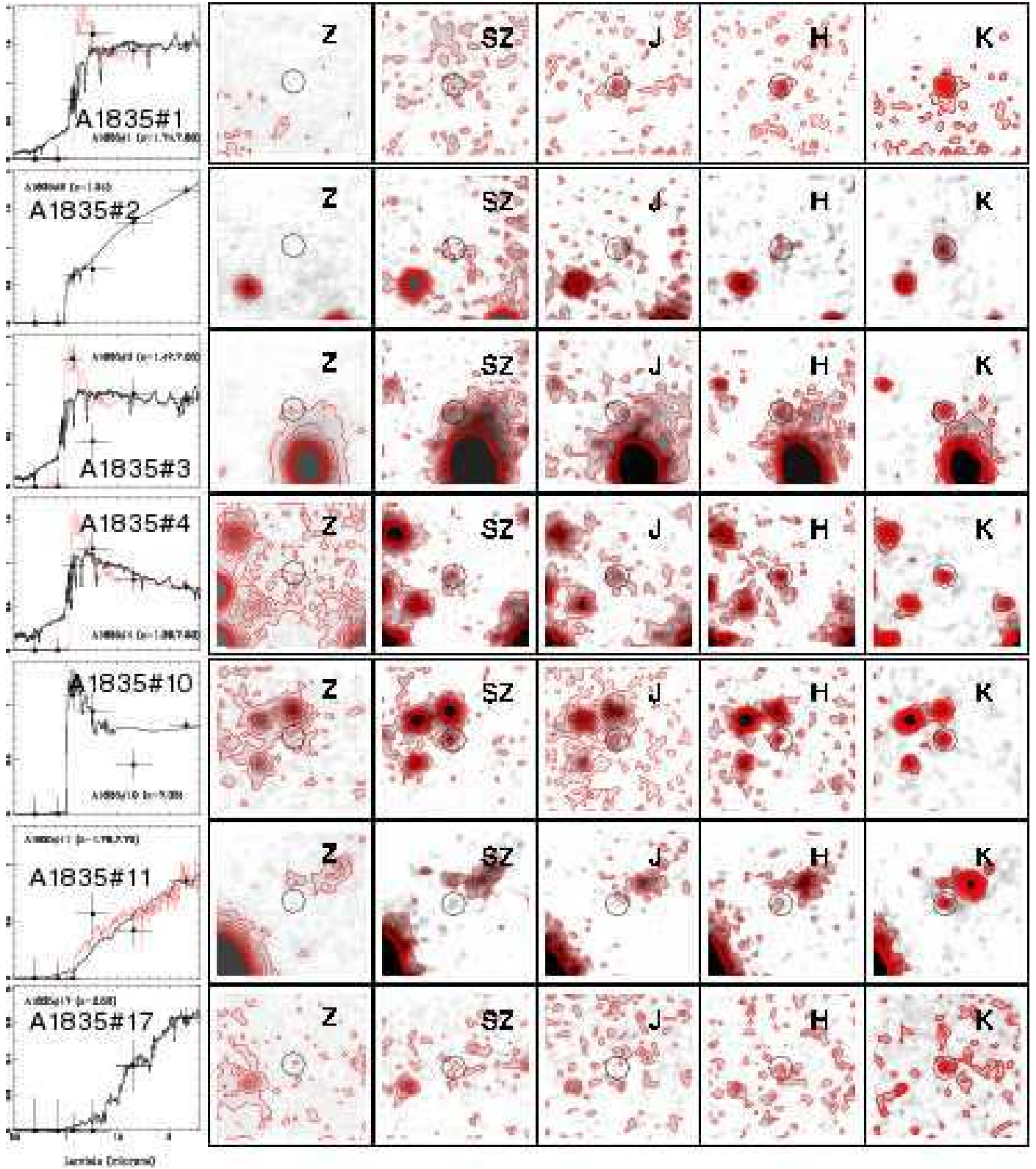,width=18.5cm,angle=0}
\caption{\label{trombino_a1835} (2 pages) Close-up of the best candidates in
Abell 1835, showing the object and their 
surrounding 10 $\times$ 10 arcsecs field. 
Objects satisfying the ERO criterion ($R-Ks>5.6$) are presented above, and 
other first or well-detected second-category candidates are given
in the next page. The
FORS-$Z$ band (non-detection criterion), and the ISAAC-NIR bands are displayed in linear
scale, from $-3\sigma$ to $6\sigma$ levels. Contours are for background level
$+$ 1, 2, 3, ... $\sigma$ respectively. 
On the left is the SED in the $RzSZJHK$ bands and the best photometric redshift solutions obtained. When two possible
solutions coexist, the higher redshift fit is displayed with a red dotted line. Fluxes values are given in $f_\lambda$, with units multiple of 10$^{-19}$ ergs\ s$^{-1}$\ cm$^{-2}$.
}
\end{figure}
\clearpage
\begin{figure}[ht!]
\psfig{figure=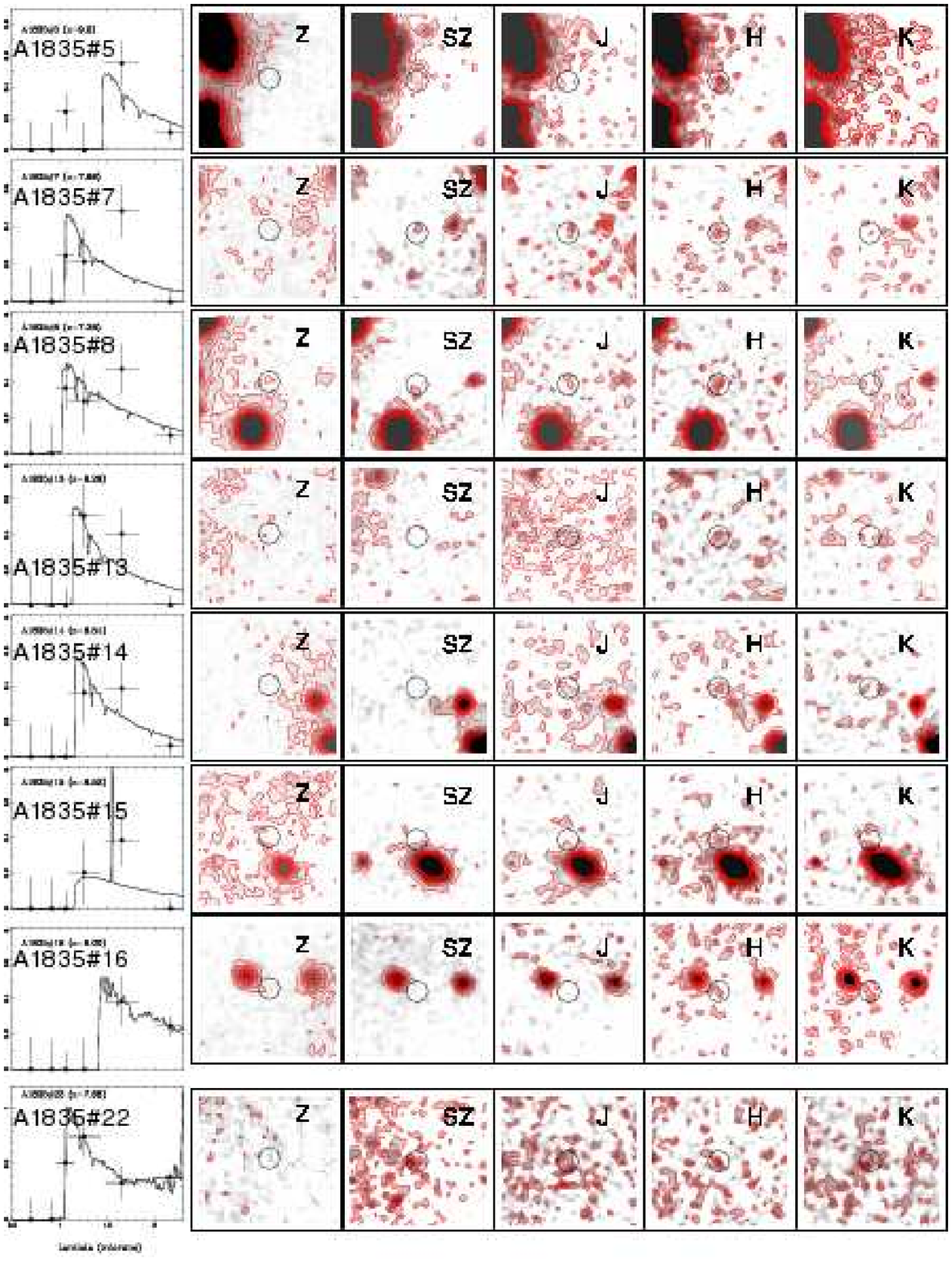,width=18.5cm,angle=0}
\end{figure}
\clearpage
\begin{figure}[ht!]
\psfig{figure=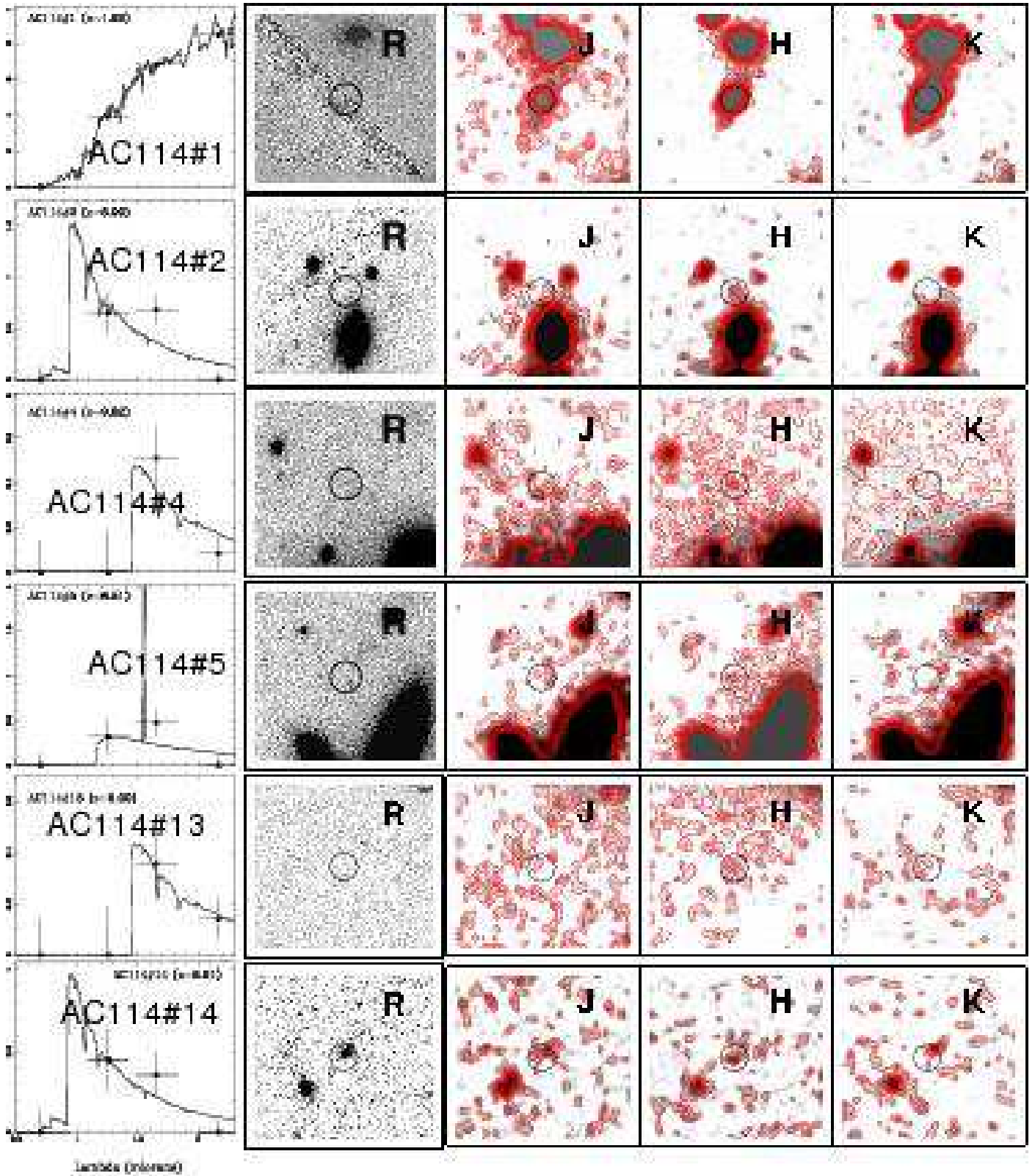,width=18.5cm,angle=0}
\caption{\label{trombino_ac114} Same figure as \ref{trombino_a1835}, 
for best-detected first and second-category candidates in AC114.
Close-ups correspond to the HST-$R$ and ISAAC-NIR bands.
}
\end{figure}
\clearpage
\onecolumn
\landscape
\begin{table}[ht]
{\scriptsize
\begin{tabular}{lllcccccccccccccl}
\hline
ID & RA & DEC & $SZ$ & $J$ & $H$ & $K$ & $\phi\ z$ & $z$ range & $\tilde{z}$ & $\mu_6$ & $\mu_{10}$ & $\tilde{\mu}$ & $L_{1500}$ & $SFR$ & notes\\
 &(14:)&(02:)& & & & & & & & & & & $10^{40} erg\ s^{-1}\ $\AA$^{-1}$ & $M_{\odot}\ yr^{-1}$ & \\
\hline\hline\\
\multicolumn{4}{l}{First-category dropouts}\\
\hline\\
\#1& 0:58.278&50:26.65& $24.56\pm 0.18$ & $23.14\pm 0.11$ & $22.22\pm 0.10$ & $21.10\pm 0.04$ & A & [6.3-8.1] & 7.80 & 1.27&1.27&   1.28 & $79.5$ & $83.5$ & Ex. ERO\\
 &  & & &  &  &  &  & [1.3-1.9] & 1.74 & - & - & - & - & - &   \\
\#2& 0:57.538&52:49.85& $24.80\pm 0.22$ & $24.05\pm 0.27$ & $22.29\pm 0.11$ & $20.95\pm 0.03$ & A & [1.18-1.64] & 1.34 & - & - & - & - & - & Ex. ERO (1)\\
\#3& 1:01.484&51:03.63& $24.03\pm 0.11$ & $24.54\pm 0.42$ & $22.69\pm 0.16$ & $21.71\pm 0.07$ & A & [6.7-7.3] & 7.05 & 1.64&1.66&   1.65 & $12.7$ & $13.3$ & Ex. ERO\\
& & & &  &  &  &  & [1.1-1.7] & 1.47 & - & - & - & - & - &   \\
\#4& 1:01.733&51:05.26& $24.31\pm 0.14$ & $23.50\pm 0.16$ & $22.82\pm 0.18$ & $21.90\pm 0.08$ & A & [6.2-7.8] & 7.53 & 1.67&1.7&   1.69 & $39.6$ & $41.6$ & Ex. ERO\\
& & & &  &  &  &  & [1.2-1.8] & 1.57 & - & - & - & - & - &   \\
\#5& 1:07.034&51:35.71& $25.82\pm 0.52$ & $ > 25.60$ & $23.24\pm 0.28$ & $23.91\pm 0.55$ & B & [8.0-10.0] & (9.00)  & 1.92&1.96&   1.96 & $28.6$ & $30.1$ & \\
\#7& 1:05.067&50:57.52& $25.81\pm 0.51$ & $25.34\pm 0.89$ & $23.39\pm 0.32$ & $ > 24.70$ & B & [6.4-8.0] & 7.66 & 1.53&1.55&   1.54 & $10.1$ & $10.6$ & \\
\#8& 1:00.058&52:44.08& $25.36\pm 0.34$ & $24.99\pm 0.64$ & $23.40\pm 0.32$ & $24.00\pm 0.60$ & B & [6.1-8.0] & 7.38 & 110.81&86.53& $>25$ & $< 0.6$ & $< 0.7$ & Ex. (2)\\
\#10& 0:59.890&50:57.59& $24.18\pm 0.12$ & $23.74\pm 0.20$ & $23.45\pm 0.33$ & $21.77\pm 0.07$ & A & [6.8-7.7] & 7.33 & 1.49&1.51&   1.50 & $33.1$ & $34.8$ & Ex. ERO\\
& & & &  &  &  &  & [1.1-1.7] & 1.47 & - & - & - & - & - &   \\
\#11& 1:06.182&50:27.74& $ > 26.90$ & $24.29\pm 0.33$ & $23.54\pm 0.36$ & $21.72\pm 0.07$ & C & [7.0-8.5]  & (7.75) & 1.31&1.32&   1.32 & $25.4$ & $26.7$ & Ex. ERO\\
& & & &  &  &  &  & [0.8-2.5] & 1.78 & - & - & - & - & - &   \\
\#13& 1:03.125&51:28.81& $ > 26.90$ & $24.41\pm 0.38$ & $23.58\pm 0.38$ & $ > 24.70$ & A & [8.0-9.2] & 8.29 & 2.2&2.26&   2.24 & $14.9$ & $15.6$ & \\
\#14& 1:04.209&51:54.55& $ > 26.90$ & $24.77\pm 0.52$ & $23.63\pm 0.39$ & $24.53\pm 0.97$ & B & [7.3-10.0] & 8.54 & 3.35&3.5&   3.45 & $ 9.7$ & $10.2$ & \\
\#15& 1:02.540&51:12.84& $ \underline{> 26.90}$ & $25.40\pm 0.94$ & $23.63\pm 0.40$ & $ \underline{> 24.70}$ & B & [7.5-10.5] & 8.52 & 1.81&1.84&   1.84 & $17.9$ & $18.8$ & \\
\#16& 1:03.657&52:54.83& $ > 26.90$ & $ > 25.60$ & $23.64\pm 0.40$ & $23.08\pm 0.25$ & C & [8.0-10.0] & (9.00) & 119.99&55.04& $>25$ & $< 1.6$ & $< 1.6$ & \\
\#17& 1:05.013&50:27.11& $ \underline{> 26.90}$ & $ > 25.60$ & $23.71\pm 0.43$ & $22.06\pm 0.10$ & C & [1.48-5.05] & 2.53 & - & - & - & - & - & Ex. ERO\\
\#22& 1:02.551&51:30.06& $25.00\pm 0.24$ & $23.99\pm 0.25$ & $23.81\pm 0.47$ & $22.59\pm 0.16$ & A & [5.5-8.1] & 7.68 & 2.28&2.34&   2.31 & $17.8$ & $18.7$ & \\
\#23& 1:05.699&51:52.92& $ > 26.90$ & $24.93\pm 0.61$ & $23.85\pm 0.48$ & $24.03\pm 0.61$ & C & [8.0-10.0] & (9.00) & 2.61&2.69&   2.67 & $12.0$ & $12.7$ & \\
\#24& 0:58.036&51:29.09& $ > 26.90$ & $25.16\pm 0.75$ & $23.88\pm 0.50$ & $ > 24.70$ & C & [8.0-10.0] & (9.00) & 1.6&1.62&   1.62 & $19.3$ & $20.2$ & \\
\#27& 1:04.299&51:57.19& $ > 26.90$ & $ > 25.60$ & $23.93\pm 0.53$ & $24.57\pm 1.01$ & C & [8.0-10.0] & (9.00) & 3.54&3.71&   3.68 & $ 8.1$ & $ 8.5$ & \\
\hline\\
\multicolumn{4}{l}{Second-category dropouts}\\
\hline\\
\#6& 0:59.659&50:54.73& $ > 26.90$ & $ > 25.60$ & $23.37\pm 0.31$ & $ > 24.70$ & B & [8.0-10.0] & (9.00) & 1.46&1.47&   1.47 & $33.8$ & $35.5$ & \\
\#18& 0:58.890&51:02.47& $ > 26.90$ & $ > 25.60$ & $23.72\pm 0.43$ & $ > 24.70$ & C & [8.0-10.0] & (9.00) & 1.47&1.48&   1.49 & $24.4$ & $25.6$ & \\
\#19& 1:00.138&52:05.20& $ \underline{> 26.90}$ & $ > 25.60$ & $23.72\pm 0.43$ & $ \underline{> 24.70}$ & C & [8.0-10.0] & (9.00) & 5.48&6.03&   5.92 & $ 6.1$ & $ 6.4$ & \\
\#20& 0:58.860&51:23.85& $ > 26.90$ & $ > 25.60$ & $23.72\pm 0.43$ & $ > 24.70$ & C & [8.0-10.0] & (9.00)  & 1.67&1.69&   1.69 & $21.3$ & $22.4$ & \\
\#21& 0:58.732&51:53.86& $ > 26.90$ & $ > 25.60$ & $23.76\pm 0.44$ & $ > 24.70$ & C & [8.0-10.0] & (9.00) & 2.13&2.18&   2.17 & $16.1$ & $16.9$ & \\
\hline\\
\#35& 1:00.693&52:09.58& $ > 26.90$ & $ > 25.60$ & $24.00\pm 0.56$ & $24.25\pm 0.75$ & C & --- & \textbf{1.68} & - & - & - & - & - & (3)\\
\hline\\
\end{tabular}
}

\caption{\label{table_a1835}Photometric properties of Abell 1835 optical-dropouts. 
From left to right : identification number, astrometric position, near-IR photometry, photometric redshift quality (see Sect. \ref{hyperz} for details),
redshift range $[z1-z2]$ and applied redshift $\tilde{z}$, magnifications at $z=6$ ($\mu_{6}$), at $z=10$ ($\mu_{10}$),
applied magnification ($\tilde{\mu}$), unlensed luminosity at rest-frame $\lambda=1500$ \AA\ .}
Objects noted ``ERO'' satisfy the $(R-K>5.6)$ criterion for Extremely Red
Objects. For EROs with a best-fit solution at high-$z$, the low-$z$ solution is also provided. 
``Ex'' is mentioned in the notes for objects excluded from the discussion, and underlined photometric 
entries correspond to forced undetections in a given near-IR band after visual inspection 
(see text for details).
All undetections are displayed with photometric upper limits, using the 1
$\sigma$ limiting magnitudes inside a 1.5 \arcsec \ diameter aperture given in
table ~\ref{images}. 

Bracketed values of $\tilde{z}$ are rough values based on the color-color selection diagrams.\\

$^1$ Near-IR counterpart of the sub-mm source SMMJ14009+0252 (Ivison et al.\ 2000, Smail et al.\ 2002).\\
$^2$ Also named A1835\#1916 (Pell\'o et al.\ 2004a)\\
$^3$ Also named A1835\#2582 (Richard et al.\ 2003)

\end{table}
\clearpage
\begin{table}[ht]
{\scriptsize
\begin{tabular}{lllcccccccccccccl}
\hline
N & RA & DEC & $J$ & $H$ & $K$ & $\phi\ z$  & $z$ range & $\tilde{z}$ & $\mu_6$ & $\mu_{10}$ & $\tilde{\mu}$ & $L_{1500}$ & $SFR$ & notes\\
 &(22:)&(-34:)& & & & & & & & & & $10^{40} erg\ s^{-1}\ $\AA$^{-1}$ & $M_{\odot}\ yr^{-1}$ & \\
\hline\hline\\
\multicolumn{4}{l}{First-category dropouts}\\
\hline\\
\#1& 58:49.777&46:54.95 & $22.19\pm 0.05$ & $20.52\pm 0.02$ & $19.23\pm 0.00$ & A & [1.58-1.89] & 1.62 & - & - & - & - & - & Ex., ERO\\
\#2& 58:49.040&47:21.94 & $24.14\pm 0.32$ & $23.01\pm 0.22$ & $ \underline{> 24.30}$ & B & [6.3-9.2] & 6.90 & 2.75&2.86&   2.78 & $10.2$ & $10.7$ & \\
\#4& 58:46.829&47:43.83 & $ \underline{> 25.50}$ & $23.33\pm 0.30$ & $24.23\pm 0.99$ & B & [8.7-11.8] & 9.82 & 13.68&10.98&  11.20 & $ 6.1$ & $ 6.4$ & \\
\#5& 58:46.505&47:25.96 & $24.90\pm 0.64$ & $23.41\pm 0.32$ & $ > 24.30$ & B & [6.0-10.0] & 8.51 & 10.1&12.62&  12.10 & $ 3.3$ & $ 3.5$ & \\
\hline\\
\multicolumn{4}{l}{Second-category dropouts}\\
\hline\\
\#13& 58:44.972&49:17.27 & $ > 25.50$ & $23.72\pm 0.43$ & $23.67\pm 0.59$ & C & [8.0-10.0] & (9.00)  & 1.54&1.57&   1.57 & $23.1$ & $24.2$ & \\
\#14& 58:53.511&48:37.85 & $24.55\pm 0.46$ & $23.73\pm 0.44$ & $ > 24.30$ & A & [5.2-9.6] & 6.91 & 2.73&2.84&   2.76 & $ 7.1$ & $ 7.4$ & \\
\#16& 58:50.243&48:35.75 & $24.55\pm 0.46$ & $23.90\pm 0.51$ & $ \underline{> 24.30}$ & A & [5.2-9.6] & 6.88 & 23.21&15.42&  21.35 & $ 0.9$ & $ 0.9$ & \\
\#19& 58:43.844&47:35.31 & $ > 25.50$ & $23.98\pm 0.55$ & $23.69\pm 0.60$ & C & [8.0-10.0] & (9.00)  & 3.31&3.51&   3.49 & $ 8.1$ & $ 8.6$ & \\
\hline\\
\multicolumn{4}{l}{Third-category dropouts}\\
\hline\\
\#15& 58:53.529 &49:13.45 & $ > 25.50$ & $23.75\pm 0.45$ & $ > 24.30$ & C & [8.0-10.0] & (9.00)  & 4.88&5.31&   5.24 & $ 6.7$ & $ 7.1$ & \\
\#17& 58:47.551 &48:53.51 & $ > 25.50$ & $23.93\pm 0.53$ & $ > 24.30$ & C & [8.0-10.0] & (9.00) & 2.85&2.97&   2.96 & $10.1$ & $10.6$ & \\
\hline\\
\end{tabular}
}

\caption{\label{table_ac114}Photometric properties of AC114 candidates. Caption is the same as in Table \ref{table_a1835}. }
\end{table}
\endlandscape
\end{document}